\documentclass[twocolumn, preprint, times]{aastex62}

\pdfoutput=1 

\usepackage{amsmath, amssymb}
\usepackage{hyperref}
\usepackage{soul}
\usepackage{tabularx}
\usepackage{amsmath}
\usepackage{upgreek}

\usepackage[latin1]{inputenc}
\usepackage{tikz}
\usetikzlibrary{shapes,arrows}

\usepackage{capt-of}
\usepackage{graphicx}
\usepackage{subfigure}

\usepackage{scalerel}
\usepackage{tikz}
\usetikzlibrary{svg.path}

\definecolor{orcidlogocol}{HTML}{A6CE39}
\tikzset{
  orcidlogo/.pic={
    \fill[orcidlogocol] svg{M256,128c0,70.7-57.3,128-128,128C57.3,256,0,198.7,0,128C0,57.3,57.3,0,128,0C198.7,0,256,57.3,256,128z};
    \fill[white] svg{M86.3,186.2H70.9V79.1h15.4v48.4V186.2z}
                 svg{M108.9,79.1h41.6c39.6,0,57,28.3,57,53.6c0,27.5-21.5,53.6-56.8,53.6h-41.8V79.1z M124.3,172.4h24.5c34.9,0,42.9-26.5,42.9-39.7c0-21.5-13.7-39.7-43.7-39.7h-23.7V172.4z}
                 svg{M88.7,56.8c0,5.5-4.5,10.1-10.1,10.1c-5.6,0-10.1-4.6-10.1-10.1c0-5.6,4.5-10.1,10.1-10.1C84.2,46.7,88.7,51.3,88.7,56.8z};
  }
}

\newcommand\orcidicon[1]{\href{https://orcid.org/#1}{\mbox{\scalerel*{
\begin{tikzpicture}[yscale=-1,transform shape]
\pic{orcidlogo};
\end{tikzpicture}
}{|}}}}

\usepackage{hyperref}

\begin{document}

\title{A Polarization Pipeline for Fast Radio Bursts Detected by CHIME/FRB}


\newcommand{\mcgillphysics}{Department of Physics, McGill University, 3600 rue University, Montr\'eal, QC H3A 2T8, Canada}
\newcommand{\msi}{McGill Space Institute, McGill University, 3550 rue University, Montr\'eal, QC H3A 2A7, Canada}
\newcommand{\cita}{Canadian Institute for Theoretical Astrophysics, 60 St. George Street, Toronto, ON M5S 3H8, Canada}
\newcommand{\dunlapinstitute}{Dunlap Institute for Astronomy \& Astrophysics, University of Toronto, 50 St. George Street, Toronto, ON M5S 3H4, Canada}
\newcommand{\dunlapdep}{David A. Dunlap Department of Astronomy \& Astrophysics, University of Toronto, 50 St. George Street, Toronto, ON M5S 3H4, Canada}
\newcommand{\mitkavli}{MIT Kavli Institute for Astrophysics and Space Research, Massachusetts Institute of Technology, 77 Massachusetts Ave, Cambridge, MA 02139, USA}
\newcommand{\mitphysics}{Department of Physics, Massachusetts Institute of Technology, 77 Massachusetts Ave, Cambridge, MA 02139, USA}
\newcommand{\ubc}{Department of Physics and Astronomy, University of British Columbia, 6224 Agricultural Road, Vancouver, BC V6T
23 1Z1 Canada}
\newcommand{\sidrat}{Sidrat Research, PO Box 73527 RPO Wychwood, Toronto, Ontario, M6C 4A7, Canada}

\author{R. Mckinven \orcidicon{0000-0001-7348-6900}}
\affiliation{\dunlapdep}
\affiliation{\dunlapinstitute}
\author{D. Michilli \orcidicon{0000-0002-2551-7554}}
\affiliation{\mcgillphysics}
\affiliation{\msi}
\author{K. Masui \orcidicon{0000-0002-4279-6946}}
\affiliation{\mitkavli}
\affiliation{\mitphysics}
\author{D. Cubranic \orcidicon{0000-0003-2319-9676}}
\affiliation{\ubc}
\author{B. M. Gaensler \orcidicon{0000-0002-3382-9558}}
\affiliation{\dunlapinstitute}
\affiliation{\dunlapdep}
\author{C. Ng \orcidicon{0000-0002-3616-5160}}
\affiliation{\dunlapinstitute}
\author{M. Bhardwaj \orcidicon{0000-0002-3615-3514}}
\affiliation{\mcgillphysics}
\affiliation{\msi}
\author{C. Leung \orcidicon{0000-0002-4209-7408}}
\affiliation{\mitkavli}
\affiliation{\mitphysics}
\author{P. J. Boyle \orcidicon{0000-0001-8537-9299}}
\affiliation{\mcgillphysics}
\affiliation{\msi}
\author{C. Brar \orcidicon{0000-0002-1800-8233}}
\affiliation{\mcgillphysics}
\affiliation{\msi}
\author{T. Cassanelli \orcidicon{0000-0003-2047-5276}}
\affiliation{\dunlapdep}
\affiliation{\dunlapinstitute}
\author{D. Li \orcidicon{0000-0001-7931-0607}}
\affiliation{\cita}
\author{J. Mena-Parra \orcidicon{0000-0002-0772-9326}}
\affiliation{\mitkavli}
\author{M. Rahman \orcidicon{0000-0003-1842-6096}}
\affiliation{\dunlapinstitute}
\affiliation{\sidrat}
\author{I. H. Stairs \orcidicon{0000-0001-9784-8670}}
\affiliation{\ubc}

\correspondingauthor{R. Mckinven}
\email{mckinven@astro.utoronto.ca}

\begin{abstract}
Polarimetric observations of Fast Radio Bursts (FRBs) are a powerful resource for better understanding these mysterious sources by directly probing the emission mechanism of the source and the magneto-ionic properties of its environment. We present a pipeline for analysing the polarized signal of FRBs captured by the triggered baseband recording system operating on the FRB survey of The Canadian Hydrogen Intensity Mapping Experiment (CHIME/FRB). Using a combination of simulated and real FRB events, we summarize the main features of the pipeline and highlight the dominant systematics affecting the polarized signal. We compare parametric (QU-fitting) and non-parametric (rotation measure synthesis) methods for determining the Faraday rotation measure (RM) and find the latter method susceptible to systematic errors from known instrumental effects of CHIME/FRB observations. These errors include a leakage artefact that appears as polarized signal near $\rm{RM\sim 0 \; rad \, m^{-2}}$ and an RM sign ambiguity introduced by path length differences in the system's electronics. We apply the pipeline to a bright burst previously reported by \citet[FRB 20191219F;][]{Leung2021}, detecting an $\mathrm{RM}$ of $\rm{+6.074 \pm 0.006 \pm 0.050 \; rad \, m^{-2}}$ with a significant linear polarized fraction ($\gtrsim0.87$) and strong evidence for a non-negligible circularly polarized component. Finally, we introduce an RM search method that employs a phase-coherent de-rotation algorithm to correct for intra-channel depolarization in data that retain electric field phase information, and successfully apply it to an unpublished FRB, FRB 20200917A, measuring an $\mathrm{RM}$ of $\rm{-1294.47 \pm 0.10 \pm 0.05 \; rad \, m^{-2}}$ (the second largest unambiguous RM detection from any FRB source observed to date).  

\end{abstract}

\section{Introduction}


Fast radio bursts (FRBs) are
microsecond to millisecond radio transients with integrated free electron column densities (quantified by the dispersion measure, DM) that significantly exceed the maximum value expected by Galactic models \citep{Cordes2002, Yao2017}. Since the discovery of this class of radio transient \citep{Lorimer2007}, various dedicated surveys have collectively amassed a public record of over one hundred confirmed FRB detections \citep{Petroff2020}\footnote{https://wis-tns.weizmann.ac.il/}\footnote{This sample will increase by $\sim 500$ events with the forthcoming release of the CHIME/FRB catalog (submitted).}, with a handful of these detected with interferometers with sufficiently long baselines for host galaxies to be determined, confirming their extragalactic origins \citep{Chatterjee2017,Bannister2019,Marcote2020,Macquart2020}\footnote{http://frbhosts.org/}. Despite these observations, the underlying mechanism driving FRB emission remains a mystery and has motivated a wide variety of emission models \citep{Platts2018}\footnote{https://frbtheorycat.org}. The recent discovery of FRB-like bursts from Galactic magnetar SGR 1935+2154 \citep{chime2020b, Bochenek2020} has demonstrated that at least some fraction of FRBs detected to date may have extragalactic magnetars as their central engine. 

The polarized signal of FRBs contain additional information not captured by total intensity data, potentially elucidating the nature of the FRB source through its intrinsic properties and the imprint imposed on the signal from magneto-ionic properties of the intervening plasma \citep[e.g.,][]{Masui2015}.     
Many of the early FRB detections were conducted in observing modes that did not record polarization information, thus, only a small fraction of the current published sample reports polarization information. 
This subsample is generally found to be highly linearly polarized. 
Exceptions include the significant circular \citep{Petroff2015,Caleb2018} or marginal linear \citep{Petroff2015,Keane2016} polarized fractions observed in some FRBs. The Faraday rotation measures (RMs), with the exception of the extremely high RM observed from FRB 20121102A \citep{Michilli2018}, are generally moderate (i.e., $\lesssim$ several hundred $\mathrm{\; rad \, m^{-2}}$), but are systematically larger than expected for randomly distributed inclination angles through the ISM of a host galaxy\footnote{For instance, assuming a Milky Way like host galaxy requires ISM properties many orders of magnitude greater than current constraints.}. This suggests a supplemental source of Faraday rotation from a dense magnetized medium associated with the FRB population as has been established for individual sources \citep[e.g.][]{Masui2015,Michilli2018}.   
Meanwhile, a wide variety of behavior has been observed in the intrinsic polarization angle ($\psi_0$) over the burst duration, referred to elsewhere as polarization position angle curves. Notably, repeating sources seem to display a preference for a flat $\psi_0$ curve \citep{Michilli2018,chime2019,Fonseca2020}
that is contrasted by the significant evolution seen in (apparently) non-repeating sources \citep[e.g.,][]{Masui2015,Cho2020,Day2020}. Recently, \citet{Luo2020} observed changes in $\psi_0$ across individual bursts from repeating source FRB 20180301A \footnote{An anomalously discrepant RM is also reported here relative to an ambiguous RM detection from the original discovery burst of this source \citep{Price2019}.}, challenging the notion that $\psi_0$ curves could function as a discriminant between repeating and non-repeating samples.  

The extent to which these anomalous features are a product of random variance of a single population or indicative of multiple populations with different intrinsic properties and source environments remains an open question. 
The enhanced statistical analysis enabled by a large sample of observations of FRB polarization should greatly increase our understanding of the population.
Indeed, anticipating the added benefit of polarized information, it is now common practice for most radio-transient surveys to observe in modes that retain polarized information. The FRB project of the Canadian Hydrogen Intensity Mapping Experiment \citep[CHIME/FRB;][]{chime2018} is one such survey, consisting of a real-time detection system that searches 1,024
formed sky beams for dispersed single pulses in the frequency range $\mathrm{400-800 MHz}$, with time resolution 0.983 ms and 16k frequency channels. The CHIME instrument is composed of four 20 m $\times$ 100 m semicylindrical paraboloid reflectors, with each reflector possessing 256 dual-polarization feeds suspended along the N-S axis. A triggered baseband recording system is capable of recording the electric field as measured by each of these feeds in the CHIME array. The phase information contained in the baseband data can be used to phase-reference detected waves to any direction within the field of view of the telescope prior to coadding over feeds, significantly improving localization capabilities down to sub-arcminute precision \citep{Michilli2021}. More relevant to this paper's focus, the complex voltage signal recorded in each of the dual, linear feeds retain the full polarization information and is natively recorded at a much higher time resolution than the intensity data used in the real-time search pipeline. 

With a detection rate of several FRBs per day (CHIME/FRB Collaboration 2021, in prep.), CHIME/FRB should detect several thousand FRBs over the next few years of operation. A large fraction of these events will have corresponding baseband data. Such a large sample requires an automated pipeline for reducing baseband recordings into useful summary statistics for different FRB properties.
In this paper we summarize the pipeline for processing beamformed baseband data into polarized observables. Section~\ref{sec:pol_rev} summarizes the definitions of important polarization observables, Section~\ref{sec:data} briefly describes the input data, Section~\ref{sec:methods} describes different RM detection methods and how they perform under different circumstances. Section~\ref{sec:pipeline} provides an overview of the pipeline responsible for processing the polarized signal of FRBs, Section~\ref{sec:examples} illustrates the pipeline using real CHIME/FRB observations and is followed by a discussion in Section~\ref{sec:discussion} and conclusion in Section~\ref{sec:conclusion}.

\section{Review of Derived Polarized Quantities}
\label{sec:pol_rev}

\subsection{Stokes Parameters and Polarization Angle}

The polarization of an electromagnetic wave relates to the preferred geometric orientation of its oscillating electric and magnetic fields. By convention, the polarization of an electromagnetic wave is determined by the direction of the electric field. In the case of fully linearly polarized radiation, oscillations in the electric field occur entirely along a single direction that,
combined with the axis of propagation, defines the plane of polarization of the emission. Circular polarization, meanwhile, refers to a the case where the fields rotate in the plane perpendicular to the direction of propagation, with the direction of rotation determining the ``handedness" of the polarization. In this way, unlike linear polarization, circular polarization can have either a negative or positive sign. 

A convenient way of representing the different geometries of the polarized emission is to transform the complex electric field into Stokes parameters. The geometry of CHIME's feed design is consistent with the IAU/IEEE convention \footnote{https://www.iau.org/static/archives/announcements/pdf/ann16004a.pdf} where the $X$ and $Y$ linear feeds point towards the east and
north, respectively. In this convention, Stokes $I, Q, U$ and $V$ parameters can be obtained by applying the transformations,

\begin{align}
\begin{split}
&I=\langle|X|^2+|Y|^2\rangle \\
&Q=\langle|X|^2-|Y|^2\rangle \\
&U=\langle2 \, real(XY^*)\rangle \\
&V=\langle-2 \, imag(XY^*)\rangle. 
\end{split}
\label{eqn:stokes}
\end{align}
Here, Stokes $I$ refers to the total intensity of the emission, Stokes $Q$ and $U$ correspond to the linearly polarized component and Stokes $V$ refers to the circularly polarized component.

The observed polarization angle, $\psi$, can be expressed in terms of Stokes $Q$ and $U$ parameters, such that,

\begin{equation}
\psi(t, \nu) = \frac{1}{2}\tan^{-1}\frac{U(t,\nu)}{Q(t,\nu)} \qquad [\rm{rad}].
\label{eqn:chi}
\end{equation}
Units here and elsewhere are denoted by $[\;]$. 

Equation~\ref{eqn:chi} has been expressed in terms of time ($t$) and frequency ($\nu$). This allows for the possibility of a change in $\psi$ over the burst envelope (see Section~\ref{sec:PPA}) or across the spectrum that can either be intrinsic to the source or introduced later as a propagation effect, such as Faraday rotation (see Section~\ref{sec:faraday rotation}). Intrinsic variations in $\psi$ may be produced by a radius-to-frequency mapping (RFM) similar to what has been posited for pulsars, where emission occurs at different altitudes within the magnetosphere \citep[e.g.,][]{Thorsett1991,Mitra2002,Noutsos2015}. Although there has been some work done exploring the applicability of RFM in describing certain FRB phenomena \cite[e.g., FRB frequency drifts;][]{Lyutikov2020}, the validity of such a model remains uncertain.  

\subsection{Faraday Rotation}
\label{sec:faraday rotation}

Faraday rotation (quantified by the rotation measure, RM) is a magneto-optical propagation effect observed as a rotation of the plane of polarization that is linearly proportional to the square of the wavelength, such that,

\begin{equation}
\mathrm{RM}=\frac{d\psi}{d\lambda^2} \qquad  [\mathrm{rad \; m^{-2}}].
\label{eqn:RMexp}
\end{equation}   
Here, $\mathrm{RM}$, $\psi$ and $\lambda$ are the rotation measure, polarization angle and observing wavelength, respectively. The RM is proportional to the magnetic field parallel to the line-of-sight (LOS) weighted by the free electron density and integrated along the path between the source and observer. Specifically,
for an FRB located at a redshift $z = z_i$, the RM in the observer's frame is,

\begin{equation}
\mathrm{RM}=C_R\int_{z_i}^{0} \frac{n_e(z) B_{\parallel}(z)}{(1+z)^2} \frac{dl}{dz} dz \qquad  [\mathrm{rad \; m^{-2}}],
\label{eqn:RMth}
\end{equation}
where $C_R=811.9 \; \mathrm{rad\;m}^{-2} / (\mu \mathrm{G \, pc\, cm}^{-3})$, $z$ is redshift, $n_e$ is the free electron density, $B_{\parallel}$ is the magnetic field strength parallel to the LOS, and $dl(z)$ is the LOS line element at $z$.


The RM, therefore, is an integrated quantity that when combined with the DM, can be used to estimate the average magnetic field strength of intervening plasma \citep[e.g.,][]{Akahori2016}. The extragalactic nature of FRBs implies contributions to the RM from not only the Milky Way's interstellar medium (ISM) and the surrounding Galactic halo but also the intergalactic medium (IGM), intervening systems such as individual galaxies and/or groups/clusters, and finally, the host galaxy and local circum-burst environment. 

\subsection{Polarization Position Angle}
\label{sec:PPA}

The polarization position angle (PPA) corresponds to the polarization angle of the emission at the source as a function of time. The PPA is commonly measured in radio pulsars where a characteristic S-shaped PPA curve is often observed and interpreted within the popular rotating vector model of pulsar emission \citep{Radhakrishnan1969}. In this way, it is different from the observed polarization angle (see Equation~\ref{eqn:chi}) in that it characterizes the geometry of the polarized signal \textit{prior} to being modulated by Faraday rotation. The effect of Faraday rotation can be removed by using the measured RM to de-rotate the spectrum through a multiplicative phase factor such that,

\begin{multline}
\rm{[Q+iU]_{int}(\lambda,t)}=\rm{[Q+iU]_{obs}(\lambda,t)} \\
\times \exp{[\rm{2i(RM(\lambda^2- \lambda_{0}^2)+\psi_{0}(t))]}}
\end{multline}

Here, $\rm{[Q+iU]_{obs}}$ is the observed spectrum, $\rm{[{Q+iU}]_{int}}$ is the intrinsic polarization vector at the source, while RM and $\psi_0$ are fitted parameters. $\psi_0$ is the polarization position angle at a reference wavelength $\lambda_0$ (i.e., at infinite frequency or zero wavelength). In the case of calibrated polarized observations, $\psi_0$ is often referenced at infinite frequency where Faraday rotation is zero. In principal, any time dependence of $\psi_0$ can be determined by fitting the polarized signal through the burst duration. In practice, S/N limitations complicate this time-resolved analysis and are, in any case, unsuitable for an automated pipeline where robust methods of characterizing the polarized signal take precedence. An alternative method for characterizing time dependence in $\psi_0$ is to apply Equation~\ref{eqn:chi} to the burst profiles of the de-rotated Stokes $Q$, $U$ parameters such that,

\begin{equation}
\psi_0(t)=\frac{1}{2}\tan^{-1}\left( \frac{U_{\mathrm{derot}}(t)}{Q_{\mathrm{derot}}(t)}\right) \qquad [\rm{rad}].
\end{equation}
Here, $Q_{\mathrm{derot}}$ and $U_{\mathrm{derot}}$ are integrated over frequency to optimize the signal-to-noise of the $\psi_0$ measurement under the assumption that there is no frequency dependence in the intrinsic polarization angle at the source. Calculating the $\psi_0(t)$ curve in this way makes it less sensitive to measurement errors associated with Stokes $Q$ and $U$, yielding a more stable curve through the burst duration. 

\section{Baseband Data}
\label{sec:data}

As outlined by \citet{chime2018} and further elaborated on by \citet{Michilli2021}, the CHIME/FRB system possesses a baseband backend capable of recording the channelized voltages from each of the 1024 dual linear feeds. Channelization occurs through a Field Programmable Gate Array (FPGA) that implements a 4-tap polyphase filter bank \citep{Price2016} to produce a spectrum with 1024 channels (each 390 kHz wide) every
2.56 $\mu s$. A programmable gain and phase offset are applied to each frequency channel, and the data are rounded to 4 + 4 bit complex numbers.
The system is configured to automatically record baseband data for events detected by the real-time system through implementation of a memory buffer that, after accounting for system latency, allows storage of $\sim20$ seconds worth of baseband data. At CHIME frequencies and bandwidth, this roughly corresponds to a maximum DM of $\mathrm{\sim 1000 \, pc \, cm^{-3}}$ for full baseband callbacks. Triggered events with larger DMs result in incomplete recordings with missing data at the top of the band. 

Shortly after baseband data is recorded, a processing pipeline is launched and are composed of \textit{refinement}, \textit{localization} and \textit{analysis} stages \citep{Michilli2021}. Products from the pipeline include a refined DM and localization that maximize the event's signal-to-noise. A single, tied-array beam is formed in the direction of the refined localization and is used as input in the analysis stage of the pipeline along with other information from the preprocessing of the event (e.g., radio-frequency interference (RFI) channel mask, spectral window, etc.). Input data of the analysis stage therefore correspond to a matrix of complex voltages in frequency, polarization and time and are fed into a variety of scientific pipelines tailored to investigating different properties of detected bursts. A major component of the polarization pipeline is dedicated to characterizing the Faraday rotation. In the following section we summarize the RM detection methods currently implemented in the CHIME/FRB polarization pipeline and provide further details in Section~\ref{sec:pipeline}.    

\section{RM Detection Methods}
\label{sec:methods}

\begin{figure*}
\centering     
\subfigure[Stokes $I, Q, U$ \& $V$ dedispersed waterfall plots displaying Faraday rotation induced modulation in $Q$ and $U$. Red lines indicate window region over which Stokes parameters are integrated, corresponding to $20\%$ of of the bursts' peak value.]{\label{fig:a}\includegraphics[width=0.40\textwidth]{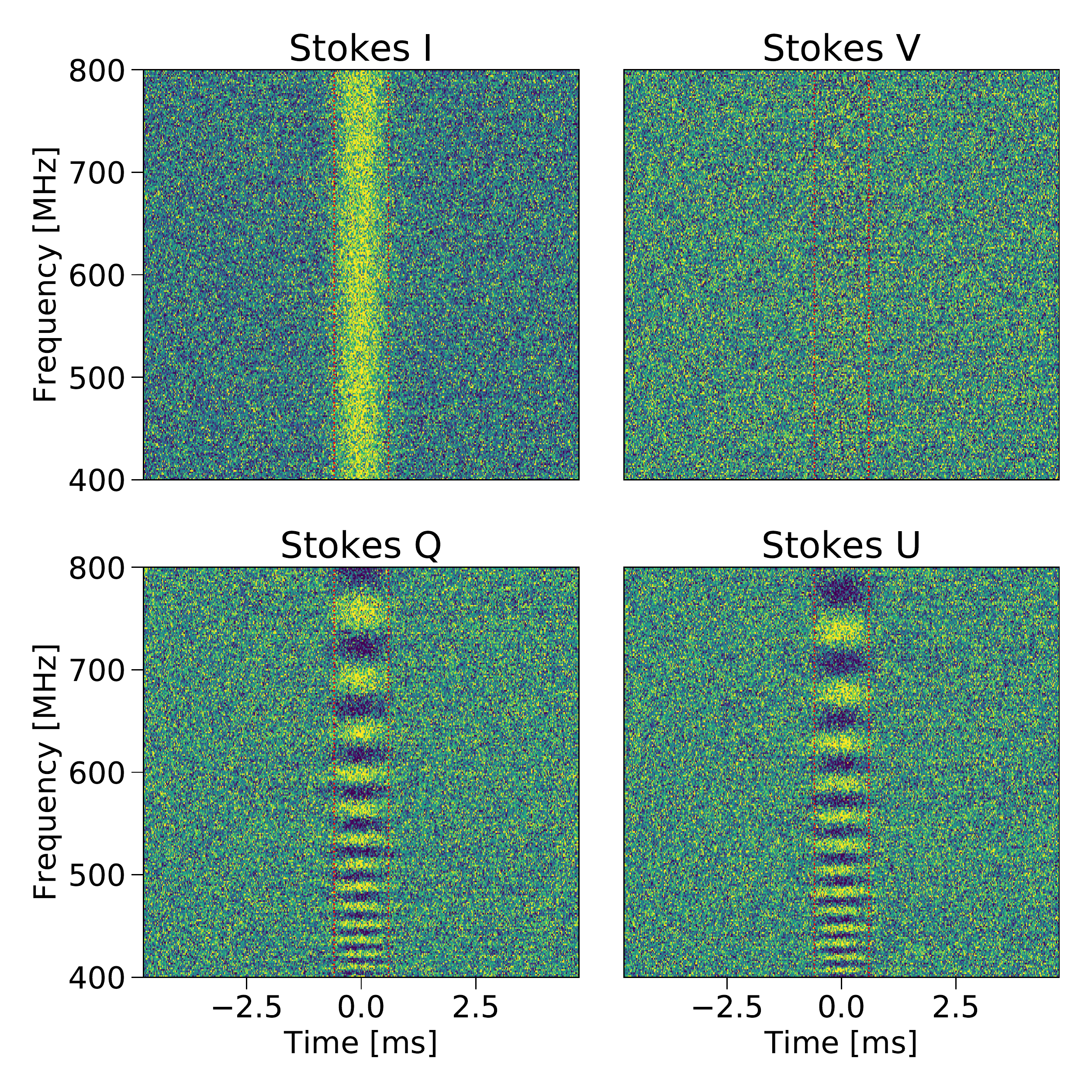}}
\hspace{0.05\textwidth}
\subfigure[The FDF over the full range of trial RMs (top) and over a constrained region near the nominal RM (bottom). Orange and blue lines correspond to the dirty and clean FDF, respectively, cleaned to a level of 8$\sigma$ (dashed line). ]{\label{fig:b}\includegraphics[width=0.40\textwidth]{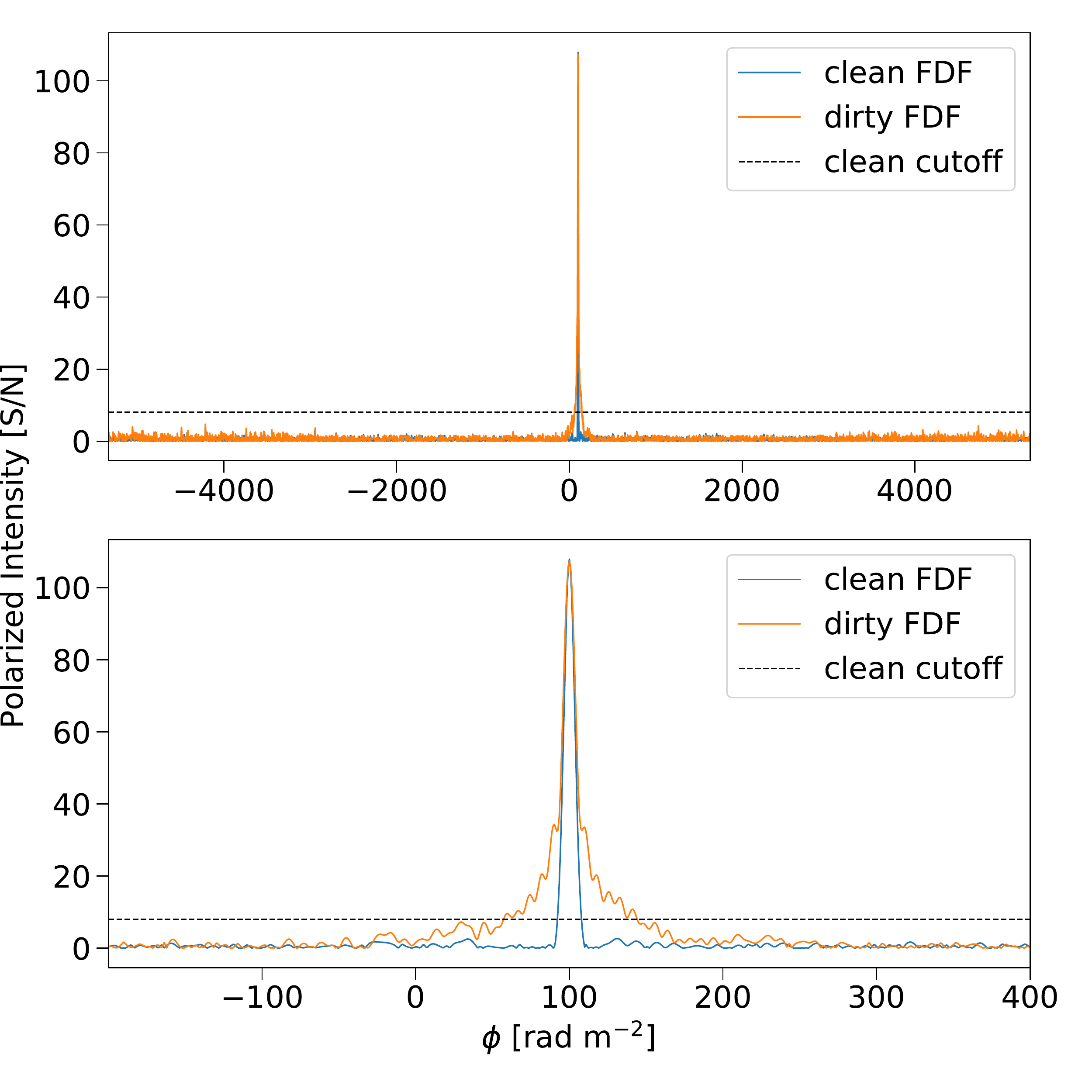}} \\
\subfigure[Stokes $I$, $Q$, $U$, $V$ and polarization angle, $\psi$, as a function of frequency. Model fits for the Stokes spectrum correspond to black lines while the model fit for $\psi(\nu)$ is indicated by the red line. These models are determined by substituting best fit parameter values (see panel d) into Equations~\ref{eqn:QUmodel}.]{\label{fig:c}\includegraphics[width=0.40\textwidth]{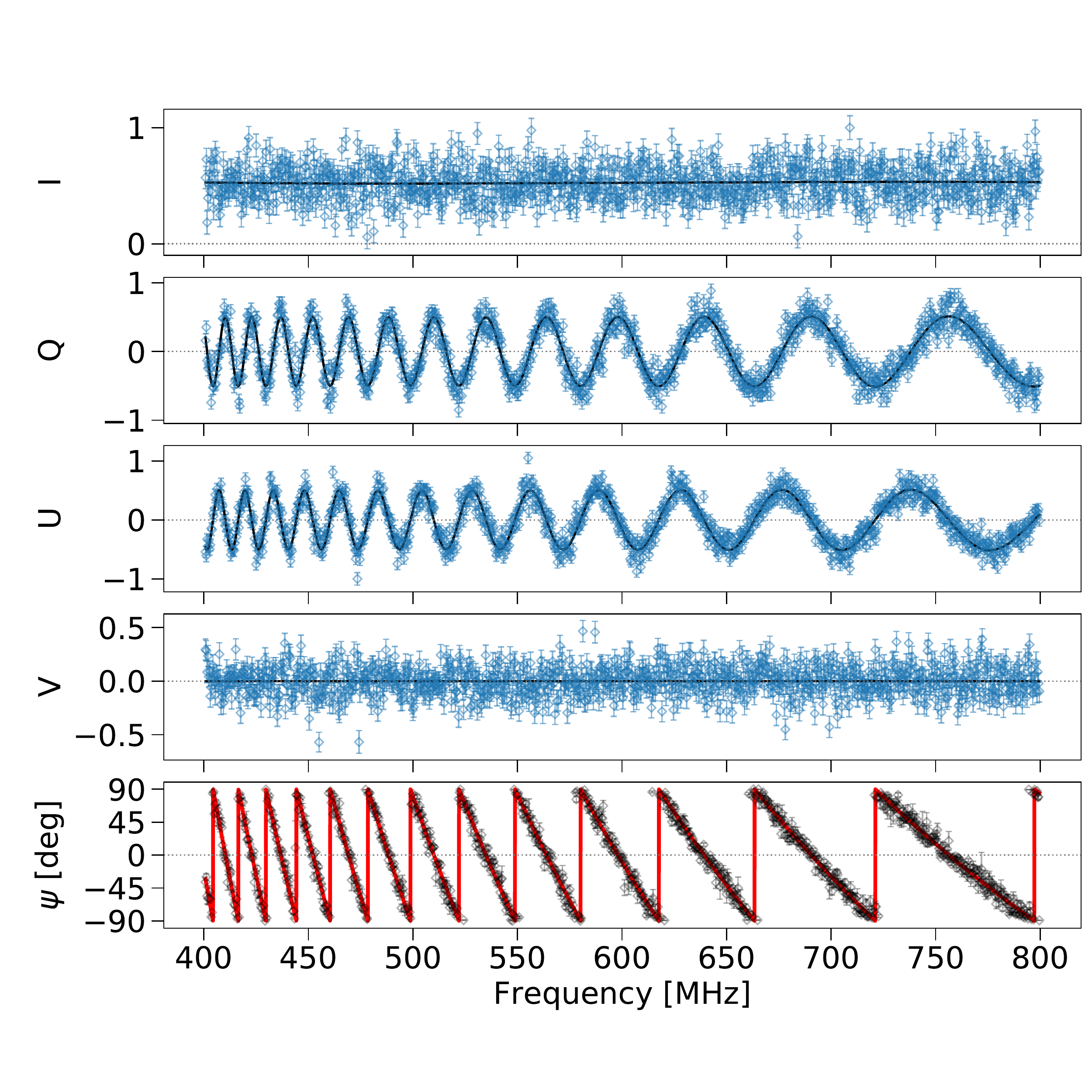}}
\hspace{0.05\textwidth}
\subfigure[Posterior distribution of fit parameters: $p$, $\psi_0$ and RM determined from the multinest sampling algorithm. Panels correspond to probability density functions for different projections through the parameter samples. Best fit values and $1\sigma$ uncertainties are indicated by vertical blue lines and black lines of corresponding marginal probability distributions.]{\label{fig:d}\includegraphics[width=0.40\textwidth]{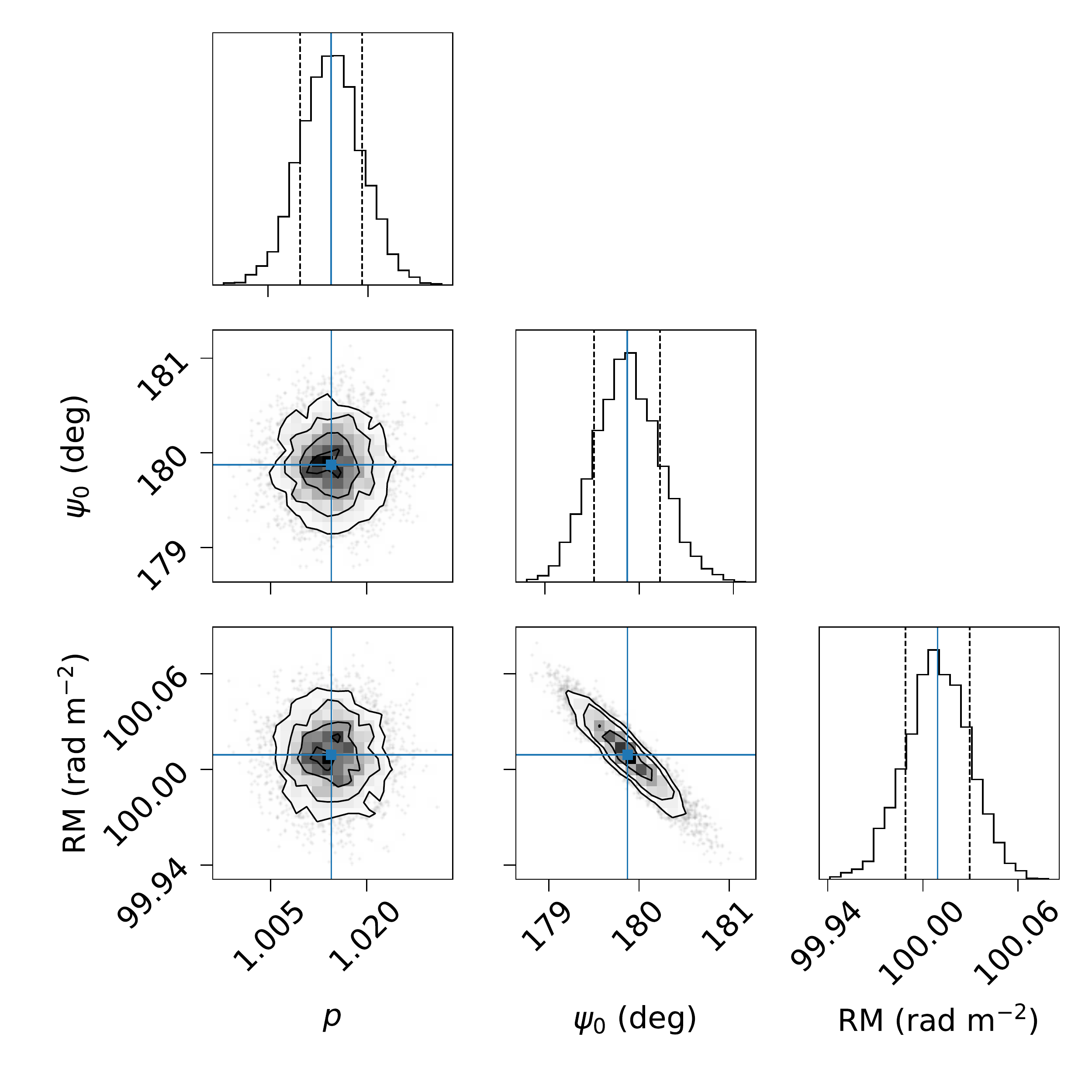}}
\caption{A simulated burst with input parameters: $p=1.00$ (linear polarization fraction), $\psi=180^\circ$ and $\rm{RM=+100 \; rad \, m^{-2}}$.}
\label{fig:sim_example1}
\end{figure*}

There are several methods for measuring the Faraday rotation or RM of a polarized astrophysical signal \citep[see][for a review]{Sun2015}. Although these methods can have different use cases, in the Faraday thin regime, all methods capitalize on the strict $\lambda^2$ scaling of the polarization angle ($\psi$; Equation~\ref{eqn:RMexp}). This property enables trial RM values to be evaluated by either directly fitting the modulation in the polarized signal or by applying a simple transformation that corrects for Faraday rotation across the bandpass. In this section, we review the popular RM detection methods of \textit{RM-synthesis} (Section~\ref{sec:RM-synth}) and \textit{QU-fitting} (Section~\ref{sec:qufitting}) currently implemented in the CHIME/FRB polarization pipeline. Both these methods are effective at detecting $\mathrm{|RM|}$ up to several thousand $\mathrm{rad \; m^{-2}}$. 
At values beyond this range, intra-channel or bandwidth depolarization becomes significant, leading to a partial loss of polarized signal. In Section~\ref{sec:coherent_derot} we review this effect and recapitulate a phase coherent technique that corrects for it in data that retain electric field phase information \citep{vanStraten2002}, effectively extending sensitivity to polarized signal out to very large $\mathrm{|RM|}$ values. 

\subsection{RM-synthesis}
\label{sec:RM-synth}

RM-synthesis \citep{Burn1966, Brentjens2005} is a robust technique for measuring Faraday rotation that amounts to a Fourier like transformation, such that,  

\begin{equation}
\mathbf{F}(\phi)=\int_{-\infty}^{\infty} P(\lambda^2)e^{-2i\phi\lambda^2} d\lambda^2.
\label{FDF}
\end{equation}
Here, $\phi$ is referred to as the Faraday depth and is an extension of RM for scenarios where the polarized signal is Faraday rotated by different amounts. $|\mathbf{F(\phi)}|$ is the total linearly polarized intensity across the bandpass after de-rotating the complex vector representing the observed linearly polarized intensity, $P(\lambda^2)=Q(\lambda^2)+iU(\lambda^2)$. 

Carrying this procedure over multiple $\phi$ values results in a \textit{Faraday Dispersion Function} (FDF), representing the polarized intensity at different trial values. Applying RM-synthesis to emission that occurs over an extended region of space often yields a complex FDF with substantial polarized emission at multiple Faraday depths \citep[e.g.,][]{Anderson2016, Dickey2019}. With FRB emission, the short millisecond time-scales strongly limit the amount of differential Faraday rotation that can occur within such a small emitting volume. In cases such as this, the FDF will appear as a single peak in polarized signal at a single Faraday depth. This regime is known as ``Faraday thin", where $\phi$ and RM are interchangeable terms.

Figure~\ref{fig:sim_example1} shows a simulated burst with RM = +100 rad m$^{-2}$. 
The effect of Faraday rotation can clearly be seen in the plot showing the burst as a function of frequency and time (waterfall plot) for Stokes $Q$ and $U$ of panel (a). Applying RM-synthesis to this spectrum produces the FDF of panel (b). The orange curve is the ``dirty" FDF for the event and includes both contributions from the signal as well as sidelobes introduced by the bandpass limitations of the observation, known as the rotation measure transfer function (RMTF). Sidelobes can be cleaned by applying an {\tt RM-CLEAN} algorithm \citep{Heald2009} that deconvolves the RMTF from the observed FDF in a manner analogous to the { \tt CLEAN} deconvolution routines applied in aperture synthesis radio imaging \citep{Hogbom1974,Clark1980}. The blue curve of panel (b) corresponds to the clean FDF. In the case where polarized emission is well described by a single RM, the best estimate of the RM will correspond to the $\phi$ value at which the FDF peaks. 


\subsection{QU-fitting} 
\label{sec:qufitting}

An alternative method for extracting an RM value is to invoke a model that can fit the oscillations in Stokes $Q$ and $U$ introduced by Faraday rotation. In the case where all polarized emission is Faraday rotated by a single RM value, the methods of Stokes QU-fitting and RM-synthesis are essentially identical. This is highlighted in Figure~\ref{fig:sim_example1}, where the RM determined by fitting the Stokes spectrum of panel (c) results in a fitted RM value, shown in panel (d), that is consistent with that independently determined via RM-synthesis. Panel (c) shows the Stokes $I$, $Q$, $U$ and $V$ spectra along corresponding models fits, obtained from a modified version of the {\tt RM-tools}\footnote{https://github.com/CIRADA-Tools/RM-Tools} software \citep{Purcell2020} that implements a Nested Sampling algorithm \citep{Skilling2004} to find the best-fitting parameters. 
Parameters in this simple benchmark model are the $\mathrm{RM}$, the polarization angle at infinite frequency, $\psi_0$, and the linear polarization fraction, $p$. In this simple model, Stokes $V$ is assumed to be zero while Stokes $I$ is fit by a fifth order polynomial ($\rm{I_{mod}}$) and is used as input in the fitting procedure applied to Stokes $Q$ and $U$. Models for Stokes $Q$ and $U$ can therefore be expressed as,

\begin{equation}
\begin{split}
    &Q_{\mathrm{mod}} = I_{\mathrm{mod}} p \cos(\alpha), \\
    &U_{\mathrm{mod}} = I_{\mathrm{mod}} p \sin(\alpha),
\label{eqn:QUmodel}
\end{split}
\end{equation}
where $\alpha = 2(\mathrm{RM} \lambda^2 + \psi_0)$ corresponds to the frequency dependent phase introduced by Faraday rotation. 

Optimal values are determined numerically through Nested Sampling, a Monte Carlo method for Bayesian analysis that simultaneously calculates both Bayesian evidences and posterior samples. This method benefits from more efficiently sampling the parameter space than conventional Markov Chain Monte Carlo (MCMC) based methods and is particularly useful for degenerate, multi-modal likelihoods. 
Parameter estimation seeks to optimize the likelihood function given a model and the data. Each data point in the fit is weighted by the inverse square of the RMS noise of the frequency channel. In this sense, model and data are compared such that $Q_i=Q_{\mathrm{mod},i}+n_i$ and $U_i=U_{\mathrm{mod},i}+n_i$, where $n_i$ is the Gaussian noise for channel i. Following the prescription of \citet{O'Sullivan2012}, the prior likelihood of particular RM and $\psi_0$ values for an observation of a single channel, $d_i$, under the assumption of Gaussian noise is,

\begin{multline}
P_i(d_i|RM,\psi_0)=\\
    \frac{1}{\pi\sigma_{Q_i}\sigma_{U_i}}\exp\left(-\frac{(Q_i-Q_{\mathrm{mod},i})^2}{2\sigma_{Q_i}}+\frac{(U_i-U_{\mathrm{mod},i})^2}{2\sigma_{U_i}}\right).
\end{multline}

Here, $\sigma_{Q,U}$ is the single channel RMS. For N frequency channels, the prior likelihood becomes,

\begin{equation}
P(d|RM,\psi_0)=\prod_{n=1}^{N}P_i(d_i|RM,\psi_0).
\label{eqn:likelihood}
\end{equation}
This formulation of the likelihood function ensures that parameter estimation is carried out by comparing absolute values of $Q$ and $U$ to model predictions. This results in more robust fit values compared to fitting the fractional polarization (i.e., $Q/I$, $U/I$), particularly for low S/N events where normalizing by total intensity can introduce substantial departures from Gaussianity in the noise. 

Parameter estimation is done through {\tt Multinest} software \citep{Feroz2009} that numerically searches for optimal parameter values that optimize the log-likelihood function. The resulting posterior distributions are shown in panel (d) of Figure~\ref{fig:sim_example1}. Along the diagonal are histograms of the estimated 1D marginal posterior probability distribution for each parameter. The best fit value and $1\sigma$ uncertainty region for each parameter are indicated by vertical blue lines and black dotted lines, respectively.    

A Faraday thin model, expressed mathematically in Equation~\ref{eqn:QUmodel}, is generally adequate for describing the polarized signal of an FRB and is employed in the automated polarization analysis pipeline of CHIME/FRB. Indeed, as with pulsars, FRBs are not likely to display Faraday complexity due to the small presumed size of the emission region over which very little internal Faraday rotation is likely to occur. However, in certain scenarios this may not be the case and the parametric nature of Stokes QU-fitting can be leveraged to fit for effects not contained in the simple Faraday rotation model. These excursions from a simple Faraday model can be produced by astrophysical or instrumental effects. The application of QU fitting to astrophysical excursions from a Faraday simple model are further discussed in Section~\ref{sec:discussion}. More relevant to the automated pipeline are the effects introduced by instrumental systematics, which strongly affect a significant fraction of FRBs detected by CHIME. The specifics of the QU-fitting implementation in the polarization pipeline and contaminant systematics are discussed further in Section~\ref{sec:pipeline}. 

\subsection{Coherent de-rotation}
\label{sec:coherent_derot}

In cases where $\rm{|RM|}$ values are large, a significant change in the polarization angle, $\psi$, can occur within a single frequency channel such that \citep[see Equation 4.12 of][]{Burke2014},

\begin{equation}
\delta \psi = \frac{-2\mathrm{RM_{obs}} c^2 \delta \nu}{\nu_c^3} \qquad [\rm{rad}].
\label{eqn:chan_rotation}
\end{equation}

Here, $\delta \psi$ corresponds to the degree of intra-channel Faraday rotation, $\rm{RM_{obs}}$ is the observed RM, $\delta \nu$ is the channel width and $\nu_c$ is the central frequency of the channel. 
Due to the strong frequency dependence in Equation~\ref{eqn:chan_rotation}, this effect becomes more pronounced at longer wavelengths. If $\rm{|RM|}$ and $\delta \nu$ are large enough, $\psi$ undergoes a large rotation within a frequency channel. The net effect is depolarization within each channel, with the level of depolarization dependent on observing frequency. This effect, known as intra-channel or bandwidth depolarization, limits the range of RM values to which any any instrument is sensitive, with detections of larger $\rm{|RM|}$ values either requiring finer frequency resolution or higher observing frequencies. The fractional depolarization can be approximated within each channel using \citep{Schnitzeler2015, Michilli2018},

\begin{equation}
f_{depol} = 1-\Bigg(\frac{\sin(\delta\psi)}{\delta\psi}\Bigg).\footnote{The factor of two difference with Equation 4 of \citet{Michilli2018} arises from different definitions of the $\delta\psi$ quantity.}
\label{eqn:depol}
\end{equation}

\begin{figure}[t]
\centering
\includegraphics[width=0.45\textwidth]{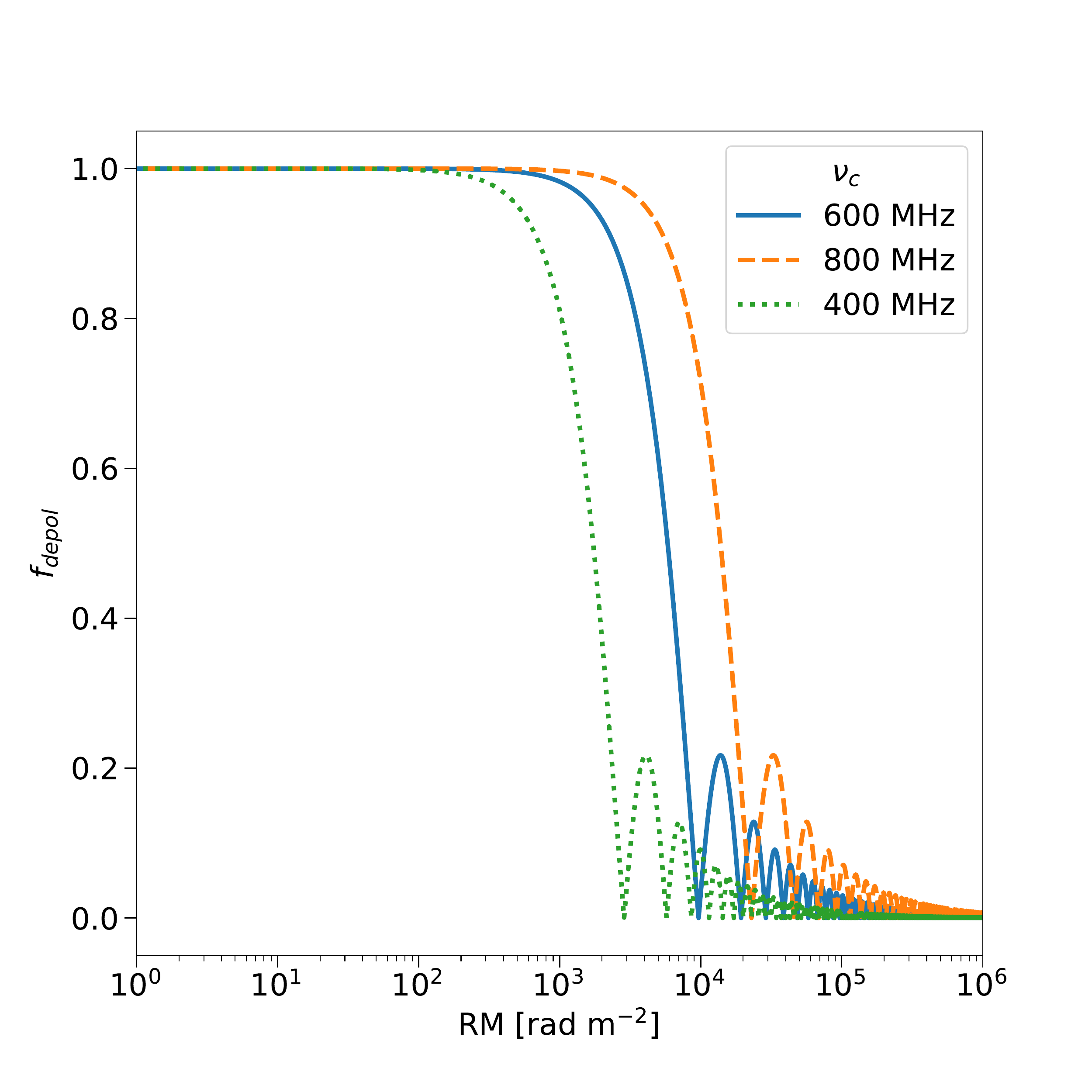}
\figcaption{The expected fractional depolarization ($f_{depol}$) as a function of RM, at the bottom (dotted green: $\nu_c=400$ MHz) middle (solid blue: $\nu_c=600$ MHz) and top (dashed orange: $\nu_c=800$ MHz) of the CHIME bandpass. The depolarization curves are calculated from Equations~\ref{eqn:chan_rotation} and \ref{eqn:depol} at the native spectral resolution of CHIME/FRB baseband data ($\delta\nu\approx 0.39$ MHz).}
\label{fig:upchan}
\end{figure}

In the case of CHIME, the relatively low observing band of 400-800 MHz and its modest frequency channel resolution of $\delta \nu$ = 390 kHz (i.e., 1024 channels) limits sensitivity to RM detections of several thousand rad m$^{-2}$, with the precise value depending on the S/N and spectrum of the event. Figure~\ref{fig:upchan} shows the expected fractional depolarization as a function of $\rm{|RM|}$. At $\nu_c=600$ MHz there is an approximately 50$\%$ drop in sensitivity to polarized emission at $\mathrm{RM\approx 5000 \; rad \, m^{-2}}$, effectively putting an upper limit on the RM range detectable at the native spectral resolution of CHIME/FRB baseband data. The exact level of bandwidth depolarization is dependent on the precise spectrum of each burst. This frequency dependence is highlighted by the depolarization shown in Figure~\ref{fig:upchan} where lower frequencies ($\nu=400$ MHz) are seen to be generally more significantly depolarized than higher frequencies ($\nu=800$ MHz) for a given RM.

Fortunately, baseband data retain the phase information of the incident electric field. This allows the limitations imposed by the native spectral resolution to be overcome, by re-sampling such that time resolution can be swapped for enhanced frequency resolution, a process we have dubbed ``up-channelization". Alternatively, the electric field phase allows us to correct for the frequency dependent phase offsets introduced by Faraday rotation. Formulating Faraday rotation as the result of the differing group velocities of the left and right circular polarization states allows us to express it as an additional dispersive effect operating differentially on the two circular bases. Expressed in this form, the correction for Faraday rotation is analogous to coherent dedispersion \citep{Hankins1971}, in which a transfer function is invoked that corrects for the phase change within frequency channels \citep{vanStraten2002}.

This method of coherently correcting for Faraday rotation amounts to a frequency dependent phase factor that is applied to the circular polarization basis pair ($\rm{|R\rangle,|L\rangle}$) such that,

\begin{equation}
\begin{split}
&|R^{'} \rangle = e^{-i\beta} |R\rangle \\
&|L^{'} \rangle = e^{i\beta} |L\rangle
\end{split}
\label{eqn:highrm}
\end{equation}
where $\rm{|R^{'}\rangle,|L^{'}\rangle}$ are the right and left circular polarized components, respectively, after correcting for the phase offset, $\beta$, introduced by Faraday rotation,
\begin{equation}
\beta  = \mathrm{RM}  \frac{c^2}{\nu^2}.
\label{eqn:transfer}
\end{equation}

\begin{figure}
\centering     
\subfigure[Dedispersed Stokes $I$, $Q$, $U$ \& $V$ waterfall plots uncorrected for intra-channel depolarization. Splitting of the left and right circular polarization modes can be seen in the panel displaying Stokes $V$.]
{\label{fig2:a}\includegraphics[width=0.35\textwidth]{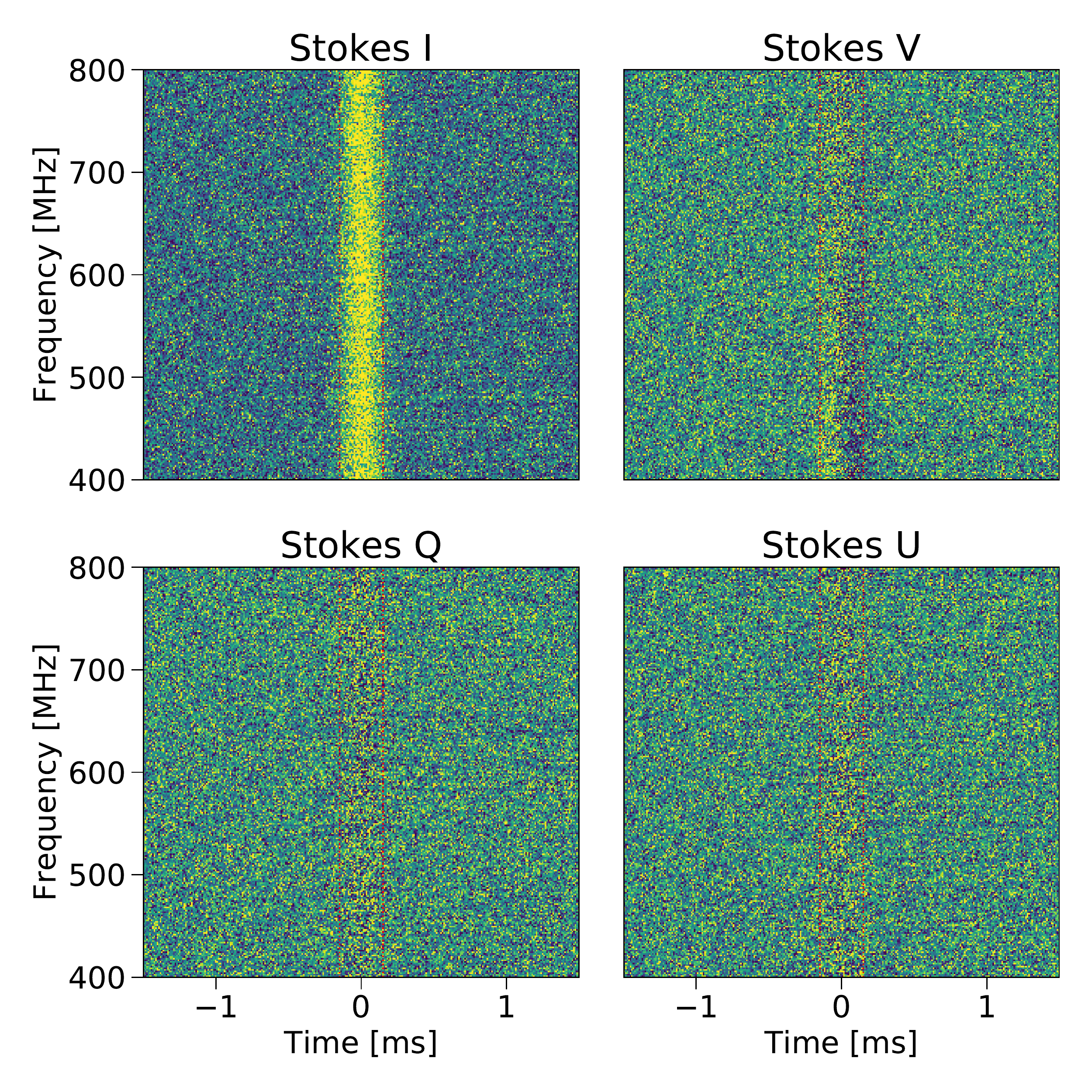}}
\subfigure[Corresponding waterfall plots corrected for intra-channel depolarization. The polarized signal is coherently de-rotated into Stokes $Q$.]{\label{fig2:b}\includegraphics[width=0.35\textwidth]{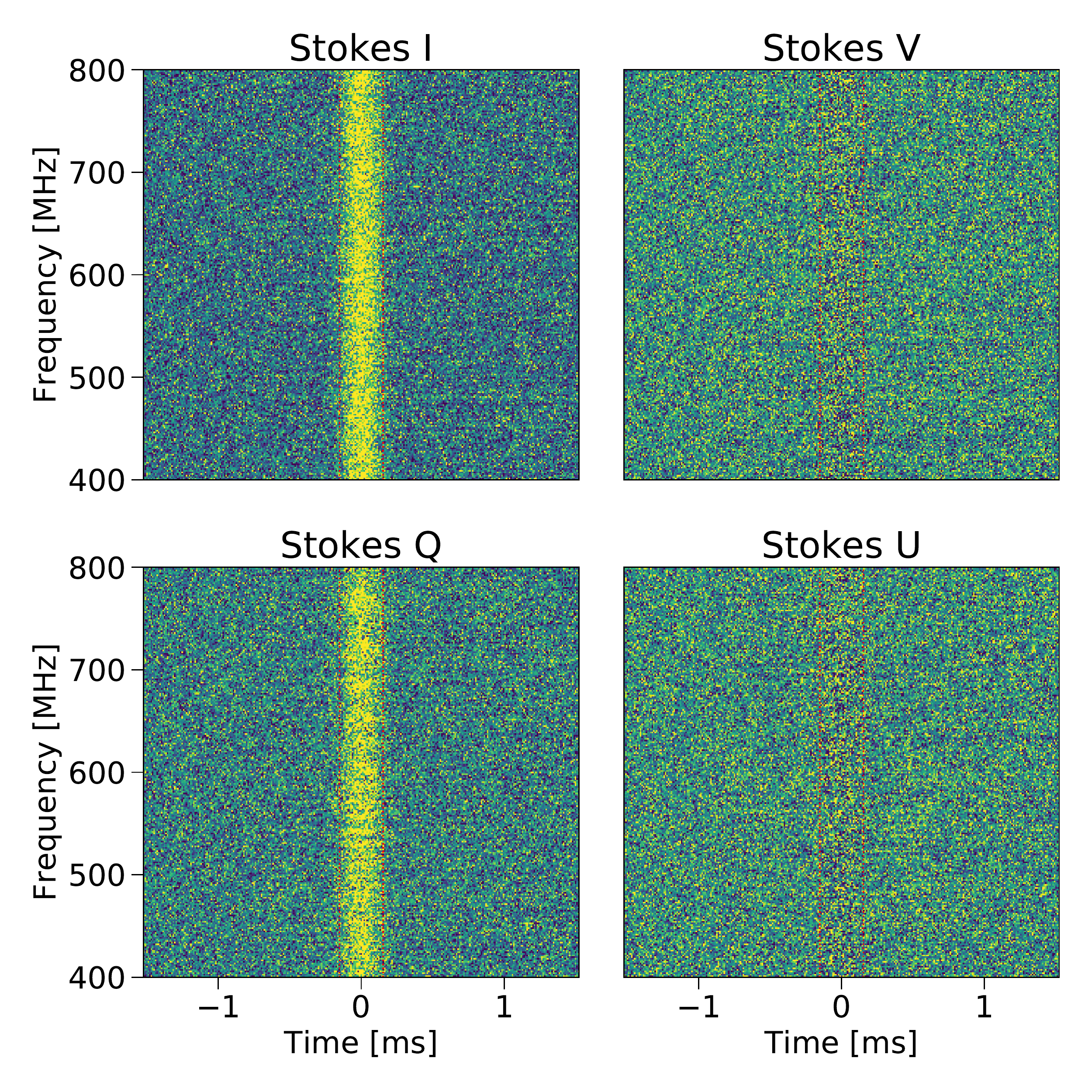}} 
\subfigure[The depolarization corrected FDF (green line) and its uncorrected counterpart (gray line). The peak of the coherently de-rotated FDF will peak near $\phi\approx 0 \; \mathrm{rad \; m^{-2}}$. Here, we have translated the corrected peak by $+200,000 \; \mathrm{rad \; m^{-2}}$ for ease of comparison with the uncorrected FDF.]{\label{fig2:c}\includegraphics[width=0.35\textwidth]{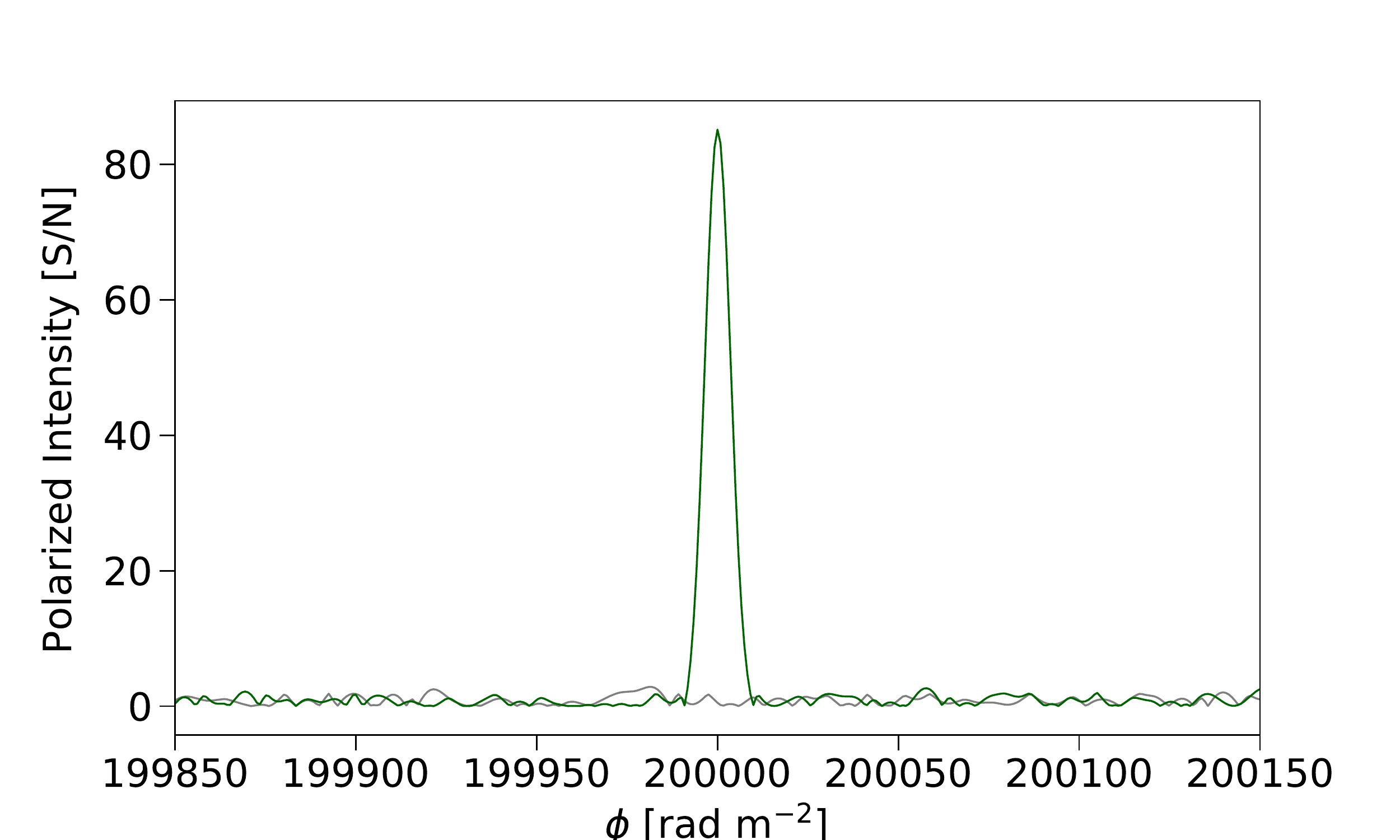}}
\caption{A simulated burst with $\rm{RM=+200,000 \; rad \, m^{-2}}$.}
\label{fig:sim_highRM_derot}
\end{figure}

Figure~\ref{fig:sim_highRM_derot} shows an example of a simulated burst with a very large RM of $+200,000 \; \mathrm{rad \, m^{-2}}$. As expected, omitting the increased RMS noise introduced by the burst, there is an absence of polarized signal in the Stokes $Q$ and $U$ waterfall plot (panel a) due to the extreme level of intra-channel depolarization. In addition, the burst appears to split into the two circular bases at the bottom of the band; a product of the differing group velocities of the two bases that are ``resolved out" for sufficiently narrow bursts with extreme RMs \citep{Suresh2019}. Since the RM is a priori known for this simulated burst, coherent de-rotation can be trivially applied to this burst by first transforming the simulated baseband data from linear to circular basis and then applying Equations~\ref{eqn:highrm} and \ref{eqn:transfer} and transforming back to linear bases. Panel (b) shows the Stokes waterfall plots after correcting for the deleterious effects of the intra-channel Faraday rotation and rotating all of the recovered polarized signal into Stokes $Q$.

A comparison of the depolarization corrected FDF (gray line) and its uncorrected counter-part (green line) in shown in panel (c). The method of coherent de-rotation effectively extends our sensitivity range to RM values far beyond what would be predicted from the native spectral resolution. Much like coherent de-dispersion, this method is resource intensive. This prevents a naive search over many RM trials, since each trial requires the computationally costly procedure of re-sampling the channelized voltages. In light of this, a semi-coherent method has been implemented in the CHIME/FRB pipeline that consists of coherent de-rotation to a sparse grid of RM trials followed by an incoherent search at neighbouring RM values. The details of this semi-coherent RM search method is presented in Section~\ref{sec:semi-coherent}. 

\section{Pipeline: Description}
\label{sec:pipeline}

Here, we summarize the various stages of the the polarization analysis pipeline implemented in the CHIME/FRB baseband system. A schematic of the pipeline is shown in Figure~\ref{fig:flow}. The various stages are described in further detail below. 

\subsection{Polarization Pipeline Stages}

\subsubsection{Pre-processing}

A single tied-array beam, phase referenced in the direction of the refined localization, is fed into the polarization pipeline. These data correspond to a complex voltage time-stream, channelized into 1024 frequency channels (each 390 kHz wide) with a 2.56 $\mu$s time resolution and formatted as a matrix in time, frequency and dual linear polarizations along N-S and E-W directions. These data are coherently de-dispersed to a S/N-optimizing DM determined from earlier stages of the baseband processing pipeline. A frequency channel mask is also determined at an earlier stage in the pipeline from routines that isolate narrow-band radio frequency interference (RFI). These routines include a method that iteratively isolates frequency channels displaying high off-pulse RMS noise and an RFI mask obtained from intensity variation of the signal across multiple beams \citep[see][for details]{Michilli2021}.

Burst duration is determined where the flux has dropped below 20\% of the burst's peak value. This value was determined through manual processing of several tens of events and was generally found to be near optimal for maximizing the integrated polarized signal. Spectral limits of the burst are determined by fitting a Gaussian function to the spectrum and setting a cut-off at the $3\sigma$ level. Both these time and spectral limits are used at the next stage of the pipeline to extract the Stokes spectrum for the burst.

\subsubsection{Stokes Parameter Extraction}

Equation~\ref{eqn:stokes} is used to construct the Stokes parameters from the channelized, complex voltages of the dual-linear polarized feeds (X,Y). Uniform weights are used to integrate polarized signal over the burst duration and were found to result generally in an average Stokes spectrum that is more robust to the deleterious effects of residual RFI. Events displaying significant time and frequency structure over the burst (e.g., complex frequency-time structure) can be manually rerun using a matched filter that appropriately integrates over this structure by using non-uniform weights that capture tempo-spectral evolution.

\tikzstyle{decision} = [diamond, draw, fill=green!20, 
    text width=4.5em, text badly centered, node distance=3cm, inner sep=0pt]
\tikzstyle{block} = [rectangle, draw, fill=blue!20, 
    text width=5em, text centered, rounded corners, minimum height=4em]
\tikzstyle{line} = [draw, -latex']
\tikzstyle{cloud} = [draw, ellipse,fill=red!20, node distance=3cm,
    minimum height=2em]
\tikzstyle{data} = [cylinder, shape border rotate=90, aspect=.4, draw,fill=yellow!20, 
    text width=5.5em, text badly centered, node distance=5cm, inner sep=0pt]
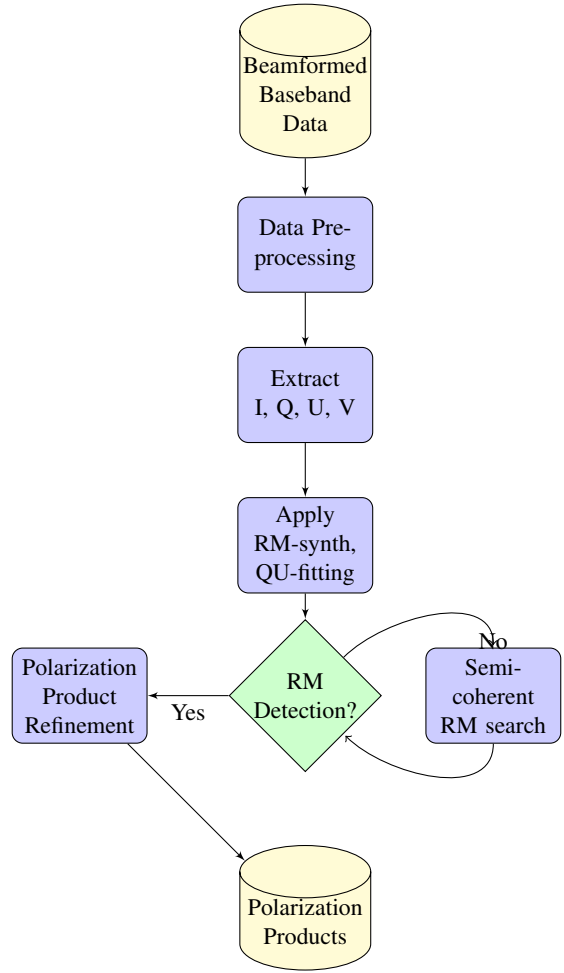
\begin{figure}
\begin{tikzpicture}[node distance = 2cm, auto]
    \node [data] (init) {Beamformed Baseband Data};
    \node [block, below of=init] (preproc) {Data Preprocessing};
    \node [block, below of=preproc] (extract) {Extract \\ I, Q, U, V};
    \node [block, below of=extract, node distance=2cm] (RM) {Apply RM-synth, QU-fitting};
    \node [decision, below of=RM, node distance=2cm] (validation) {RM Detection?};
    \node [block, left of=validation, node distance=3.0cm] (refine) {Polarization Product Refinement};
    \node [block, right of=validation, node distance=2.5cm] (coherent) {Semi-coherent RM search};
    \node [data, below of=validation, node distance=3cm] (product) {Polarization Products};
    
    \path [line] (init) -- (preproc);
    \path [line] (preproc) -- (extract);
    \path [line] (extract) -- (RM);
    \path [line] (RM) -- (validation);
    \path[line] (validation) -- node {Yes} (refine);
    \path[line] (refine) -- (product);
    \draw[->,shorten >=1pt] (validation) to [out=45,in=90] (coherent) node [above,yshift=0.5cm] {No};
    \draw[->,shorten >=1pt] (coherent) to [out=-90,in=-45] (validation);
\end{tikzpicture}
\captionof{figure}{A schematic summarizing the different stages of the CHIME/FRB polarization pipeline.}
\label{fig:flow}
\end{figure}

\subsubsection{RM detection}

RM detections are made through a modified version of the {\tt RM-tools} package \citep{Purcell2020} that contain implementations for both RM-synthesis and QU-fitting. The equation, 
\begin{equation}
|\mathrm{RM_{max}}|\approx \frac{\sqrt{3}}{\mathrm{median}(\delta\lambda^2)}
\end{equation}
\citep{Brentjens2005} is used to determine the approximate RM at which 50\% bandwidth depolarization occurs where $\delta\lambda^2$ is the channel width in units of wavelength-squared ($\mathrm{m^2}$). This value is used to set the RM search limits of both RM-synthesis and QU-fitting methods. For values exceeding this, RM search methods must be preceded by a coherent de-rotation operation (see Section~\ref{sec:semi-coherent} for details).  

\bigbreak
\textit{RM-synthesis:}
\bigbreak

RM-synthesis is a robust method for obtaining an initial RM detection. It is well-suited for implementation in an automated pipeline where low S/N events or residual RFI may stymie a parametric method that is sensitive to initial guesses of model parameters. Moreover, the resulting FDF produced by RM-synthesis is an ideal diagnostic tool for parsing an astrophysical signal from instrumental effects. In light of this, RM-synthesis is applied first in the pipeline to obtain an initial estimate of the RM that is then further refined by QU-fitting.

Performing RM-synthesis on the extracted Stokes spectrum produces a ``dirty" FDF that is then cleaned of artefacts introduced by the RMTF (see Section~\ref{sec:methods}). This cleaning procedure amounts to modelling the intrinsic FDF of the source by discrete Dirac delta functions in $\phi$ space that are then convolved with the RMTF of the observation as a best attempt at reconstructing the observed FDF \citep[see][for details]{Heald2009}. The level of cleaning is determined by the threshold relative to the RMS noise, such that $\phi$ bins where the FDF exceed this value are modelled as delta functions. Cleaning is generally advantageous in scenarios where Faraday complexity is present. This is generally not the case for FRBs, making cleaning a somewhat superfluous step for the purposes of RM determination. Instead, cleaning is implemented in the pipeline for diagnostic reasons, helping determine if complex structure in the ``dirty" FDF is an artefact of the RMTF or some other unknown systematic. For the automated pipeline, FDFs are cleaned conservatively to a level of $8\sigma$. Here, $\sigma$ refers to the noise in the FDF and is estimated from quadratic sum of the RMS in Stokes $Q$ and $U$ across all frequency channels (i.e., $\sigma = \sum_{i=0}^{N} \sqrt{\sigma_{Q_i}^2 + \sigma_{U_i}^2}$; $N =$ number of channels) over a time interval preceding the burst. 

An RM is obtained from the clean FDFs by applying a parabolic fit to the FDF peak. Measurement uncertainties are estimated in a manner analogous to radio imaging \citep{Condon1997}, using the relation $\sigma = \mathrm{FWHM/(2 \, S/N)}$. Here, the FWHM characterizes the width of the peak in Faraday depth space and S/N corresponds to the signal-to-noise ratio of the peak polarized intensity in the FDF. 
In the idealized scenario of Figure~\ref{fig:sim_example1}, RM-synthesis and QU-fitting are effectively equivalent methods. The limitations of RM-synthesis become apparent when additional polarized signal is introduced by instrumental effects. In the case of CHIME, polarized observations are dominated by two systematics: a delay in the beamformed voltages between the two polarizations and, to a much lesser extent, a differential response between them. Appendix~\ref{sec:systematics} illustrates the effect of these two systematics, highlighting how RM values obtained by RM-synthesis are vulnerable to certain systematic biases. This is in contrast to the QU-fitting, for which the model provided in Equation~\ref{eqn:QUmodel} can be extended to fit for additional instrumental effects.  

\tikzstyle{decision} = [diamond, draw, fill=green!20, 
    text width=4.5em, text badly centered, node distance=3cm, inner sep=0pt]
\tikzstyle{block} = [rectangle, draw, fill=blue!20, 
    text width=7em, text centered, rounded corners, minimum height=4em]
\tikzstyle{process} = [rectangle, draw, fill=green!20, 
text width=7em, text centered, rounded corners, minimum height=4em]
\tikzstyle{line} = [draw, -latex']
\tikzstyle{cloud} = [draw, ellipse,fill=red!20, node distance=3cm,
    minimum height=3em]
\tikzstyle{data} = [cylinder, shape border rotate=90, aspect=.4, draw,fill=yellow!20, 
    text width=5.5em, text badly centered, node distance=5cm, inner sep=0pt]
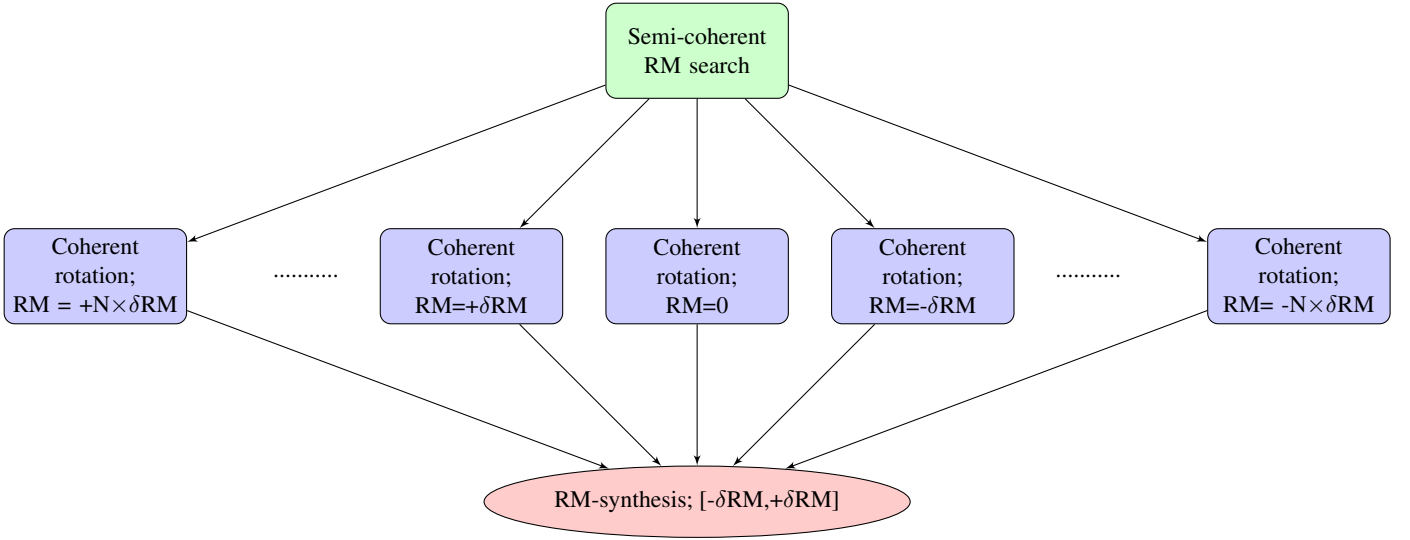
\begin{figure*}
\begin{tikzpicture}[node distance = 2cm, auto]
    \node [process, node distance=10cm] (coherent) {Semi-coherent RM search};
    \node [block, below of=coherent, node distance=3cm](test){Coherent rotation; \\ RM=0};
    \node [block, left of=test, node distance=3cm](test2){Coherent rotation; \\ RM=+$\delta$RM};
    \node [block, left of=test2, node distance=5cm](test3){Coherent rotation; \\ RM = +N$\times\delta$RM};
    \node [block, right of=test, node distance=3cm](test4){Coherent rotation; \\ RM=-$\delta$RM};
    \node [block, right of=test4, node distance=5cm](test5){Coherent rotation; \\ RM= -N$\times\delta$RM};
    
    \node[draw=none,fill=none, left of = test2, node distance=2.2 cm] {...........};
    \node[draw=none,fill=none, right of = test4, node distance=2.2 cm] {...........};
    
    \node [cloud, below of=test, node distance=3cm](RMsynth){RM-synthesis; [-$\delta$RM,+$\delta$RM]};
    
    \path [line] (coherent) -- (test);
    \path [line] (coherent) -- (test2);
    \path [line] (coherent) -- (test3);
    \path [line] (coherent) -- (test4);
    \path [line] (coherent) -- (test5);
    
    \path [line] (test) -- (RMsynth);
    \path [line] (test2) -- (RMsynth);
    \path [line] (test3) -- (RMsynth);
    \path [line] (test4) -- (RMsynth);
    \path [line] (test5) -- (RMsynth);

\end{tikzpicture}
\captionof{figure}{A schematic of the semi-coherent RM search routine that offsets the deleterious effects of intra-channel depolarization by coherently rotating to discrete trial RMs and applying RM-synthesis. The separation between trial values, $\delta RM \sim 700 \; \mathrm{rad \; m^{-2}}$, ensures that no more than $10\%$ depolarization can occur across the semi-coherent search range. This amounts to several thousand coherent de-rotation operations to search out to RM amplitudes as large as $\mathrm{10^6 \; rad \, m^{-2}}$.}
\label{fig:flow2}
\end{figure*}

\bigbreak
\textit{QU-fitting:}
\bigbreak

QU-fitting is applied to refine the initial RM detection made by RM-synthesis or indeed confirm a non-detection. 
Models that simultaneously capture the polarized astrophysical and instrumental signal are implemented into the Nested Sampling QU-fitting framework outlined in Section~\ref{sec:methods}. The default mode of the pipeline is to fit for the astrophysical parameters of the linear polarized fraction, $p$, RM, $\psi_0$, and the physical delay between the two linear polarizations, $\tau$ (cable delay). This amounts to fitting a revised model that accounts for the Stokes $U-V$ leakage introduced by a non-zero $\tau$, 

\begin{equation}
\begin{split}
& Q_{\rm{mod}}^{'} = Q_{\rm{mod}} \\
& U_{\rm{mod}}^{'} = U_{\rm{mod}}\cos(2\pi\nu\tau)-V_{\rm{mod}}\sin(2\pi\nu\tau) \\
& V_{\rm{mod}}^{'} = U_{\rm{mod}}\sin(2\pi\nu\tau)+V_{\rm{mod}}\cos(2\pi\nu\tau)
\end{split}
\label{eqn:QUVmodel}
\end{equation}
Here, $\rm{Q_{mod}}$, $\rm{U_{mod}}$ and $\rm{V_{mod}}$ refer to the models for the astrophysical polarized signal described in Equation~\ref{eqn:QUmodel}. $\rm{Q_{mod}^{'}}$, $\rm{U_{mod}^{'}}$ and $\rm{V_{mod}^{'}}$, meanwhile, are models for the observed Stokes parameters after being modified by the frequency dependent phase difference between X and Y voltages introduced by a non-zero $\tau$. Assuming that the polarized signal is dominated by the linear component, we set $\rm{V_{mod}}=0$. This condition can be relaxed to allow for non-negligible circular polarization that is intrinsic to the source and is further explored in Appendix~\ref{sec:refine}. Modifying the likelihood function Equation~\ref{eqn:likelihood} to account for the leaked signal found in Stokes $V$ allows us to once again estimate best-fitting model parameters by maximizing the modified likelihood function. In all cases, uniform priors are assumed on the fitted parameters.

\subsubsection{Semi-coherent search}
\label{sec:semi-coherent}


The semi-coherent RM search implemented in the pipeline is a two stage process, involving a coherent routine to correct for Faraday rotation over a sparse grid of trial RMs, followed by an incoherent search at neighboring RM values. Possible RM detections at neighboring values are probed by performing RM-synthesis on a coherently de-rotated spectrum, producing an FDF for each trial RM. If the resulting FDF does not produce above a detection threshold, the routine moves to the next trial RM, performing the identical operations of coherent de-rotation and incoherent search until detection is made or a full range of RM values has been explored. A rather stringent detection threshold of S/N $>$ 6 is used to avoid triggering false detections on artificial FDF peaks introduced by systematics. 

A schematic summarizing this routine is shown in Figure~\ref{fig:flow2}. While this routine can, in principle, be performed to arbitrarily large $\rm{|RM|}$ values, we limit the automated pipeline to search within the range $\mathrm{-10^6\leq RM\leq 10^6 \, rad \, m^{-2}}$ to ensure the automated pipeline processes incoming events on a reasonable timescale. This amounts to several thousand coherent de-rotation operations to explore the entire RM range. Coherent de-rotation operations over the sparse grid of trial RMs is by far the most resource-intensive stage of the polarization pipeline, taking roughly 20 minutes to complete a search out to $\rm{|RM|=10^6 \, rad \, m^{-2}}$ when running on a single core CPU. The redundant nature of the operation makes it well-suited for parallelization which is a focus of on-going work.  

The RM step size between coherent operations, $\delta \mathrm{RM}\sim 700 \; \mathrm{rad \; m^{-2}}$, is determined as the $10\%$ depolarization level, referenced at the bottom of the CHIME band ($\nu = 400.390625 \; \mathrm{MHz}$). While this omits the spectral dependence of intra-channel depolarization (i.e., bandwidth depolarization is frequency dependent) it is sufficiently conservative that RM detections from bright, highly polarized bursts are unlikely to be missed. That said, there does exist a phase space over which polarized events will evade detection under current configuration of the semi-coherent search. These problematic events include fainter bursts with intrinsically low linear polarized fractions or bursts with an $\mathrm{|RM|}$ value that exceeds the search limits of the automated pipeline. Rather than be treated by the automated pipeline, these problematic events are left to be manually processed with a tighter more extensive grid of trial $\mathrm{RM}$ values. While $\mathrm{RM}$ detection can, in principle, exist out to arbitrarily large values; upper bounds on the maximum possible $\mathrm{|RM|}$ can be deduced by the absence of a burst-splitting morphological imprint \citep{Suresh2019}. For a 1 ms burst, this morphological imprint begins to manifests at $\mathrm{|RM|}\gtrsim 2 \times 10^6 \; \mathrm{rad \; m^{-2}}$ as an apparent widening of the burst at the bottom of the CHIME band relative to the top. Meanwhile, events that continue to evade RM detection despite manual processing can be used to infer upper bounds on the linear polarized fractions given their S/N. 

\section{Pipeline: Examples}
\label{sec:examples}

In the following section, we use two real FRB detections by CHIME/FRB to illustrate the various stages of the polarization analysis pipeline. In particular, we use a bright, low RM FRB to evaluate the performance of our modified QU-fitting procedure, and a high RM event to validate our coherent de-rotation algorithm. 

\subsection{Low RM Example: FRB 20191219F}
\label{sect:example2}

\begin{figure*}
	\centering
\begin{center}
    \includegraphics[width=0.40\textwidth]{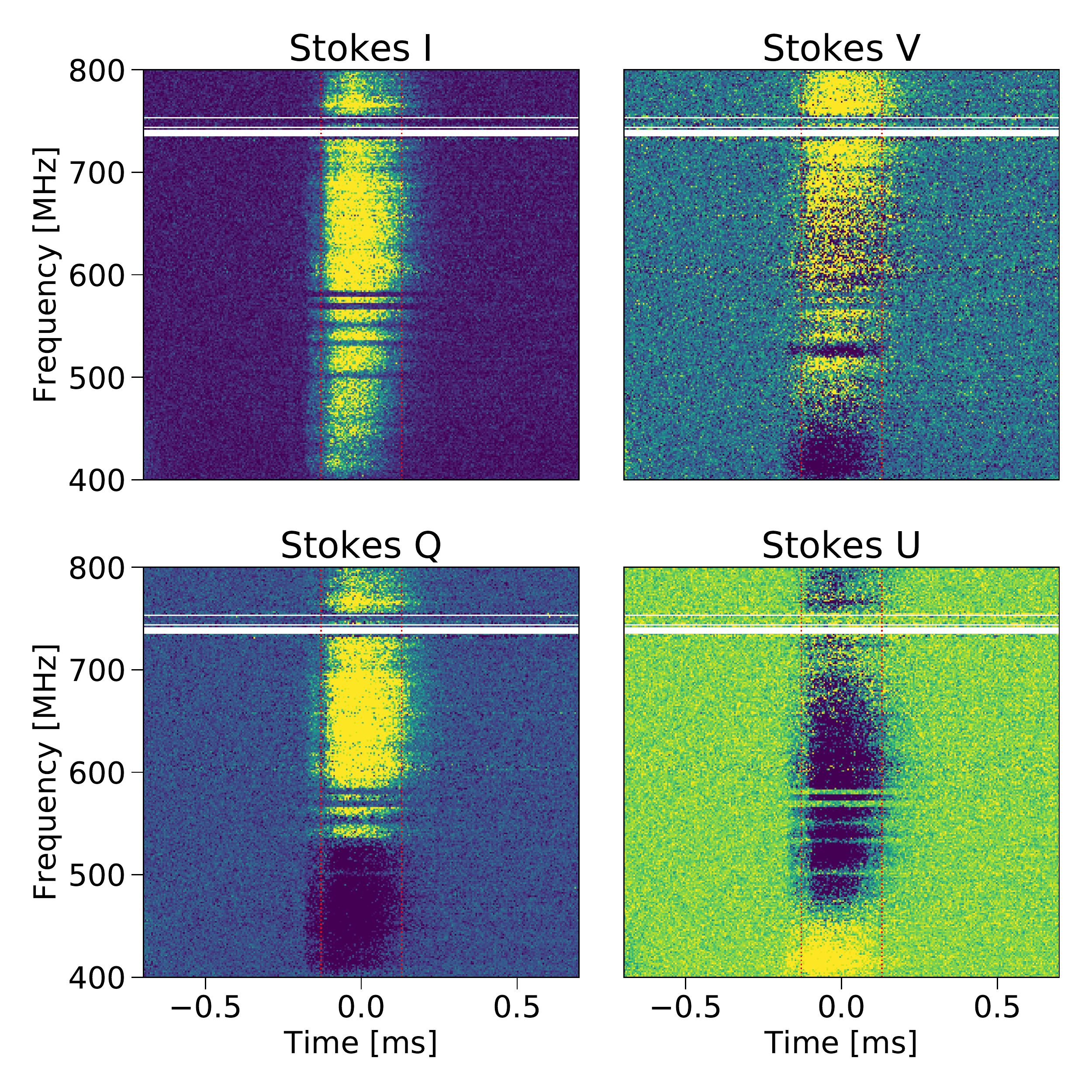} 
    \includegraphics[width=0.40\textwidth]{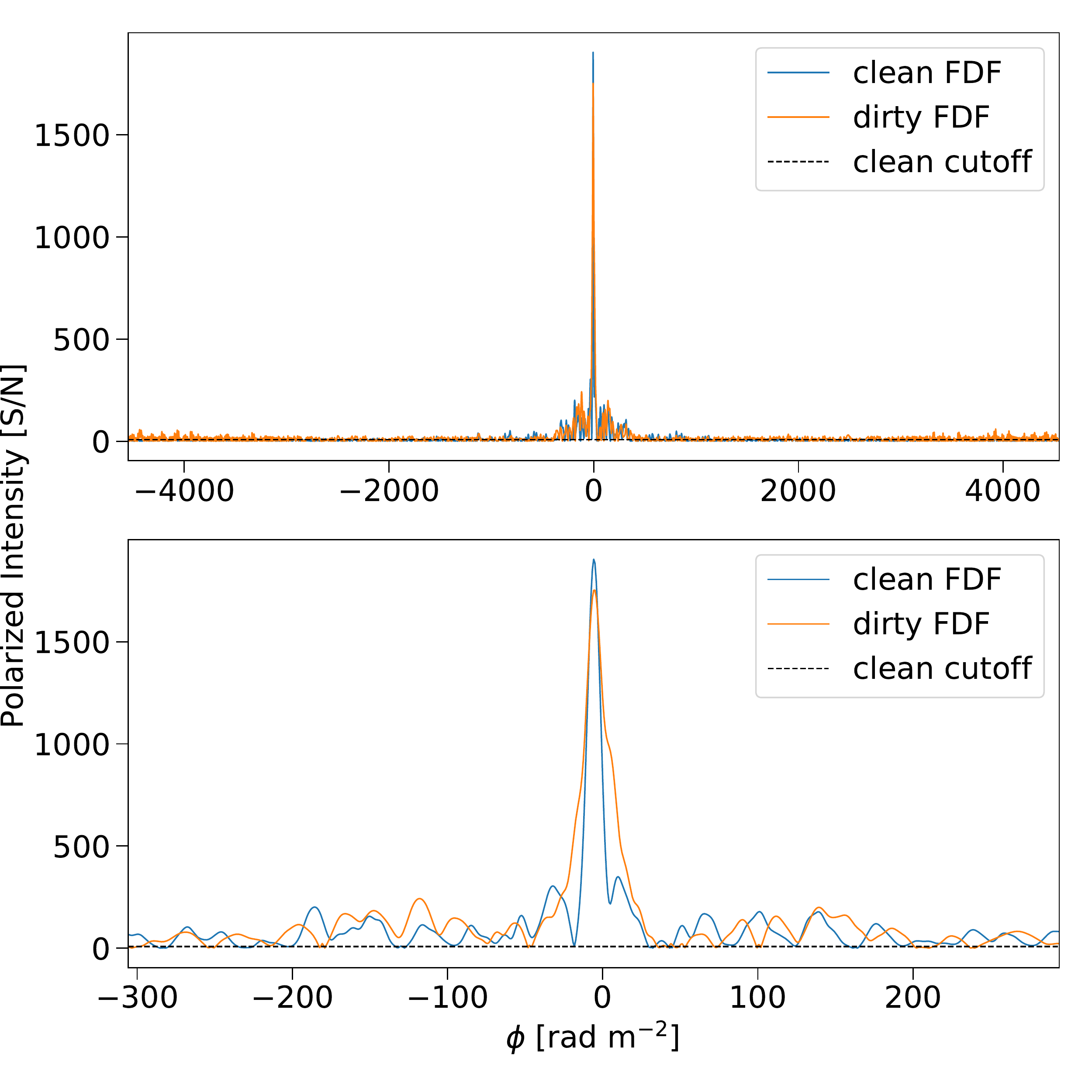} \\
    \includegraphics[width=0.40\textwidth]{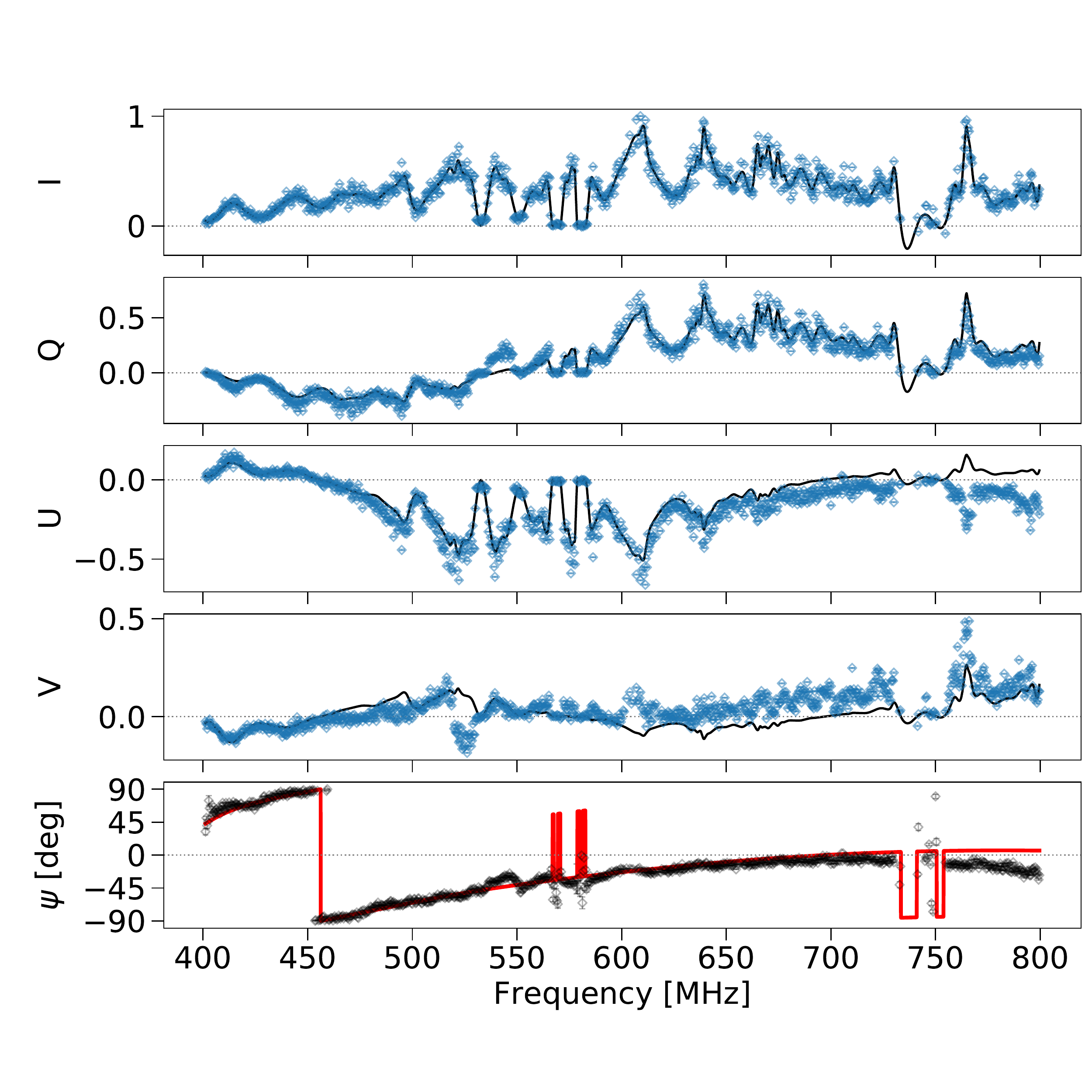}
    \includegraphics[width=0.40\textwidth]{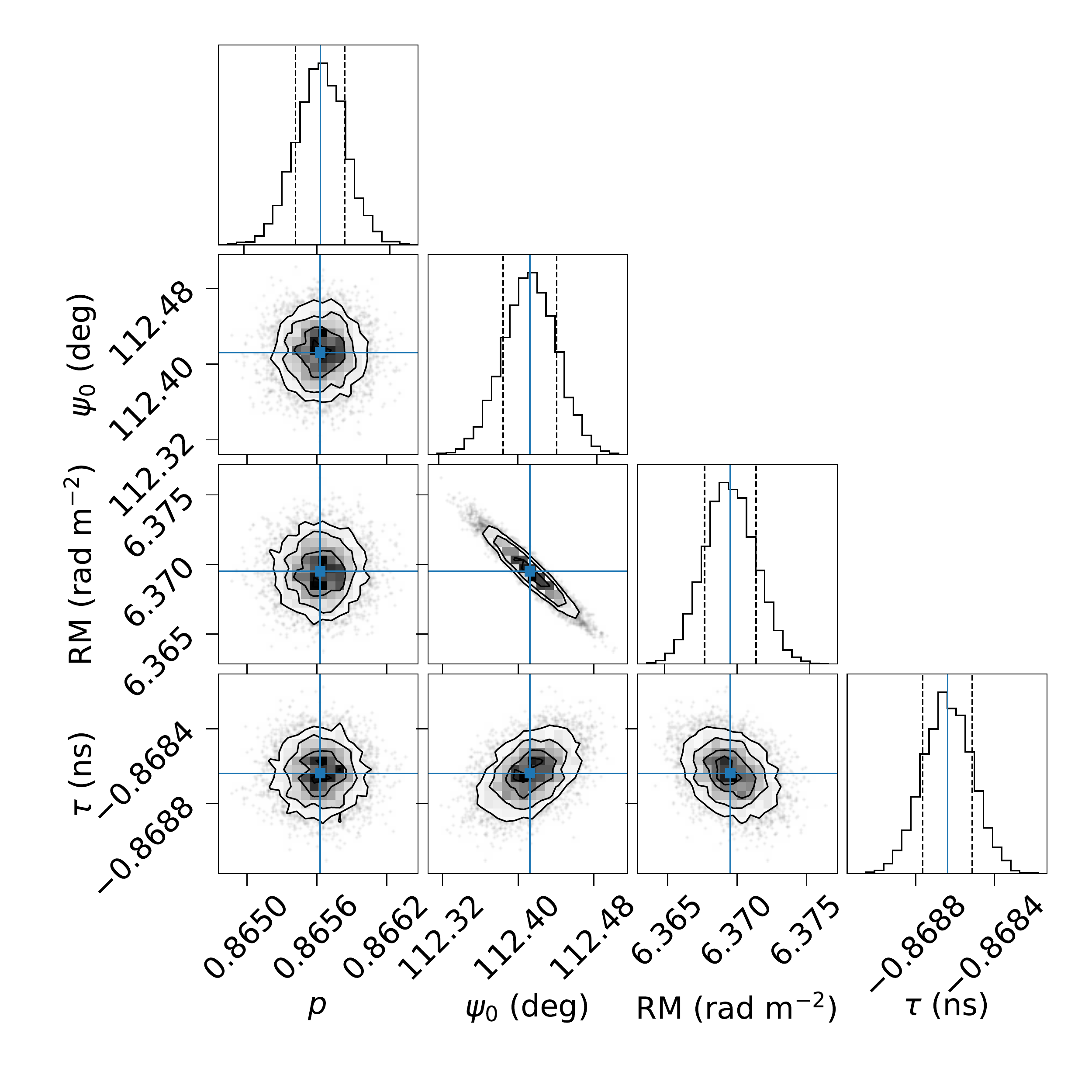}
    \caption{Diagnostic plots output by the CHIME/FRB polarization pipeline for FRB 20191219F. Empty frequency channels in the Stokes waterfall plots indicate masked RFI channels. Sharp discontinuities in model $\psi(\nu)$ (red line) are caused by a vanishing Stokes $U$ term at those frequencies. Best-fit parameters for this event are summarized in Table~\ref{tab:example2}.} 
\label{fig:example2}
\end{center}
\end{figure*}

In December 2019, CHIME/FRB detected a bright burst (S/N $>$ 170) from FRB 20191219F across six of the 1024 formed skybeams of the real-time intensity data. A trigger was initiated by the system that successfully captured baseband data for this event \footnote{Baseband data was also recorded by the CHIME pathfinder instrument allowing sub-arcminute interferometric localization previously reported by \citet{Leung2021}}. Running the baseband localization pipeline resulted in a refined localization of $\mathrm{RA, \, Dec  = (226.2034 \pm 0.0441, 85.4168 \pm 0.0037)}$ degrees (J2000) and an S/N optimizing $\mathrm{DM=464.558 \pm 0.003 \; pc \, cm^{-3}}$. The DM measurement is obtained at an earlier stage in the baseband pipeline by dedispersing to a reference value and then incoherently dedispersing over a range of trial DM values.
The small DM uncertainty quoted here is a product of the brightness of the event and the $\frac{1}{S/N}$ scaling in the Gaussian fit of the peak in S/N, DM phase space. 

The polarization pipeline was then run on the beamformed data, resulting in the diagnostic plots shown in Figure~\ref{fig:example2}. The waterfall plot has been re-binned to a time and frequency resolution of 10.24 $\mu$s/bin and 1.56 MHz/bin, respectively. Evidence of Faraday rotation is seen in the frequency dependent modulation of Stokes $Q$ and $U$ in addition to an apparent leaked polarized signal in Stokes $V$. Running RM-synthesis on the Stokes $Q$, $U$ spectrum, uncorrected for leakage, yields a clear RM detection at RM = $-5.614 \pm 0.001$ rad m$^{-2}$. This initial detection is refined by fitting Equation~\ref{eqn:QUVmodel}, applying a univariate spline to Stokes $I$ to obtain a smoothed model of the burst spectrum, $I_{\mathrm{mod}}$\footnote{The automated pipeline fits the Stokes $I$ spectrum using a fifth-order polynomial. While this spectral model performs well on most events, it often fails to fully capture features introduced by the instrument's bandpass that are particularly pronounced for high-S/N events. Here, we show results of a fit that incorporates a non-parametric method (i.e., univariate spline) that interpolates over spectral structure to capture the full complexity introduced by the bandpass.}.

Implementing this model into the QU-fitting routine yields a best-fit value for the cable delay of $\tau=-0.8686 \pm 0.0001$ ns. Interestingly, the fitted $\mathrm{RM = +6.370 \pm 0.002 \; rad \; m^{-2}}$ is of opposite sign to the initial detection. This sign ambiguity is introduced by the effects of the uncorrected cable delay (see Appendix~\ref{sec:systematics} for details). Correcting for the cable delay amounts to a multiplicative phase factor that scales with $\tau$ and $\nu$, such that, 
\begin{equation}
Y'= Y \exp(-2\pi i \nu \tau).
\end{equation}

Re-constructing the Stokes spectrum from the corrected $(X,Y')$ polarizations\footnote{n.b., The correction need only be applied to one of the polarizations ($Y'$) to correct for the \textit{relative} delay between the two polarized voltage streams.} successfully removes the frequency dependent modulation seen in the Stokes $V$ waterfall, and re-performing RM-synthesis with this corrected spectrum yields RM and $\psi_0$ values that are in agreement with those measured from QU-fitting along with a $\sim$5$\%$ boost in signal. Table~\ref{tab:example2} summarizes the fit results. An ionospheric RM contribution of $\mathrm{RM_{iono}=0.35 \pm 0.05 \; rad \, m^{-2}}$ is calculated using {\tt ionFR5}\footnote{https://github.com/csobey/ionFR} \citep[][see Section~\ref{sec:iono}]{Sotomayor2013}. Using this value to correct for the ionospheric contribution leaves us with a measurement of $\mathrm{RM=6.020 \pm 0.002 \pm 0.050 \; rad \, m^{-2}}$, where the errors represent statistical and ionospheric uncertainties, respectively.  

\begin{figure}
\centering     
\subfigure[FRB 20191219F]{\label{fig3:a}\includegraphics[width=0.40\textwidth]{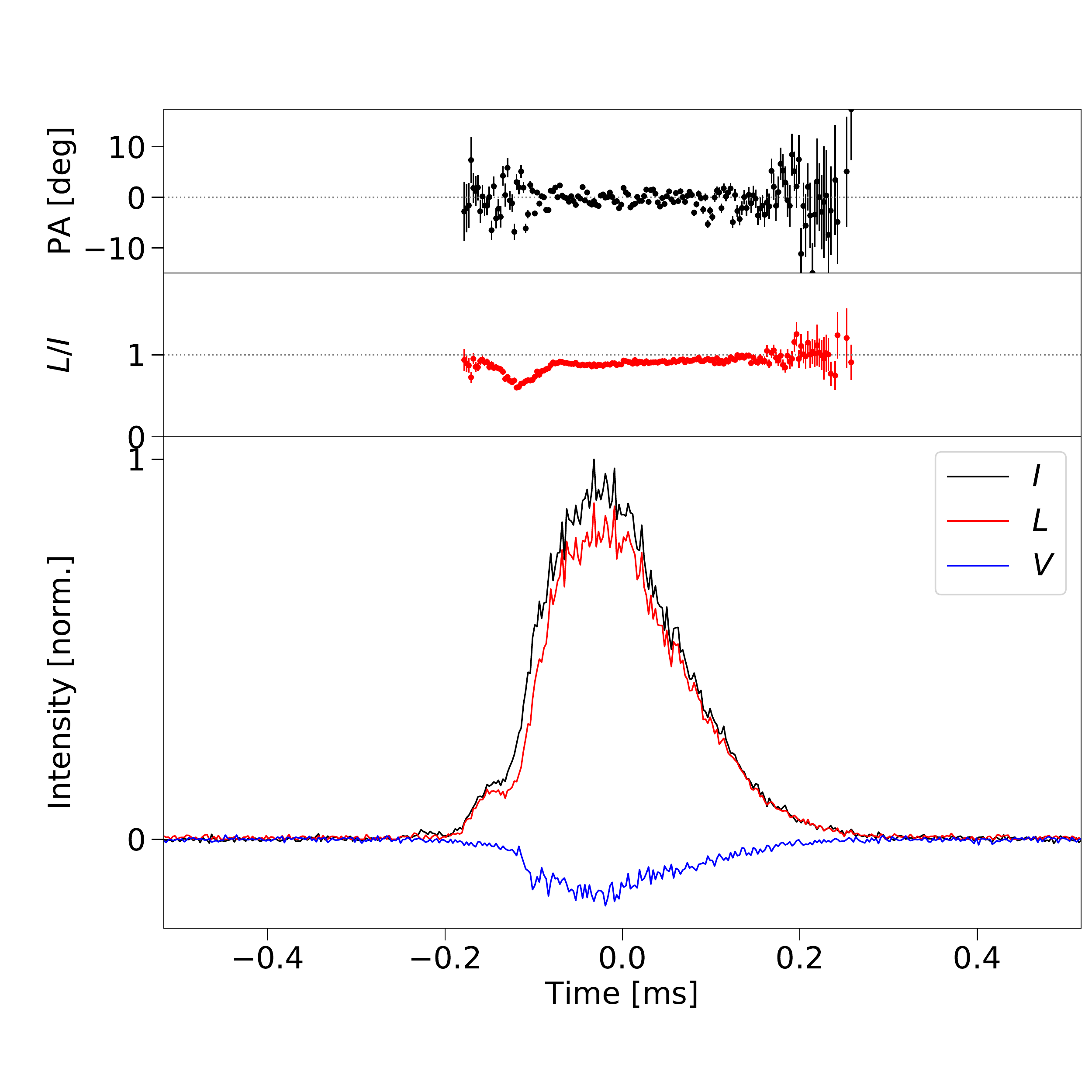}}
\subfigure[FRB 20200917A]{\label{fig3:b}\includegraphics[width=0.40\textwidth]{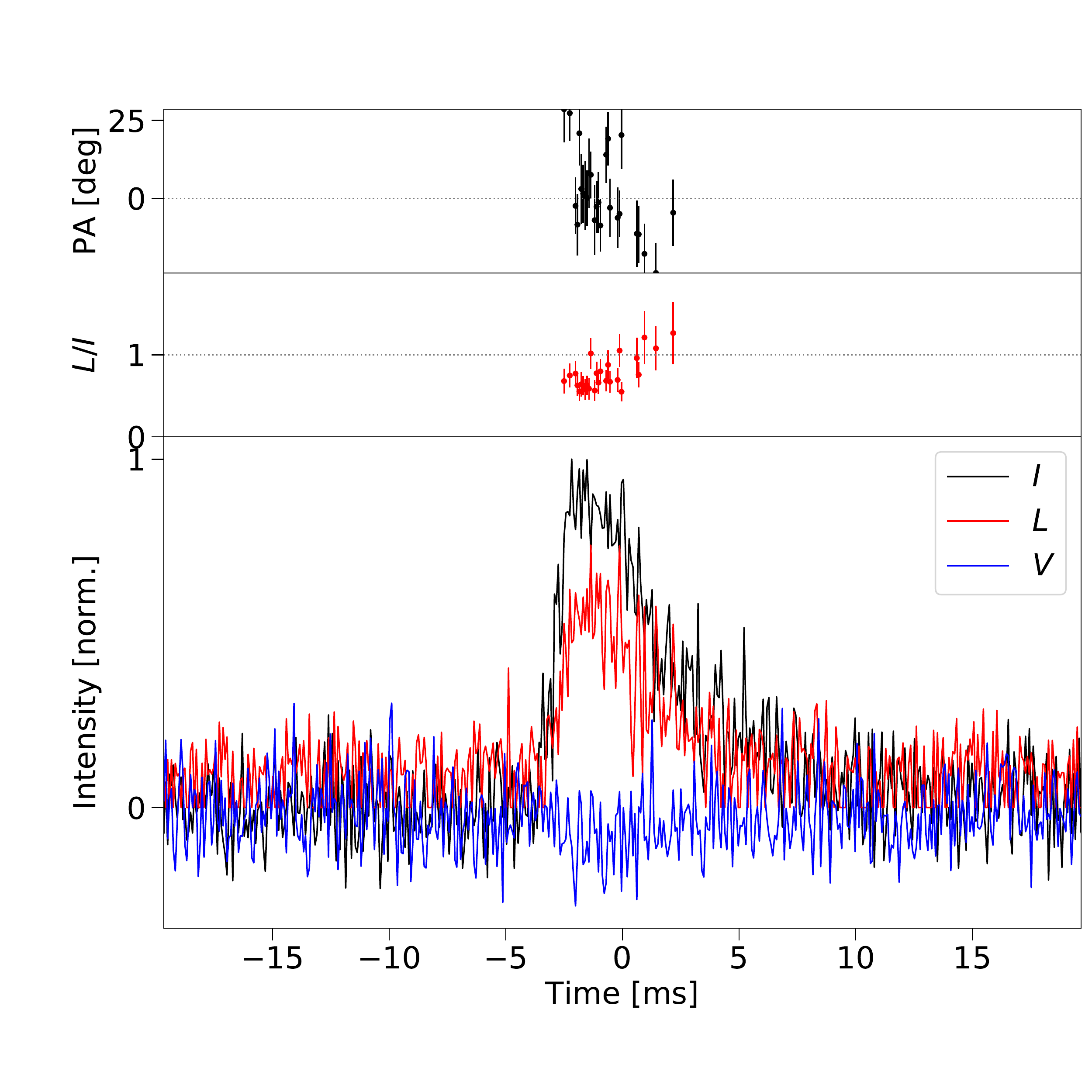}} \\
\caption{Polarized burst profiles for (a) FRB 20191219F and (b) FRB 20200917A showing the total (black), linear (red) and circular (blue) intensities as a function of time (bottom panel). The middle panel displays the linear polarized fraction (L/I) and the top panel, the polarization angle (PA) as a function of time.}
\label{fig:example_prof}
\end{figure}

Figure~\ref{fig:example_prof} shows the burst profile for the total intensity (black) along with the linear (red) and circular (blue) components at the native baseband time resolution of 2.56 $\mu$s. The burst is highly linearly polarized ($L/I>80\%$) with a slight upward trend in the linear polarized fraction that suggests some tempo-spectral evolution in the polarized component. Interestingly, the linearly polarized fraction is highest at the trailing edge of the burst, suggesting an evolution in the polarized signal across the burst. 
Substantial residual Stokes $V$ signal is present even after correcting for the cable delay. It is uncertain from this analysis whether this residual Stokes $V$ signal is a result of some as yet unknown systematic or is intrinsic to the source. In Appendix~\ref{sec:refine}, we extend the analysis of this FRB by incorporating additional parameters characterising the intrinsic properties of the linearly and circularly polarized components, finding evidence for a significant circular component. Finally, the PA is remarkably flat over the burst profile, but does appear to display some interesting correlated structure on very short time scales. 

\begin{table}[!htbp]
\centering
\caption{Fitted Polarization Parameters of FRB 20191219F}
\begin{tabular}{*4c}
\toprule
Parameter && RM-synthesis\footnote{Values reported here are from running RM-synthesis on the spectrum uncorrected for cable delay and highlight the RM sign ambiguity when not accounting for this systematic.} & QU-fitting \\ 
\toprule
 RM [rad m$^{-2}$] && $-5.614 \pm 0.001$  &  $+6.370 \pm 0.002$ \\
 $\psi_0$ [deg] && $52.97 \pm 0.02$  & $112.41 \pm 0.03$  \\
 L/I && $\approx 0.65$ & $0.8657 \pm 0.0007$  \\
 $\tau$ [ns] && N/A & $-0.8686 \pm 0.0001$ \\
 \hline
\end{tabular}
\label{tab:example2}
\end{table} 

\subsection{High RM Example: FRB 20200917A}
\label{sect:example3}

\begin{figure}
\centering     
\subfigure{\label{fig3:a}\includegraphics[width=0.45\textwidth]{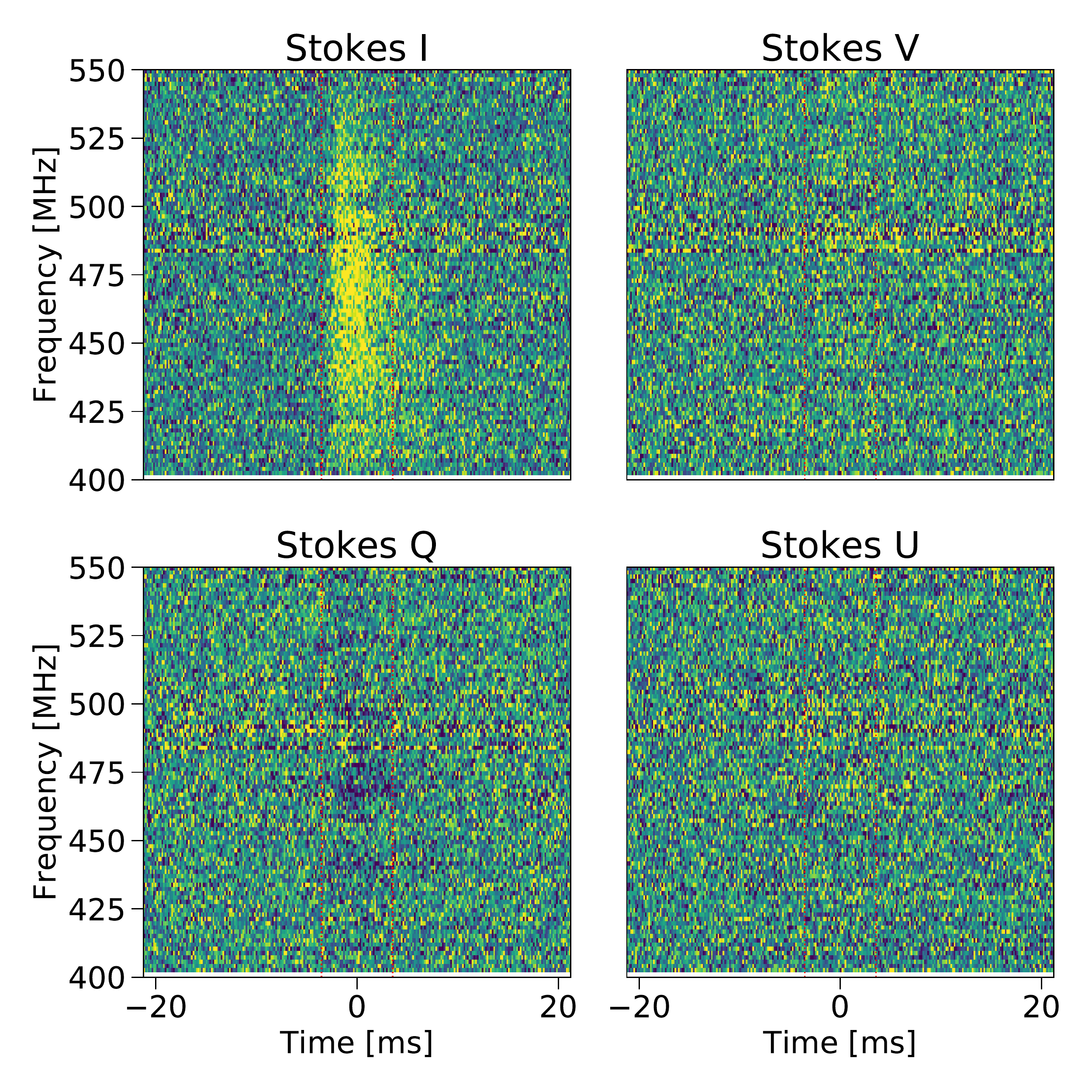}}
\caption{The Stokes waterfall plot of FRB 20200917A rebinned with a time and frequency resolution of 164.84 $\mu$s/bin and 1.56 MHz/bin, respectively. Faint polarized signal in Stokes $Q$ is $I-Q$ leakage induced by the differential gain of the primary beam for the two polarizations.}
\label{fig:wfall_high_RM}
\end{figure}

\begin{figure}
\centering     
\subfigure{\includegraphics[width=0.45\textwidth]{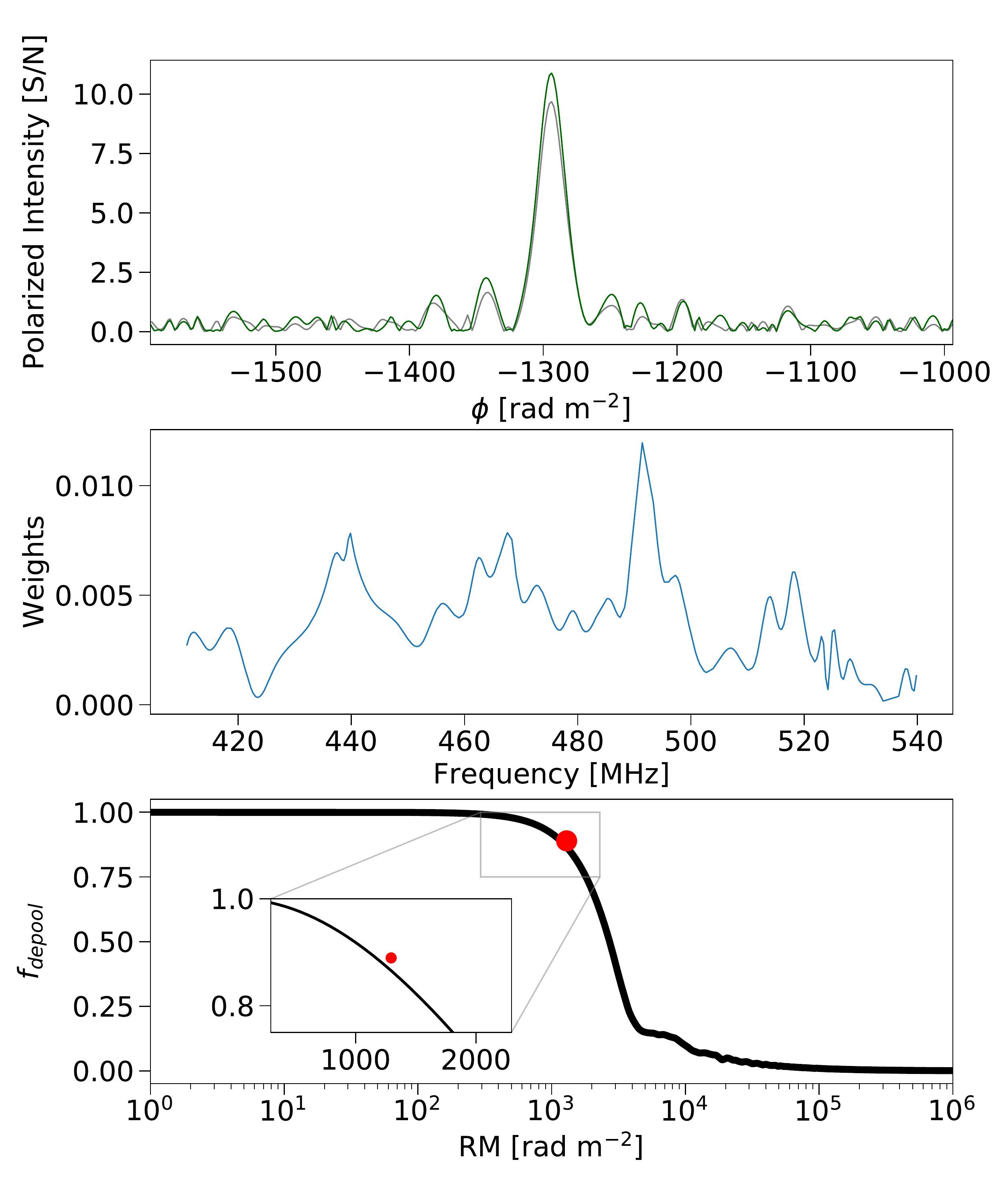}}
\caption{Validation of the coherent de-rotation algorithm using FRB 20200917A. Top: The FDF, corrected (green) and uncorrected (gray) for bandwidth depolarization. Middle: the weights applied to the depolarization curve of each frequency channel in which emission was detected. Bottom: the synthesized depolarization curve obtained by integrating the estimated depolarization of Equation~\ref{eqn:depol} across each frequency channel with the red marker indicating the ratio of the FDF peak polarized intensities, uncorrected and corrected for bandwidth depolarization.}
\label{fig:high_RM}
\end{figure}

In Section~\ref{sec:coherent_derot} we used simulated baseband data to demonstrate how the coherent de-rotation algorithm successfully retrieves bandwidth depolarized signal. While this is an encouraging check, it is possible that unknown systematics introduced by the telescope optics (e.g., coupling) or further downstream in the signal chain (e.g., channelization, spectral leakage) may pose practical limitations on the reliable application of our coherent de-rotation algorithm. The only foolproof method of ruling out deleterious effects introduced by CHIME-specific systematics is to perform a similar analysis on real FRBs with RM values far outside the nominal sensitivity range of the baseband data (see Figure~\ref{fig:upchan}). Unfortunately, running the semi-coherent search on the limited subsample of apparently unpolarized events has not returned any detections at large RMs.\footnote{Morever, CHIME/FRB has not yet captured baseband data from FRB 20121102A, or a source with a confirmed RM of similar amplitude.} While this could indicate the ineffectiveness of the coherent de-rotation algorithm on CHIME/FRB baseband data, we show here that an FRB with a moderate $\mathrm{|RM|}$ still suffers from partial depolarization and can therefore be used to validate our algorithm. 

In this Section, we perform coherent de-rotation on   
FRB 20200917A. Discovered in September 2020, this event triggered a successful recording of baseband data. A singlebeam was formed in the direction of best localization, 
$\mathrm{RA, \, Dec}$  $\mathrm{= (315.1217 \pm 0.0544, 75.8036 \pm 0.0090)}$ degrees, and dispersed to an S/N optimizing DM of $\mathrm{DM=883.3 \pm 0.1\; pc \, cm^{-3}}$. Figure~\ref{fig:wfall_high_RM} shows the resulting Stokes waterfall plot rebinned with a time and frequency resolution of 164.84 $\mu$s/bin and 1.56 MHz/bin, respectively. An initial RM detection near $\mathrm{+1300 \; rad\, m^{-2}}$ was made via RM-synthesis. This detection was subsequently refined by QU-fitting. As was the case in the previous example, accounting for cable delay in the QU-fitting results in a best-fit RM of opposite sign to the initial detection, $\rm{RM=-1294.3\pm 0.1}$ rad/m$^2$. The validity of the sign was confirmed by observing a boost in the FDF peak intensity after re-performing RM-synthesis on the cable delay corrected spectrum. Table~\ref{tab:example3} summarizes the fit results. An ionospheric RM contribution of $\mathrm{RM_{iono}=0.17 \pm 0.05 \; rad \, m^{-2}}$ was determined and used to correct for the ionospheric contribution, leaving us with a measurement of $\mathrm{RM=-1294.47 \pm 0.10 \pm 0.05 \; rad \, m^{-2}}$.  

The fact path length differences in CHIME, an interferometer with 1024 dual feeds, can be well characterized by a single delay ($\tau$) is a product of X and Y polarization being independently calibrated of one another. There is a significant difference between the best-fit values for $\tau$ of FRB 20200917A and FRB 20191219F. The source of this disagreement is associated with thermal expansion of the instrument, as previously noted in the context of CHIME/FRB localization \citep{Michilli2021}. Interestingly, this temperature dependent effect captured in polarized data offers an alternate means of characterizing the thermal expansion of the dish; independent of similar efforts through systematic offsets in the localization of known sources. 


Coherently de-rotating this cable-delay-corrected spectrum by the RM value determined from QU-fitting and re-performing RM-synthesis results in a depolarization corrected FDF shown in Figure~\ref{fig:high_RM}. Similar to that previously shown for a simulated burst (i.e., Figure~\ref{fig:sim_highRM_derot}), panel (a) highlights the S/N boost achieved by the depolarization corrected FDF (green line) over its uncorrected counter-part (gray line). Calculating the ratio of the uncorrected and corrected FDF peak intensities yields a value of 0.87. The lower plot of panel (b) compares this value, indicated by the red marker, to the depolarization curve (black line) for the burst. The depolarization curve is constructed by integrating the frequency dependent depolarization across the burst sub-band and using a Stokes $I$ fit (cubic spline) to obtain weights of the depolarization contribution at each frequency (top panel). The co-incidence of the red marker with the theoretical depolarization curve at the fitted RM indicates that the coherent de-rotation routine is indeed retrieving all of the bandwidth depolarized signal.

The coherently de-rotated polarized burst profile of FRB 20200917A is shown in panel (b) of Figure~\ref{fig:example_prof}. The burst is linearly polarized ($\mathrm{L/I>60\%}$) with no significant circular component. Like FRB 20191219F, FRB 20200917A displays a slight increase in the linear polarized fraction at the trailing edge of the burst. Meanwhile, evolution in the PA is apparent across the burst phase, displaying small but significant substructure similar to that seen in other FRBs at high time resolution \citep[e.g.,][]{Day2020,Luo2020,Nimmo2021}. This structure can possibly be explained by a time dependence of $\psi_0$ or as an artefact of a slight frequency dependence of $\psi_0$, manifesting as structure in the PPA curve from a changing spectrum through the burst phase. An additional complication is potential effects introduced by scattering, which is non-negligible for this event.
Here, the well known flattening property of scattering on the PA curve \citep[e.g.,][]{Li2003}, combined with the strong frequency dependence, can give rise to artificial PA structure by more strongly affecting lower frequencies. This explanation is somewhat at odds with the secular increase in linear polarized fraction at later times where the deleterious effects of scattering are most significant leading to partial or complete depolarization \citep[e.g.,][]{Sobey2021}. A systematic method for probing the observed PA structure and relating this analyses to different emission models and propagation effects is left for future work.

\begin{table}[!htbp]
\centering
\caption{Fitted Polarization Parameters of FRB 20200917A}
\begin{tabular}{*4c}
\toprule
Parameter && RM-synthesis\footnote{Values reported here are from running RM-synthesis on the spectrum uncorrected for cable delay and highlight the RM sign ambiguity when not accounting for this systematic (see Appendix~\ref{sec:systematics}).} & QU-fitting \\ 
\toprule
 RM [rad m$^{-2}$] && $+1292.6 \pm 0.2$  &  $-1294.3 \pm 0.1$ \\ 
 $\psi_0$ [deg] && $ 146.7 \pm 5.0$  & $7.5 \pm 4.0$  \\
 L/I && $\approx 0.4$ & $0.536 \pm 0.004$  \\
 $\tau$ [ns] && N/A & $-1.23 \pm 0.02$ \\
 \hline
\end{tabular}
\label{tab:example3}
\end{table} 

\section{Discussion}
\label{sec:discussion}

\subsection{Rotations Measures of FRB 20191219F and FRB 20200917A}

\begin{figure}
\centering     
\includegraphics[width=0.45\textwidth]{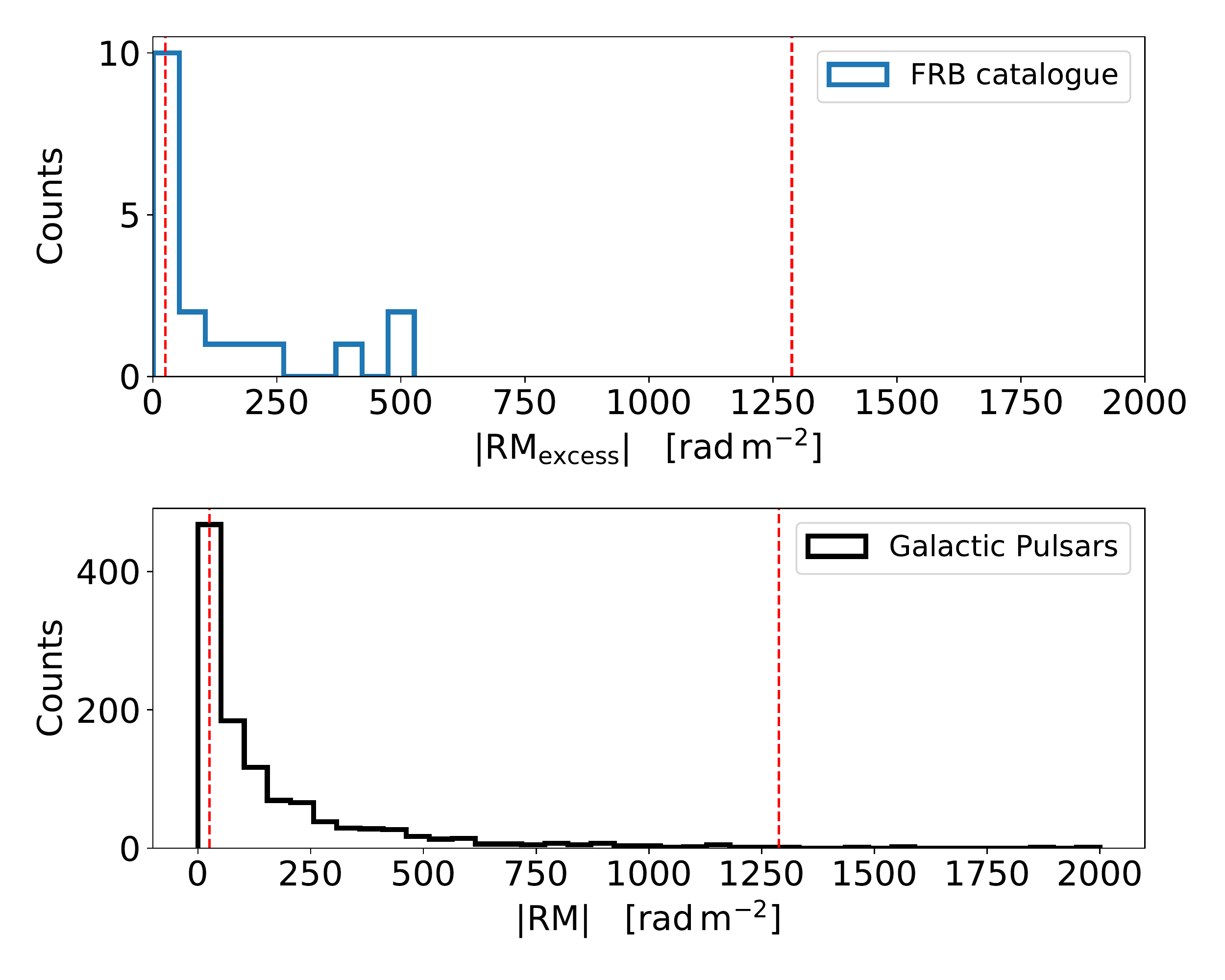}
\caption{Comparison of $\rm{RM_{excess}}$ estimates of FRB 20191219F and FRB 20200917A (red, dashed lines) with those of the FRB catalogue (top panel) and the RM distribution of the Galactic pulsar sample (bottom panel), constrained to a range $\mathrm{|RM|< 2000 \; rad \, m^{-2}}$. N.b. FRB 20121102A, whose variable RM lies in the range $67,000 \lesssim |RM| \lesssim 103,000 \; \mathrm{rad \, m^{-2}}$, has been omitted from the top panel histogram.}
\label{fig:hists}
\end{figure}

For both FRB 20191219F and FRB 20200917A, we obtain an estimate of the RM contribution of the Galactic foreground, $\mathrm{RM_{MW}}$ \citep{Hutschenreuter2021}. In the case of FRB 20191219F, we estimate $\mathrm{RM_{MW}=-20 \pm 7 \; rad \, m^{-2}}$, implying a modest excess RM ($\mathrm{|RM_{excess}|\sim 20-30 \; rad \, m^{-2}}$) from extragalactic sources of Faraday rotation. 
This is contrasted by FRB 20200917A, which has a Galactic RM contribution of $\mathrm{RM_{MW}=-12.0 \pm 18 \; rad \, m^{-2}}$, implying a large excess RM of $\mathrm{|RM_{excess}|\sim 1260-1300 \; rad \, m^{-2}}$. These results are summarized in Figure~\ref{fig:hists} which compare the $|\mathrm{RM_{excess}}|$ values of these two bursts to the corresponding sample from the published FRB catalogue and the Galactic pulsar RM sample. While the $|\mathrm{RM_{excess}}|$ value of FRB 20191219F is consistent with the published FRB sample, FRB 20200917A displays an $|\mathrm{RM_{excess}}|$ that is substantially greater than most other FRBs, with the exception of FRB 20121102A \citep{Michilli2018}. Moreover, $|\mathrm{RM_{excess}}|$ of FRB 20200917A significantly exceeds typical values of the pulsar RM sample\footnote{The discrepancy with the Galactic pulsar sample is likely even larger since: 1) the intrinsic RM is diluted by cosmological redshift, such that, $\mathrm{RM_{obs} \propto RM_{int}}/(1+z)^2$ and 2) The Galactic pulsar sample is heavily biased to sampling sightlines along the Galactic midplane where the average RM is significantly greater.} \citep{Manchester2005}\footnote{http://www.atnf.csiro.au/research/pulsar/psrcat}, suggesting a supplemental source of Faraday rotation other than the host galaxy's diffuse ISM. We emphasize here that the comparison to the Galactic pulsar sample is used to inform upper limits on the RM contribution of the host galaxy and not the Galactic RM contribution, which is more readily estimated from extragalactic sources \citep[e.g.;][]{Hutschenreuter2021}.

The significance of the offset between $\mathrm{|RM_{excess}|}$ of FRB 20200917A with that of the published FRB sample is less significant than it might appear due to a strong selection/publication bias. Specifically, this burst was selected from a much larger sample of CHIME-detected FRBs for its high-RM, such that the efficacy of our analysis on high-RM sources could be demonstrated. In fact, the sample from which it was selected contains more sources than the previously published RMs shown in Figure~\ref{fig:hists}. Hence, it is somewhat unsurprising that we should observe an FRB with such an $\mathrm{|RM_{excess}|}$ value. Moreover, the comparison of this measurement to those of the published FRB sample omits cosmological dilation, which dilutes the RM contribution of the host Galaxy by a factor $1/(1+z)^2$ (Equation~\ref{eqn:RMth}). Accounting for this effect adds significant ambiguity in assessing the significance of $\rm{RM_{excess}}$ of FRB 20200917A relative to other FRBs. Indeed, as noted by \citet{Connor2020}, the local RM of FRB 20160102A \citep{Caleb2018} could be as large as $ -2400 \; \rm{rad \, m^{-2}}$ if its $\rm{DM=2596.1\pm 0.3 \; pc \, cm^{-3}}$ is dominated by the IGM. 

We follow the analysis applied to the recently discovered repeating source, FRB 20200120E \citep{Bhardwaj2021}, to probe for intervening structures that could produce supplemental Faraday rotation. We rule out Galactic sources of Faraday rotation, finding the sightline of FRB 20200917A to be unassociated with any known foreground structures, including HII regions \citep{anderson2014}, star forming regions and stellar clusters \citep{avedisova2002}. For extragalactic sources of Faraday rotation, we do not find any nearby galaxies or galaxy clusters within 1 square degree of the localization region\footnote{We obtain a similar result for FRB 20191219F, precluding analysis similar to \citet{Prochaska2019}} \citep{Wen2018}. Following a similar line of reasoning to that of \citet{Connor2020} in the analysis FRB 20191108A, we conclude that the substantial $|\mathrm{RM_{excess}}|$ observed from FRB 20200917A likely originates within the host galaxy itself. This Faraday rotation includes a contribution from the smoothly distributed component of the diffuse ISM as well as possible contributions from intervening discrete structures displaying enhanced electron column densities and/or magnetic field strengths. 

Discrete structures may be related to the central engine, as in the case of the dense, magnetized plasma of a supernova remnant. This possibility has recently been put forth to describe the large but decreasing $\mathrm{|RM|}$ observed from FRB 20121102A \citep{Hilmarsson2021}. Alternatively, the excess $\mathrm{|RM|}$ may reflect an environmental preference of the population, such as the proximity of the Galactic center magnetar, PSR J1745-2900, to Sagittarius A$\star$ \citep[e.g.,][]{Bower2003,Desvignes2018} or a manifestation of a fortuitous alignment of the FRB sightline with a galaxy's large scale magnetic field. Indeed, a Galactic analogue of this latter scenario would be the strong Faraday rotation (several thousand $\mathrm{rad \, m^{-2}}$) observed from extragalactic sources intersecting the  Sagittarius arm tangent and attributed to the diffuse ISM rather than any discrete structures \citep{Shanahan2019}. While this scenario is in principle possible, it is disfavoured by the strong scattering and dispersion signatures imparted on the emission at such low inclinations angles. With only one observation from this source, it remains difficult to distinguish amongst these possibilities.

In the absence of additional information, the exact RM contribution of discrete, over-dense regions of the ISM and its diffuse counterpart remain entirely degenerate in describing the observed Faraday rotation. One method for distinguishing these competing sources is to incorporate additional information contained in the scattering properties of the burst. Scintillation, the variation of intensity with frequency due to multi-path interference, can be used to determine the nature and geometry of the scattering medium. \citet{Masui2015}, in their analysis FRB 20110523A, used the scattering/scintillation properties to conclude that the observed Faraday rotation originated from a dense, magnetized plasma near ($\mathrm{\lesssim 40 \, Kpc}$) the source. Carrying out similar analysis here for FRB 20200917A is promising given the strong evidence for scattering but is beyond the scope of this paper and is left for future work\footnote{A robust analysis probing for scintillation is non-trivial given the relatively low S/N of this event and the presence of uncalibrated systematics in the CHIME/FRB dataset.}

\subsection{Polarization and the CHIME Primary Beam}

There remains substantial uncertainty in CHIME's primary beam (PB) model that seeks to describe the frequency and polarization dependent response as a function of position on the sky. The absence of an accurate model complicates not only the calibration of important FRB observables such as fluence and spectral properties but also other science done with CHIME such as pulsar monitoring \citep{CHIME/PSR2020} and the 21-cm line emission survey between $0.8 \leq z \leq 2.5$ \citep{Newburgh2014}\footnote{A CHIME/cosmology instrument paper is currently in preparation.}. In the case of the latter, differences in the radiation patterns of the two polarized beams result in conversion of unpolarized signal into polarized, greatly complicating the process of isolating the unpolarized 21-cm signal from polarized contaminant signal and motivating methods for refinement of CHIME's PB (Singh et al. 2021, in prep.; Wulf et al. 2021, in prep.).  
In the case of CHIME/FRB, uncertainties in the PB complicate the analysis of polarized spectra by adding artificial features in the constructed Stokes parameters. These features result from the differential gain and phase between the X and Y polarizations. For CHIME, the phase errors are secondary to the much larger instrumental polarization produced by the differences in the absolute sensitivities of the two polarizations. 

In Appendix~\ref{sec:systematics} we illustrate the effect of differential gain, showing a simulated burst ($\mathrm{RM=+100 \; rad \, m^{-2}}$) where the Y polarization sensitivity is 50\% of the X polarization. In such cases, Stokes QU-fitting can be extended by invoking a parameter characterizing the differential gain, $\eta$, between the two polarizations. Unfortunately, a realistic PB model for CHIME cannot be characterized by a single $\eta$ value due to the chromaticity of the two polarizations. This is particularly true at large angular excursions from the main lobe where differences in polarized gains are greatest and change significantly with frequency. In the absence of an accurate PB, corrections for the instrumental leakage introduced by differential gain of the two polarizations is a challenging problem. Thankfully, FRBs for which this effect is significant can be easily identified by an FDF leakage artefact that peaks near $\mathrm{RM \approx 0 \; rad \,m^{-2}}$. While this instrumental polarized signal may lead to a sample of FRBs with incorrect RM detections near $\rm{0 \; rad \,m^{-2}}$, for the vast majority of cases the intrinsic polarized signal greatly exceeds the instrumental polarization. 

Future refinements of the polarization pipeline will include an accurate beam model that captures the frequency dependent leakage. This can be done either by using the PB to produce bandpass corrected channelized voltages prior to forming the Stokes parameters, or by including the beam model directly in the Stokes QU-fitting procedure. The latter method has the advantage of retaining flexibility, potentially allowing PB model refinements in the fitting procedure. This refined fitting procedure could also be extended to the polarization analysis of the daily pulsar monitoring program of CHIME/Pulsar \citep{CHIME/PSR2020} which tracks 400 pulsars and covering declinations down to Dec $\approx -20$ degrees. The combined FRB/pulsar data set would greatly benefit ongoing efforts to map CHIME's primary beam \citep{Berger2016} by extending the number of usable calibrator sources. While uncertainties in the intrinsic spectra of FRBs, and to a lesser extent pulsars, prevent their use as calibrators of the absolute gain of the PB, model fits to the polarized leakage from these sources can nonetheless be used to determine the relative gain between the two polarizations. 

\subsection{Ionospheric Corrections}
\label{sec:iono}


At the moment, the CHIME/FRB polarization pipeline makes no attempt to correct for the RM contribution of Earth's ionosphere, $\mathrm{RM_{iono}}$. In general, $|\mathrm{RM_{iono}}|$ will be of order few rad $\rm{m{^{-2}}}$, with the precise value dependent on several factors including: the direction in the sky, geographic location, time of day and activity cycle of the Sun \citep{Mevius2018a}. Variability of this magnitude represents a substantial contribution to the systematic error on any RM measurement. If left uncorrected, FRBs detected by CHIME will be biased by the preferential coverage of the northern hemisphere, such that Earth's bipolar magnetic field will skew the resulting RM distribution. The size of this systematic bias is likely not sufficient to effect the interpretation of overall RM distribution but may be important for certain science questions predicated on a population of FRBs with low $\mathrm{|RM_{extra}|}$ values \citep[e.g., see][]{Hackstein2019, Hackstein2020}. 

In general, ionospheric contributions will be much more important for the interpretation of individual sources rather than the sample as a whole. Specifically, the significance of RM variability observed in bursts from repeating sources will need models that accurately estimate and correct for $\mathrm{RM_{iono}}$. Accurate models allow correlations in the polarized observables and other burst properties to be probed, namely; the correlation between DM and RM can be used to constrain magnetization of the local circumburst medium as has been done for the Vela and Crab pulsars for example \citep{Hamilton1985,Rankin1988}. Accurate ionospheric modelling will only become more relevant as CHIME continues to detect more repeating sources and captures events covering a larger time span where ionospheric conditions are likely to change significantly. Moreover, the recent establishment of periodic activity from repeating sources FRB 20180916B and FRB 20121102A \citep{chime2020,Cruces2020} has motivated consideration of whether these periodicities is replicated in variability of certain burst properties like polarization.      

Estimates of $\mathrm{RM_{iono}}$ are generally obtained from combining a model for Earth's magnetic field with The IONosphere Map EXchange (IONEX) maps \citep{Schaer1999}, describing the ionized turbulent plasma layer in the upper atmosphere. 
There are numerous software packages available that attempt to accurately describe various ionospheric contributions \citep[e.g., ionFR, RMextract;][]{Sotomayor2013, Mevius2018b}.    
Another package, ALBUS\footnote{A.G. Willis; \href{https://github.com/twillis449/ALBUS_ionosphere}{https://github.com/twillis449/ALBUS\_ionosphere}}, developed at the Dominion Radio Astrophysical Observatory (DRAO) that hosts the CHIME telescope, uses readings from local GPS stations. This allows for a higher cadence of calculations and a better sampling of the local variability in the ionosphere that, in theory, should lead to more reliable $\mathrm{RM_{iono}}$ estimates. 

A systematic comparison of the performance of these software packages is planned using CHIME/Pulsar \citep{CHIME/PSR2020} and has already led to improved RM measurements for 80 pulsars \citep{Ng2020}. Preliminary testing has shown RMextract and ALBUS to be in reasonable agreement at elevations greater than 45 degrees but somewhat discrepant at lower elevations (A.G. Willis, priv. comm.). Tracing the source of this discrepancy will be important for CHIME/FRB where the instrument design and the tiling of the 1024 formed skybeams of the real-time search pipeline yield a non-negligible fraction of FRBs detected at lower elevations.


\section{Conclusion}
\label{sec:conclusion}

The baseband system operating on CHIME/FRB will capture several thousand FRB events over the next few years. This dataset will allow greatly enhanced analysis of a wide range of FRB properties. In this paper, we have reviewed the automated pipeline responsible for processing beamformed voltage data into polarized observables such as Faraday rotation measures, polarized fractions, and polarization angle profiles.  

Using a combination of simulated and real FRB events, we compare parametric (QU-fitting) and non-parametric (RM-synthesis) RM detection methods. We find RM-synthesis susceptible to systematic errors introduced by instrumental effects known to affect CHIME observations.
These effects include a delay between the X and Y channelized voltage recordings (cable delay) and a differential response of the two polarizations arising from CHIME's primary beam. We find that the mixing between Stokes $U$ and $V$, induced by cable delay, can lead to RM detections with incorrect signs. Meanwhile, mixing between Stokes $I$ and $Q$ introduced by the differential gain of the primary beam are discussed but generally found to be less significant and are not included in the default QU-fitting of the automated pipeline. 

We report on the polarization properties of two new FRBs, FRB 20191219F and FRB 20200917A. After correcting for systematics, FRB 20191219F is found to be significantly linearly polarized ($>$85\%) with a slightly increasing fraction at the trailing edge of the burst. Meanwhile, a significant circularly polarized component is present but may be the product of unknown systematics. After correcting for the ionospheric RM contribution, we measure the RM of this source to be $\mathrm{RM=6.020 \pm 0.002 \pm 0.050 \; rad \, m^{-2}}$, where the errors represent statistical and ionospheric uncertainties, respectively. Subtracting an estimate of the Milky Way $\mathrm{RM_{MW}}$ contribution yields an excess RM of $\mathrm{30-40 \; rad \, m^{-2}}$ from extragalatic sources of Faraday rotation. FRB 20200917A, meanwhile, displays a significant ($>$50\%) and a slightly increasing polarized fraction at the trailing edge of the burst. The polarization angle displays small but significant variability over the burst phase. After correcting for bandwidth depolarization and a non-zero cable delay, we calculate
$\mathrm{RM=-1294.47 \pm 0.10 \pm 0.05 \; rad \, m^{-2}}$. This is the second largest \textit{unambiguous}\footnote{\citet{Price2019} report an RM detection of $-3163\pm 20 \; \mathrm{rad \; m^{-2}}$ from FRB 20180301A but with significant systematics present.} RM from any FRB source observed to date and suggests a dense magnetized plasma associated with the source.


Finally, we introduce a semi-coherent de-rotation routine that combines phase-coherent techniques for correcting Faraday rotation  \citep[e.g.,][]{vanStraten2002} with conventional incoherent RM detection methods, finding this routine well suited for in cases where the RM of the source is large but not a priori known.
The coherent de-rotation algorithm's effectiveness in retrieving bandwidth depolarized signal is validated with simulated data, as well as with a real event displaying partial depolarization. This semi-coherent RM search routine may find wider applicability to other transient surveys where phase information of the incident polarized signal is retained.

\acknowledgements

The Dunlap Institute is funded through an endowment established by the David Dunlap family and the University of Toronto. 
R.M. recognizes support from the Queen Elizabeth II Graduate Scholarship and the Lachlan Gilchrist Fellowship. 
B.M.G. acknowledges the support of the Natural Sciences and Engineering Research Council of Canada (NSERC) through grant RGPIN-2015-05948, and of the Canada Research Chairs program.
We thank Ziggy Pleunis and Vicky Kaspi for helpful commentary on an early version of this paper.
The polarization analysis pipeline presented here makes use the RMtools package written by Cormac Purcell, and maintained by Cameron Van Eck.     

\bibliographystyle{aasjournal}
\bibliography{references}

\appendix

\renewcommand\thefigure{\thesection.\arabic{figure}}    
\setcounter{figure}{0} 

\section{Instrumental Polarization of CHIME Observations}
\label{sec:systematics}

The refinement stage seeks to improve model fits of the polarized signal by incorporating additional parameters characterizing effects introduced by instrumental systematics. The dominant systematics affecting CHIME-detected FRBs are a time delay and a differential response between the linear X,Y polarizations. The delay is a result of different path lengths of the two polarizations through the system's electronics and results in a frequency dependent phase offset between the two voltage streams. Meanwhile, the differential response is an artefact of differences in the primary beam shape for the two polarizations and is highly dependent on frequency and pointing. 

Both these effects cause mixing between the Stokes parameters that must be accounted for in the fitting routine. The formalism for accounting for these systematics is normally expressed in terms the correlator voltages through the Jones matrix \citep{Hamaker1996}. Since our fitting procedure takes place in Stokes space, we transform these systematics by following the procedure outlined by \citet{Johnston2002} and represent these systematics in Stokes space. We deal first with cable delay and then turn to beam differences.

Beginning with cable delay, the matrix, 

\begin{equation}
\begin{pmatrix}
U^{'} \\
V^{'}
\end{pmatrix}
=
\begin{pmatrix}
\cos (2\pi \nu \tau) & -\sin (2\pi \nu \tau) \\
\sin (2\pi \nu \tau) & \cos (2\pi \nu \tau)
\end{pmatrix}
\begin{pmatrix}
U \\
V
\end{pmatrix}
\label{eqn:tau}
\end{equation}
represents the mixing between Stokes $U$ and $V$ introduced by a time delay, $\tau$, between the two polarizations, and where $\rm{U^{'}}$ and $\rm{V^{'}}$ are the observed values. 

\begin{figure*}
	\centering
\begin{center}
    \includegraphics[width=0.40\textwidth]{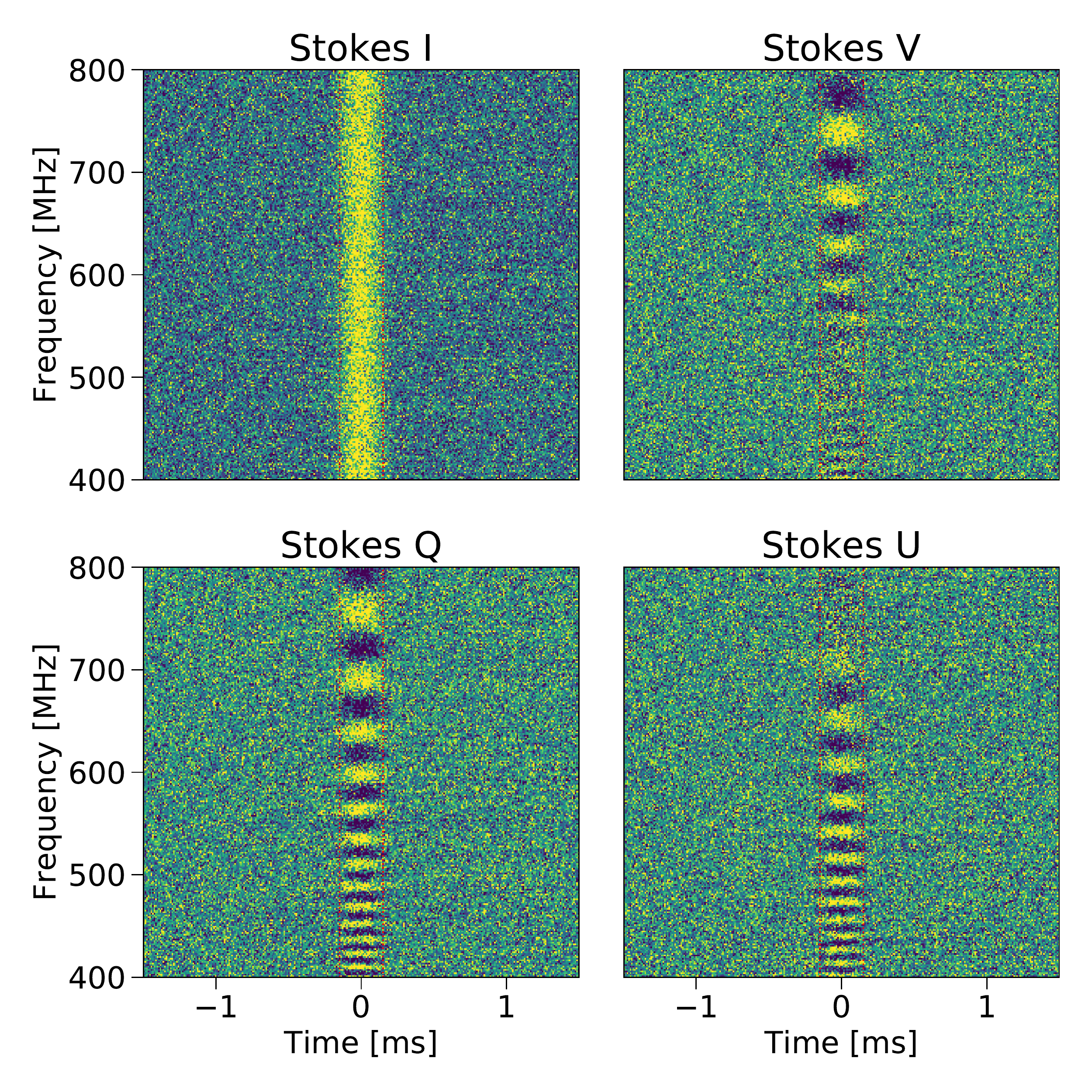} 
    \includegraphics[width=0.40\textwidth]{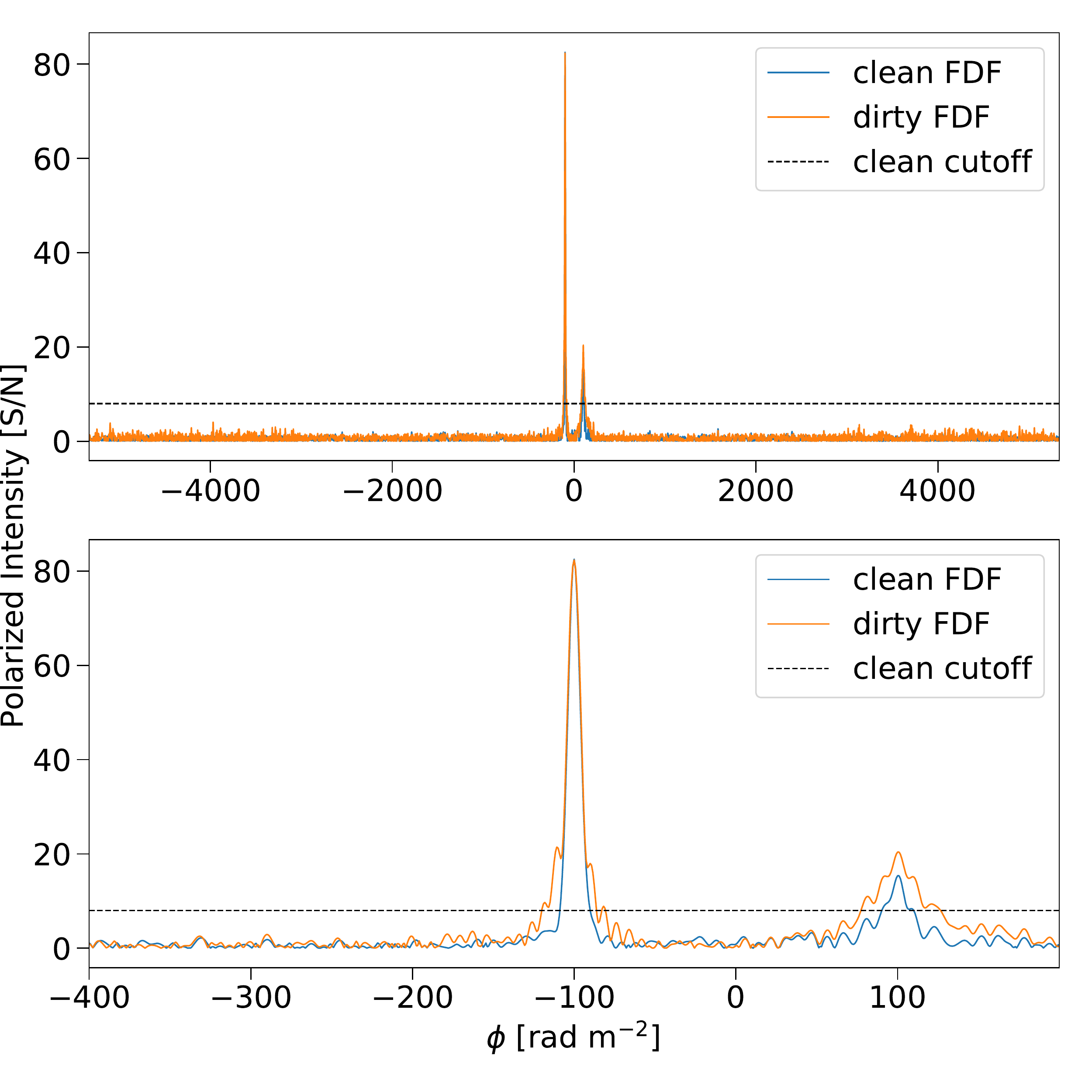} \\
    \includegraphics[width=0.40\textwidth]{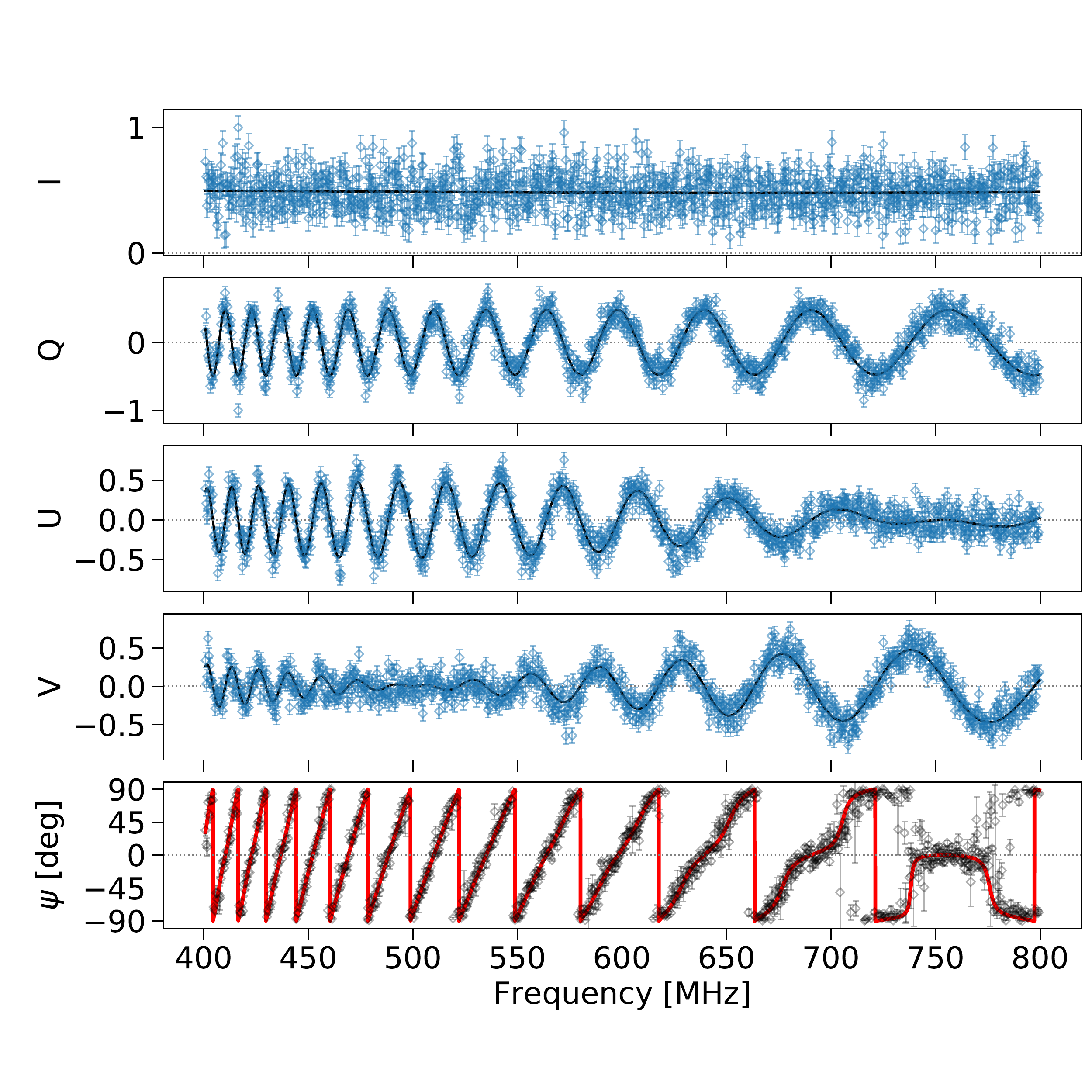}
    \includegraphics[width=0.40\textwidth]{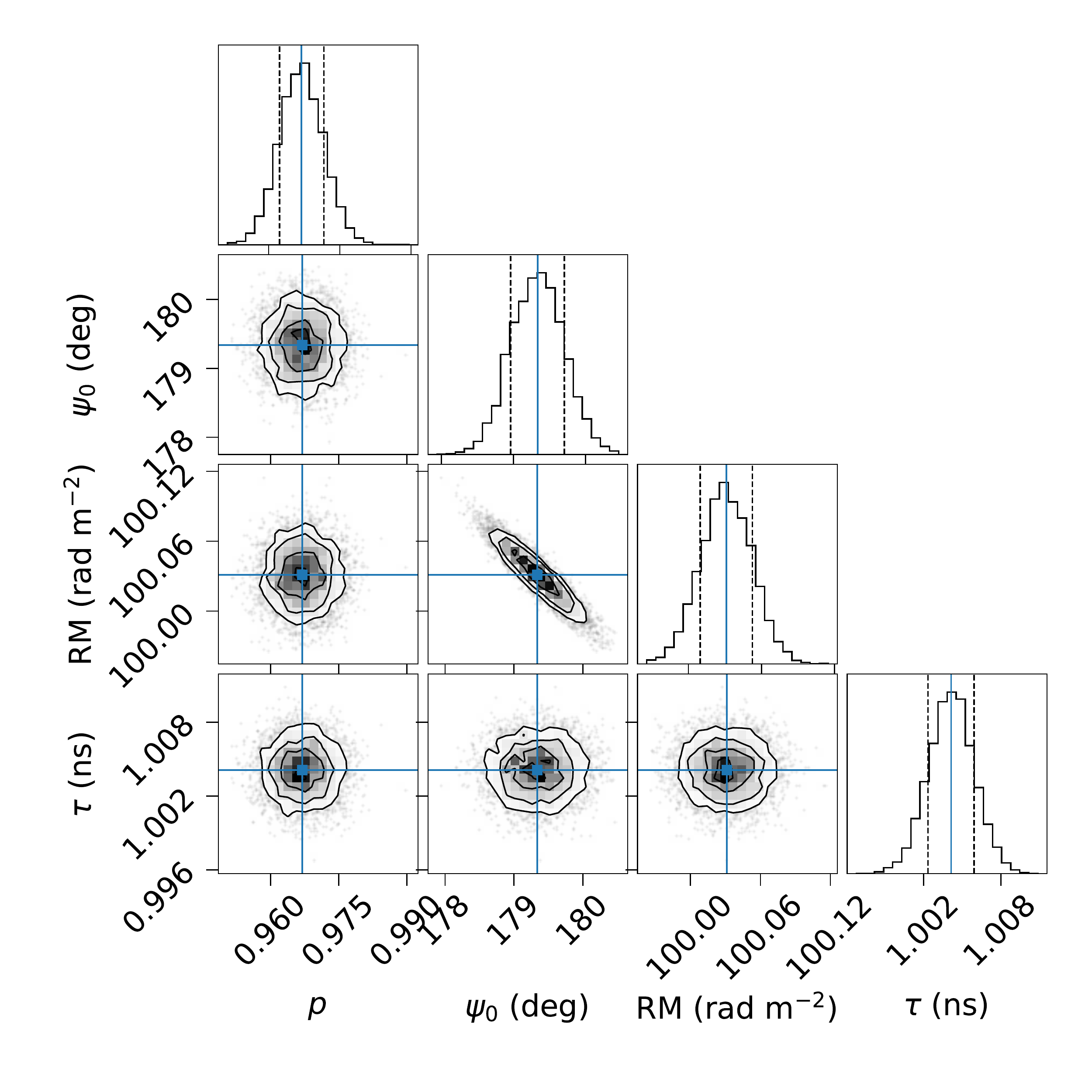}
    \caption{Same as Figure~\ref{fig:sim_example1} but with an additional fitted parameter defining the delay between the two linear voltages ($\mathrm{\tau = 1 \, ns}$) and accounting for it effect through Equation~\ref{eqn:tau}.} 
\label{fig:sim_example2}
\end{center}
\end{figure*}

\begin{figure}[!htb]
	\centering
\begin{center}
    \includegraphics[width=0.40\textwidth]{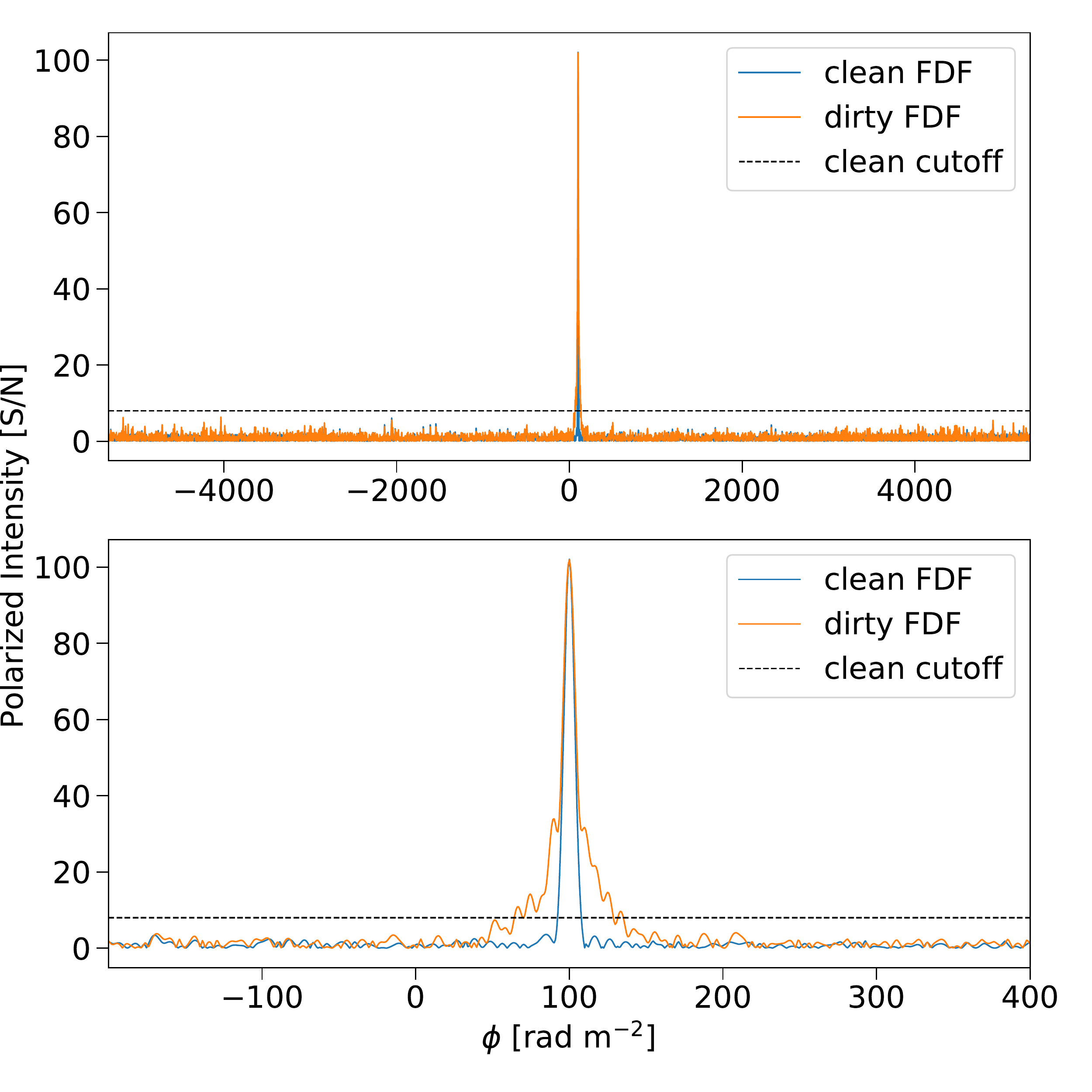} 
    \includegraphics[width=0.40\textwidth]{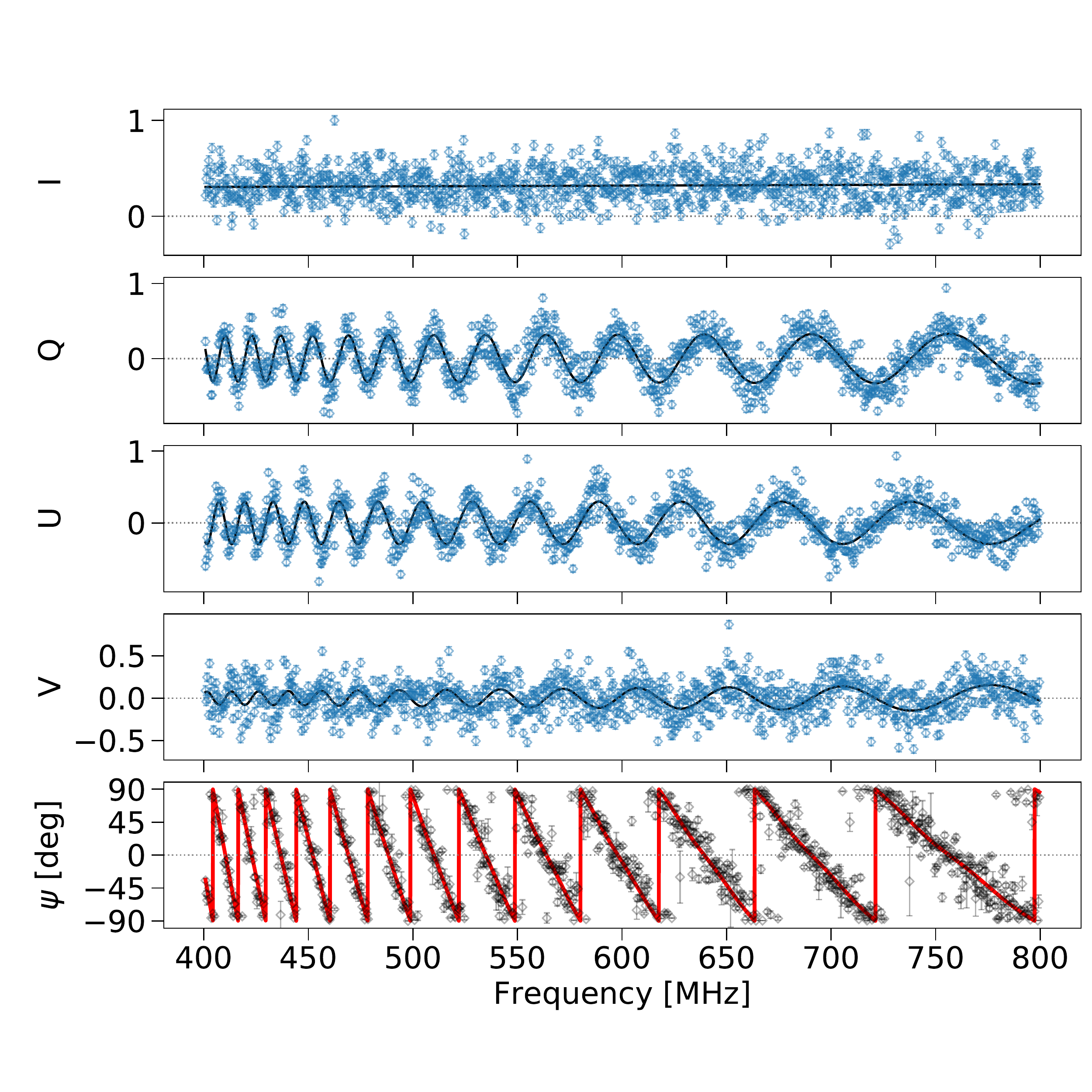} 
    \includegraphics[width=0.40\textwidth]{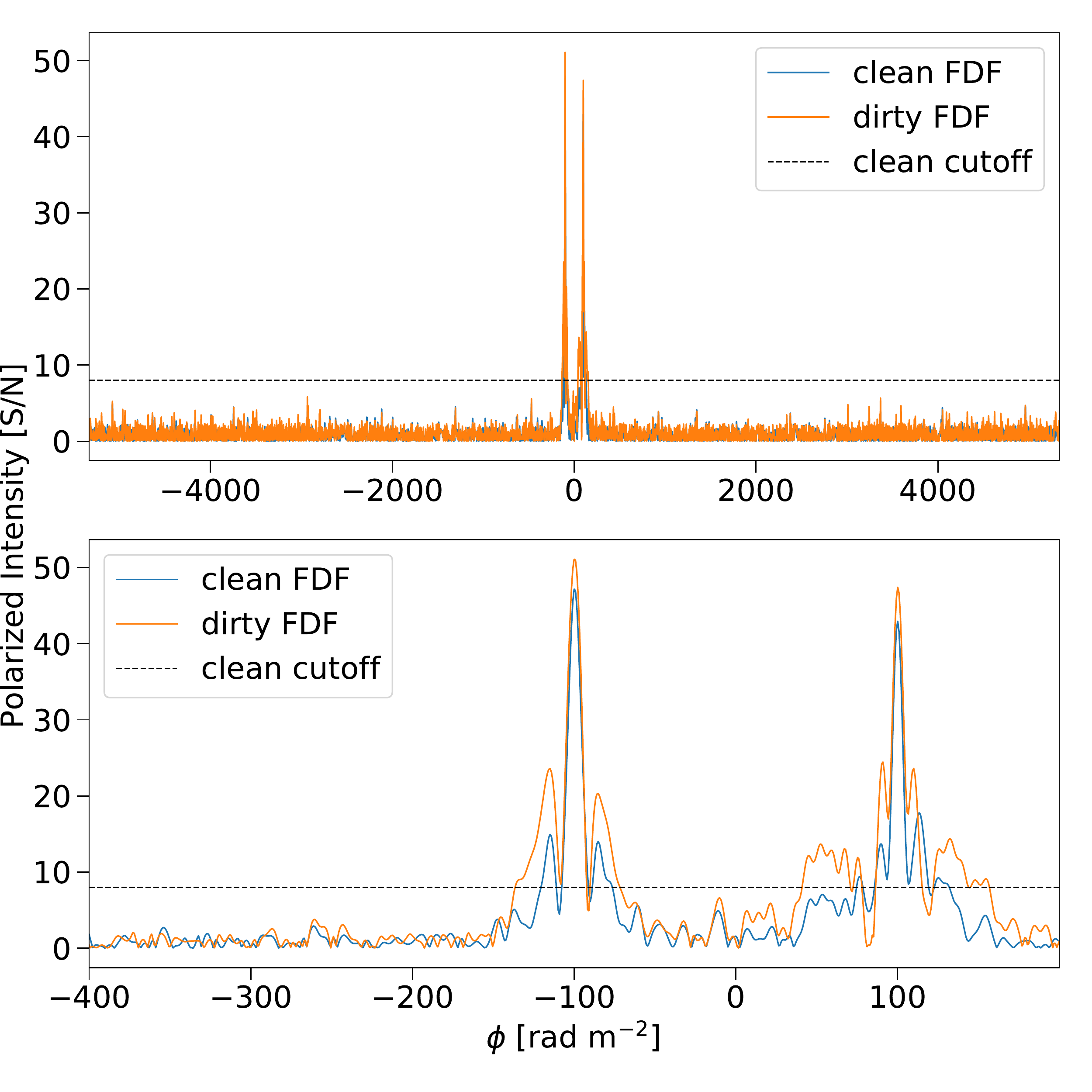} 
    \includegraphics[width=0.40\textwidth]{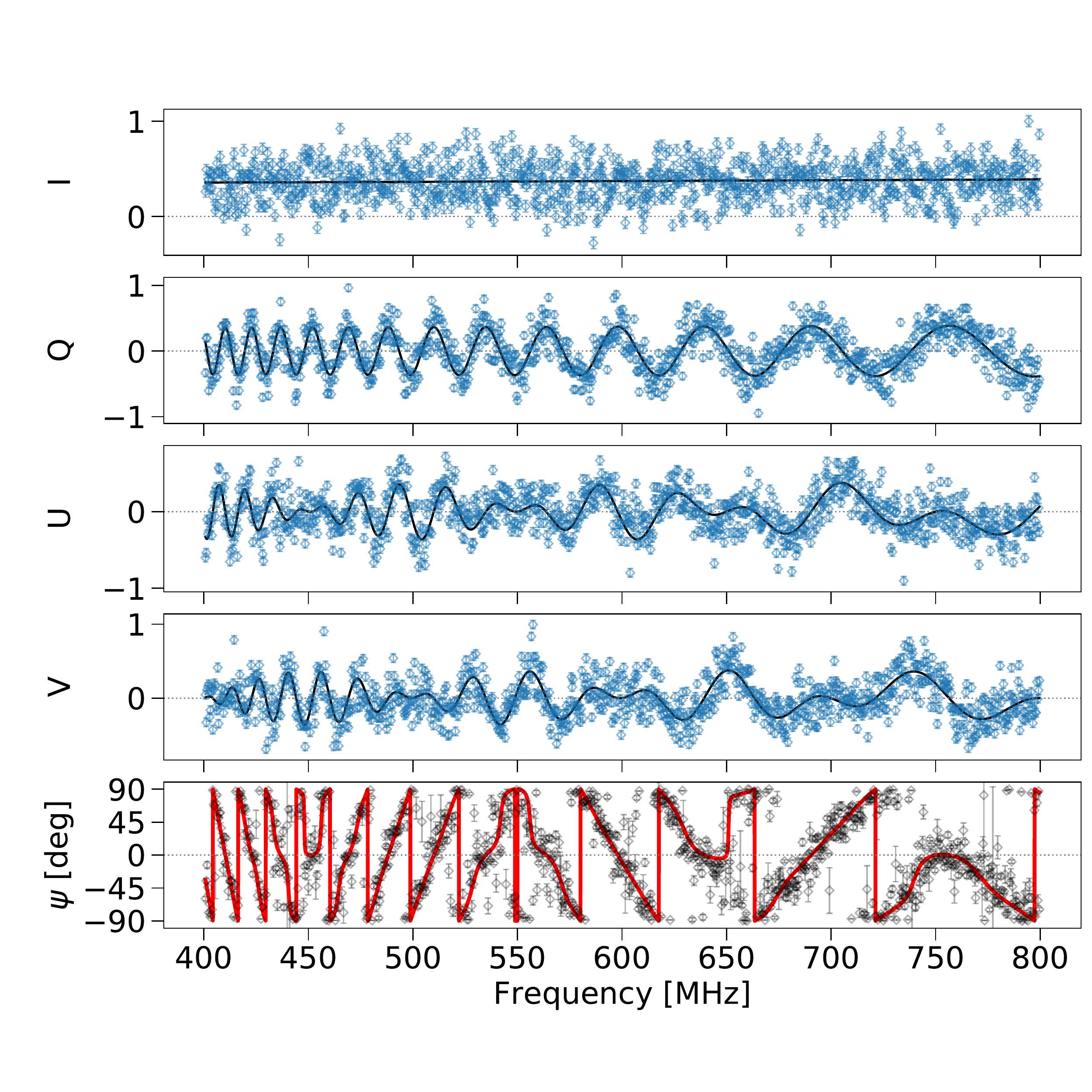} 
    \includegraphics[width=0.40\textwidth]{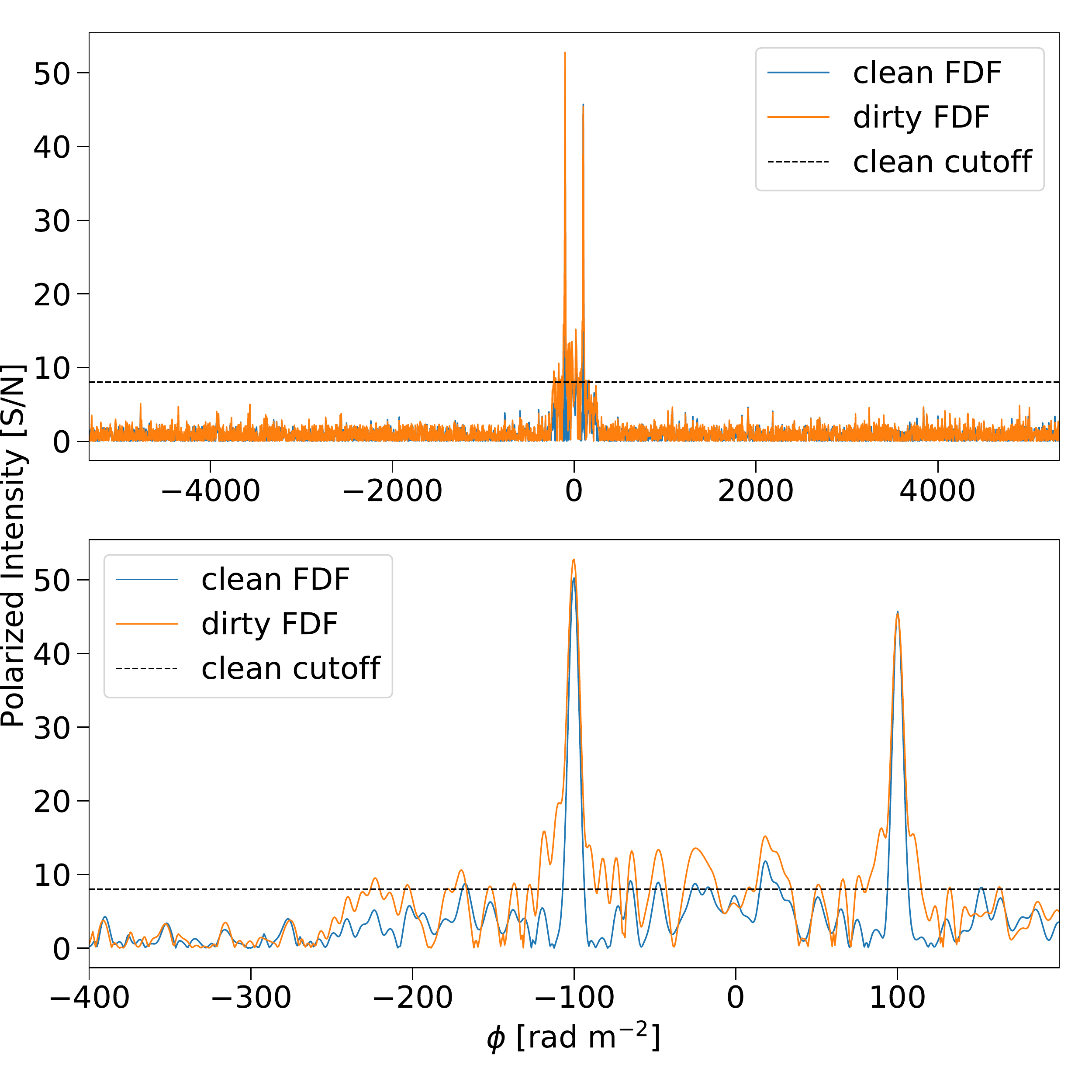} 
    \includegraphics[width=0.40\textwidth]{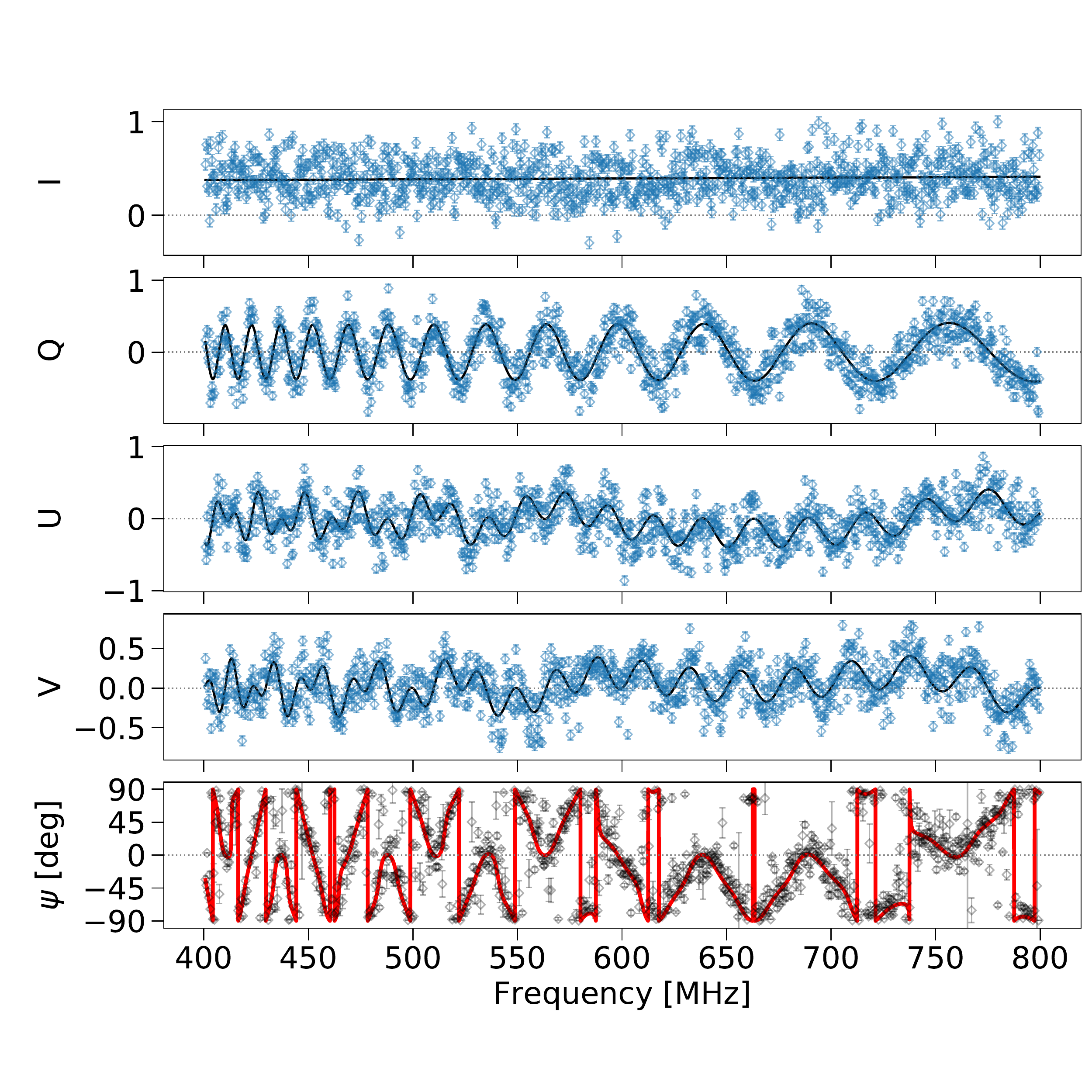} 
    \caption{The effect of cable delay on the observed polarized spectrum of a simulated burst ($p=1.00$, $\psi=180^\circ$ and $\rm{RM=+100 \; rad \, m^{-2}}$). FDFs (left column) and polarized spectra (right column) are produced assuming different cable delays: $\tau=0.1$ ns (top), $\tau=5$ ns (middle) and $\tau=20$ ns (bottom).}
 \label{fig:tautrial}
\end{center}
\end{figure}

A simulated burst with a cable delay of $\mathrm{\tau = 1 \, ns}$ is shown in Figure~\ref{fig:sim_example2}, and clearly shows mixing between Stokes $U$ and $V$ that is correctly fitted for by the model. Interestingly, applying RM-synthesis to this burst results in two distinct peaks in the FDF; one located at near the nominal $\rm{RM \approx +100 \, rad \, m^{-2}}$ and another more prominent aliased peak at $\rm{RM \approx -100 \, rad \, m^{-2}}$. In general, a non-negligible cable delay will always manifest in an aliased $\mathrm{RM}$ ``detection" at roughly the negative of the nominal $\mathrm{RM}$. The reason for this is due to cable delay causing a difference in the polarized signal contained in the real and imaginary part of the complex term, $P(\lambda^2)=Q(\lambda^2)+iU(\lambda^2)$. At the most extreme, applying Equation~\ref{FDF} to a spectrum where $U(\lambda^2)\approx0$ (i.e., complete $U-V$ leakage) results in an FDF that is nearly symmetric with peaks at $\mathrm{\pm RM}$. 
 The amount of polarized intensity that is displaced into the aliased feature is dependent on RM, $\tau$ and the bandpass of the observation. Combined these parameters determine the portion of the burst sub-band where $\frac{d\psi}{d\lambda^2}>0$ and $\frac{d\psi}{d\lambda^2}<0$. Figure~\ref{fig:tautrial} illustrates this, using simulated data at different $\tau$ values to demonstrate how increasing the cable delay effects the polarized spectrum. Specifically, when $|\tau|$ is small (top row), the induced phase shift between the $X$ and $Y$ polarizations is not sufficient to change the sign of $U$ and, thus, the $\rm{RM}$. At larger $\tau$ values (middle and bottom row), multiple phase wrapping occur over the CHIME band and obfuscate the regular $\lambda^2$ scaling of $\psi$, producing not only leakage imprint on Stokes $V$ but also a relative sign change in Stokes $U$ over discrete portions of the band. These effects manifest in the corresponding FDFs as non-negligigle polarized emission off of the modelled $\rm{RM}$ and a greater likelihood of incorrectly determining the sign of the $\rm{RM}$. While this issue can be partially circumvented by judiciously subbanding the data prior to running RM-synthesis, this technique becomes untenable for large $\tau$ values (or large fractional bandwidths) where multiple phase wrappings occur over the band.

Blindly applying RM-synthesis, therefore, in cases where cable delay is present will lead to a significant fraction of the RMs being detected with the wrong sign. Failing to correct for this effect can, therefore, introduce misleading biases in the RM distribution of the FRB population. A thorough understanding of the source of a non-zero $\tau$ parameter is on-going work and will involve mapping any time and position dependence using the FRB baseband sample. A comparison with equivalent analysis from CHIME/Pulsar measurements may reveal interesting differences that may point to phase offsets introduced by the different processing backends of the CHIME/Pulsar and CHIME/FRB experiments, for example through induced Kramers-Kronig phase shifts introduced in each instruments spectrometer \citep[see][for details]{Robishaw2018}. 

Turning to effects of beam differences, properties of the X and Y primary beams can be accounted for by introducing an additional parameter, $\eta$, that corresponds to the ratio of Y polarization sensitivity relative that X. Its effect on the Stokes parameters is to introduce mixing between Stokes $I$ and $Q$. This can be expressed as, 
\begin{equation}
\begin{split}
& I^{'} = \frac{1}{2} \Big[I (1+\eta^2) + Q(1-\eta^2)\Big] \\
& Q^{'} = \frac{1}{2} \Big[I (1-\eta^2) + Q(1+\eta^2)\Big] \\
& U^{'} = \eta U \\
& V^{'} = \eta V 
\end{split}
\label{eqn:eta}
\end{equation}
where $I^{'},Q^{'},U^{'},V^{'}$ are the Stokes parameters modified by $\eta$.

\begin{figure*}
	\centering
\begin{center}
    \includegraphics[width=0.40\textwidth]{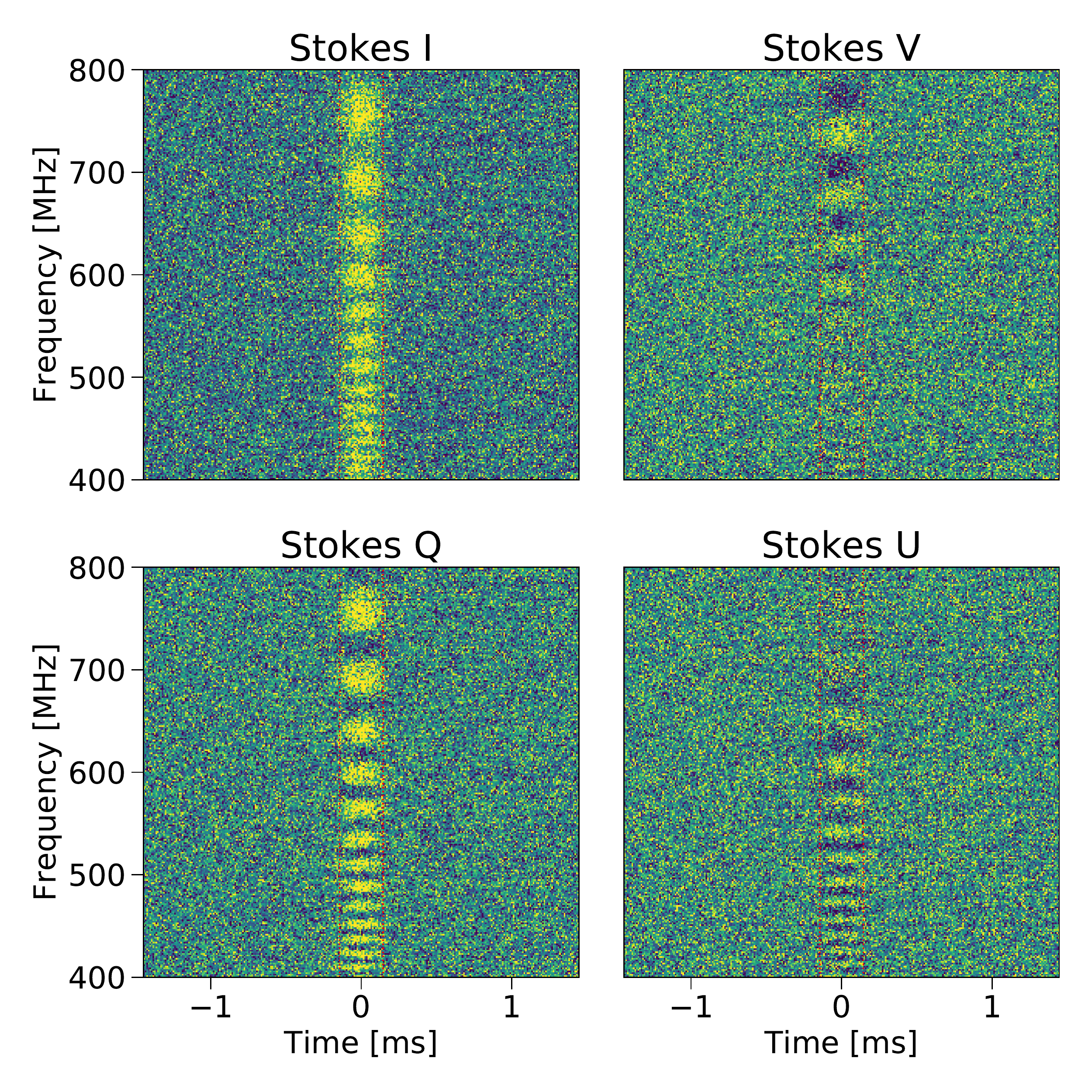} 
    \includegraphics[width=0.40\textwidth]{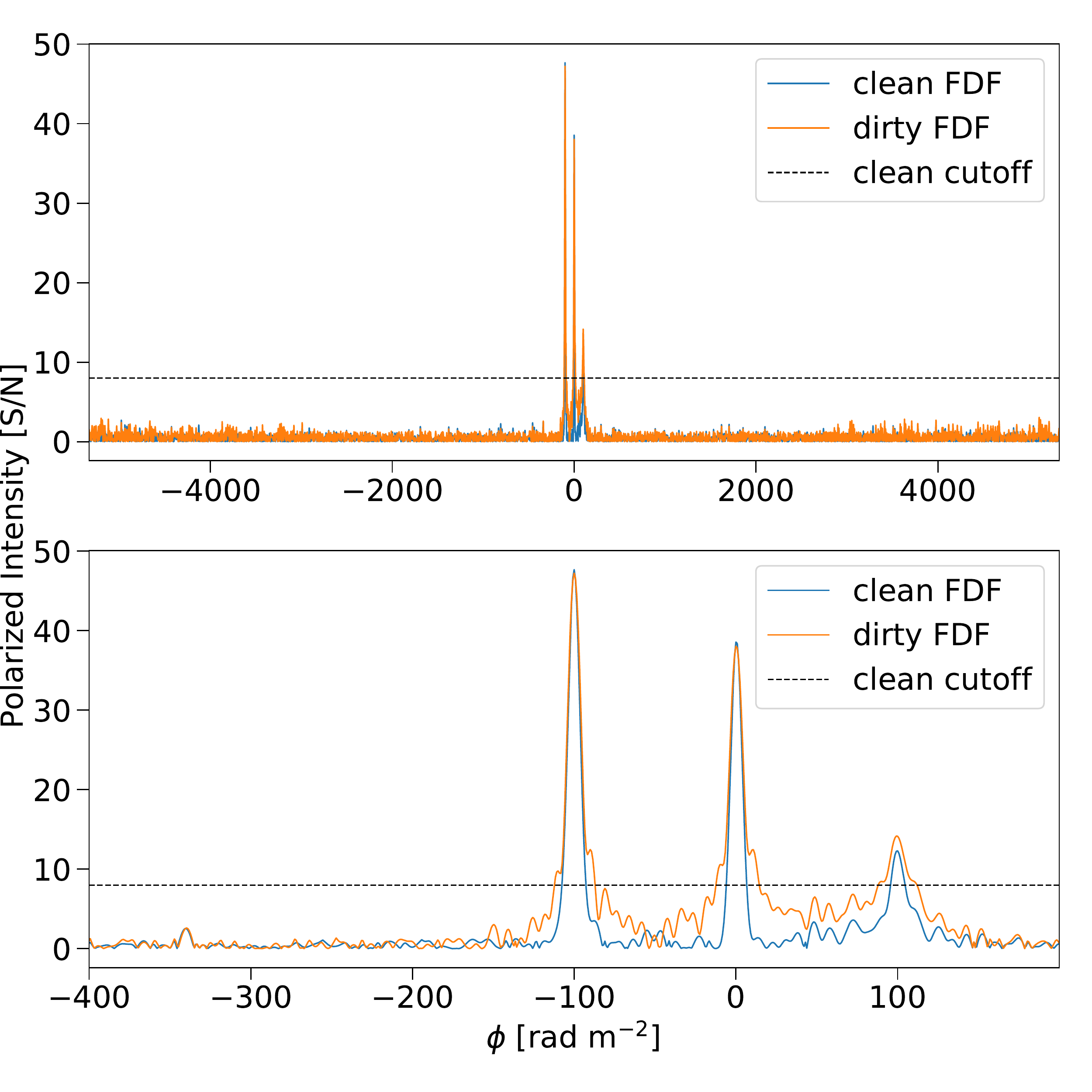} \\
    \includegraphics[width=0.40\textwidth]{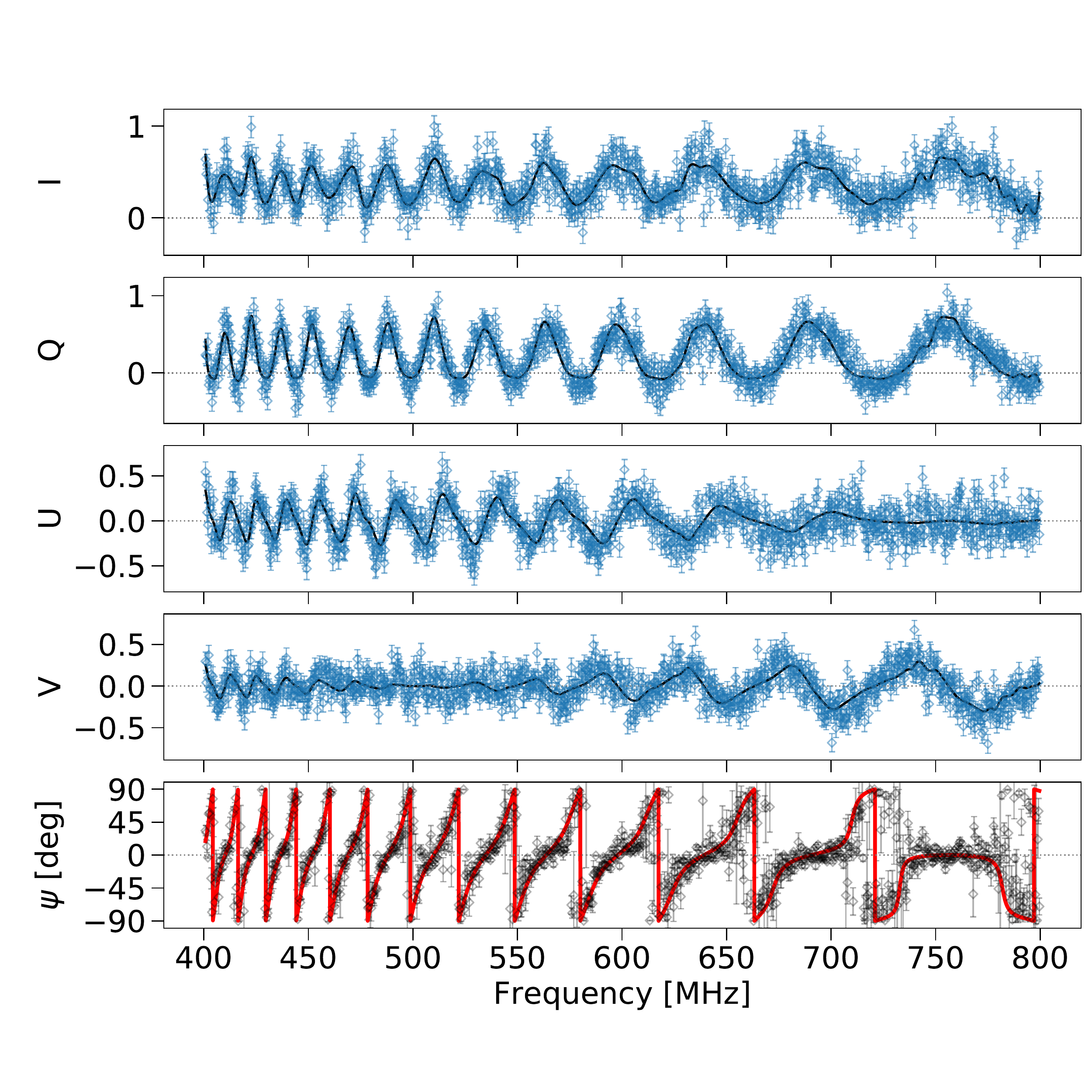}
    \includegraphics[width=0.40\textwidth]{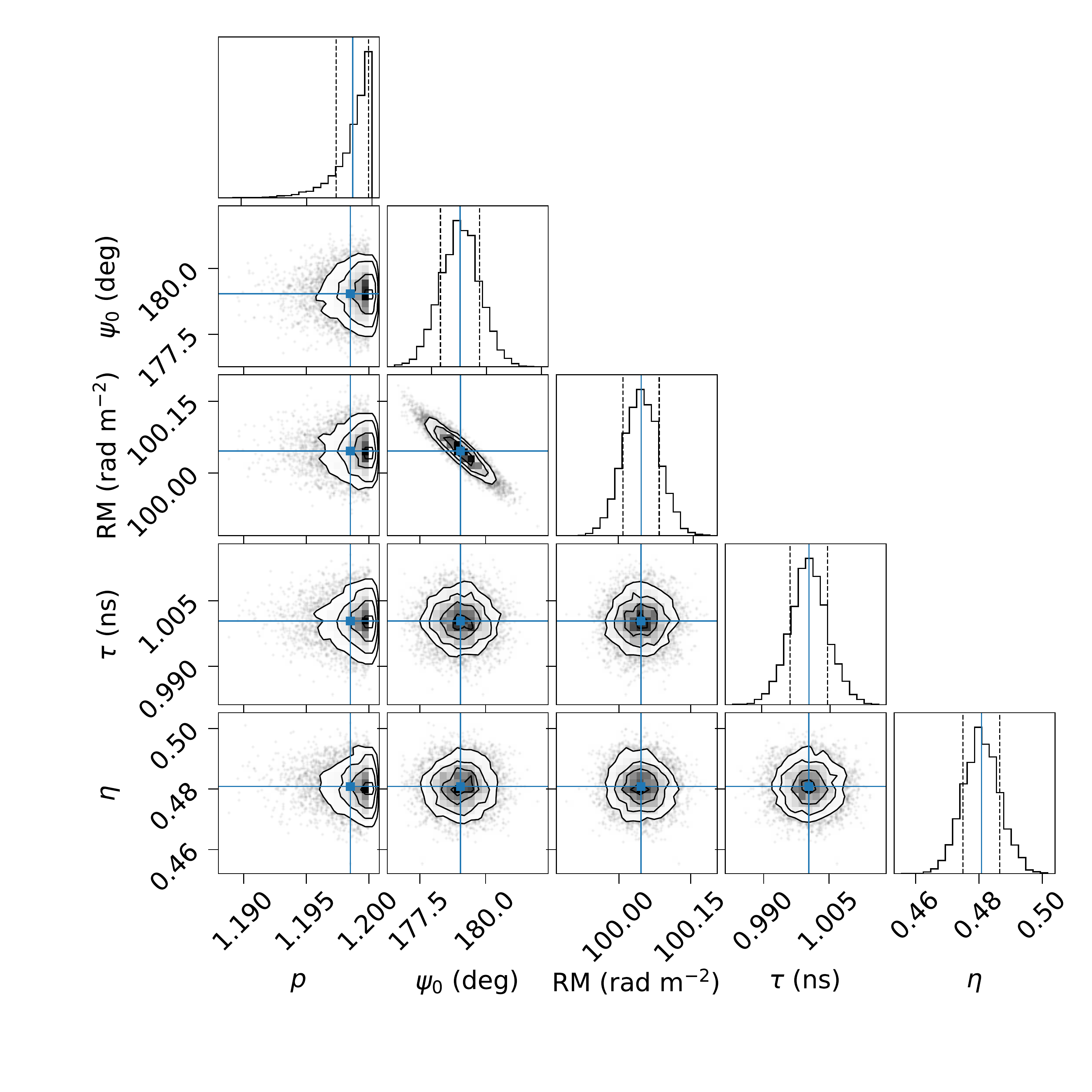}
    \caption{Same as Figure~\ref{fig:sim_example2} but with an additional fitted parameter defining the differential response of the two linear voltages ($\eta$ = 0.5) and accounting for it effect through Equation~\ref{eqn:eta}} 
\label{fig:sim_example3}
\end{center}
\end{figure*}

Figure~\ref{fig:sim_example3} shows an example of such a case where, in addition to a cable delay of $\tau$ = 1 ns, a differential response is modelled and fitted, labelled as ``gain diff"  in the posterior distributions. The corresponding FDF shows that adding a differential response adds additional Faraday complexity. In particular, a differential sensitivity between X and Y polarizations will add an offset to the Stokes $Q$ spectrum, leading to a leakage signal at $\rm{RM \approx 0 \, rad \, m^{-2}}$. This effect is particularly important for RM detections near 0 rad m$^{-2}$, where confusion with leakage can be significant if unaccounted for. Also, the differential sensitivity of the two polarizations is likely to increase far from CHIME's meridian, making events detected in side-lobes particularly vulnerable to this instrumental effect. 

\section{Refined Parameter Measurements}
\label{sec:refine}

Designed for robustness, the CHIME/FRB polarization pipeline does not attempt to characterize variability in the polarized signal as a function of burst duration or frequency. The model currently implemented in the pipeline fits four parameters: $p$, RM, $\psi_0$, $\tau$. An additional parameter, $\eta$, is invoked in cases where the I-Q leakage is significant. While this simple model does a remarkably good job of characterizing the polarized signal of most CHIME detected FRBs, there are cases where additional parameters are needed to re-construct the observed polarized signal. Moreover, even in cases where the polarized signal is well described by a simple model, the significance of any small scale variations in RM or $\psi_0$ in time and/or frequency can be systematically tested by comparing the goodness-of-fit statistics (e.g., reduced chi-squared, Bayesian information criterion) for increasing complex models. This refined analysis is a challenge to automate to the extent that would be required for implementation in the pipeline. Therefore, sections below explore the prospects for model refinement in the context of manual application of QU-fitting and model evaluation.  


\subsection{Frequency Dependence}

\begin{figure*}
	\centering
\begin{center}
    \includegraphics[width=0.45\textwidth]{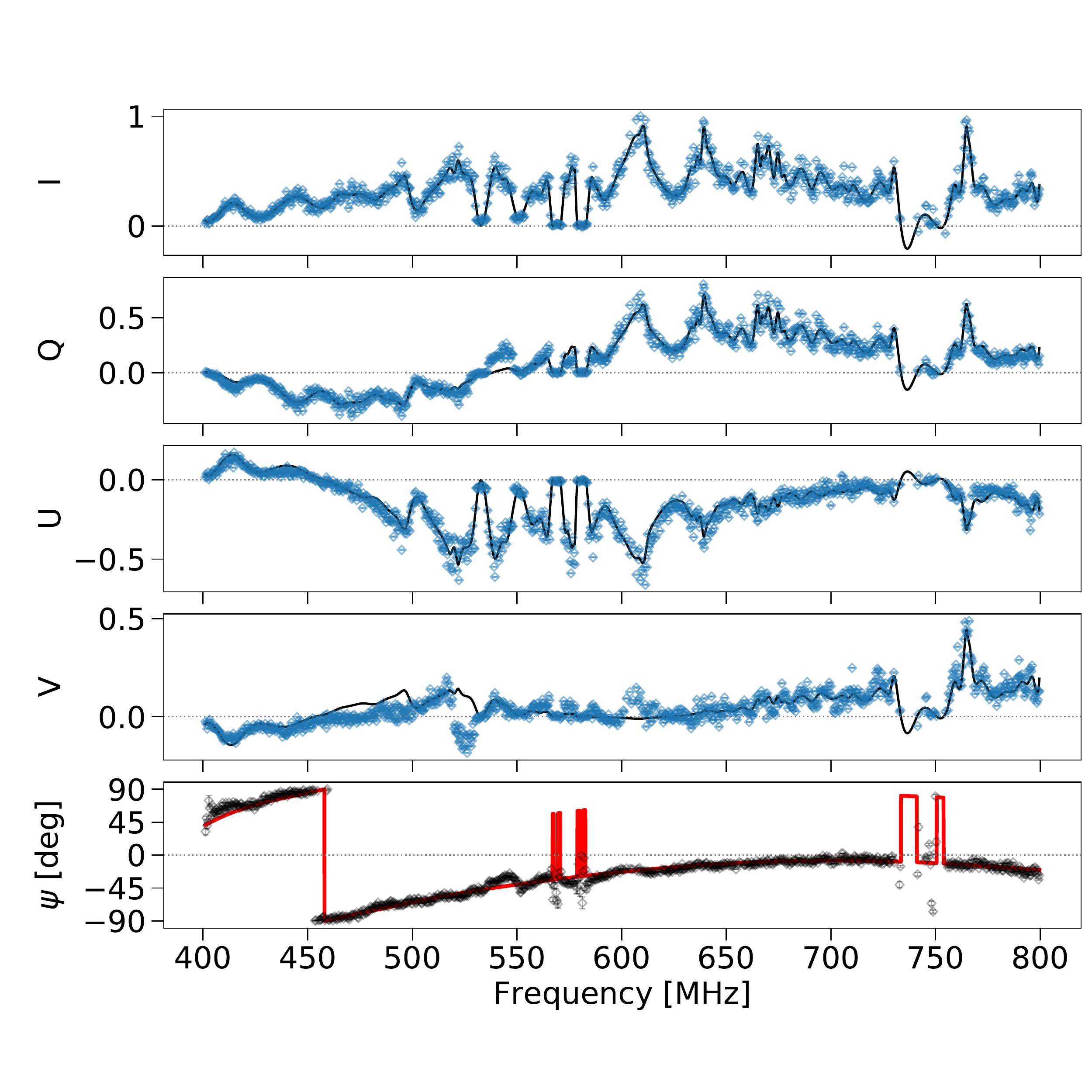}
    \includegraphics[width=0.49\textwidth]{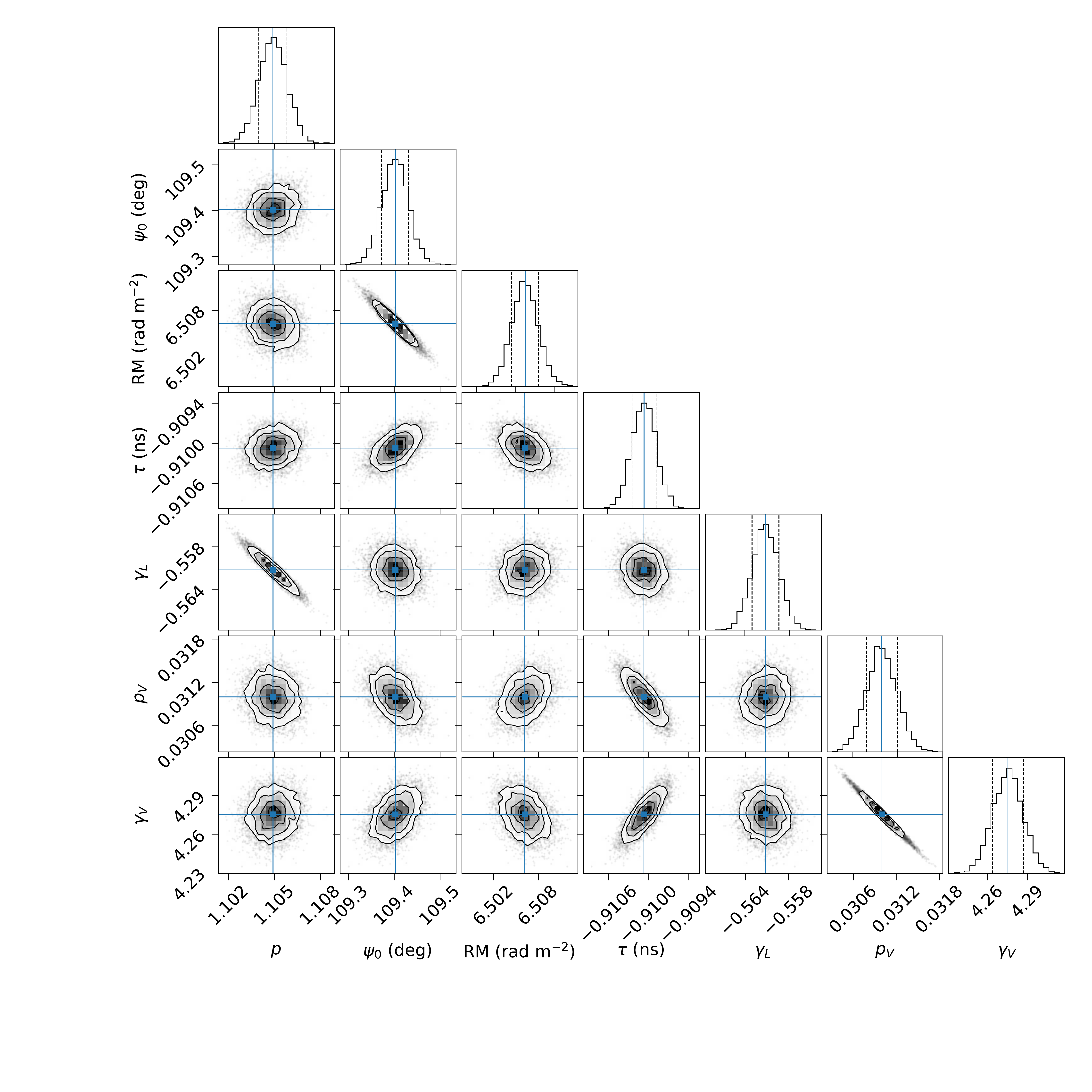}
    \caption{The results of QU-fitting applied to FRB 20191219F using a model that assumes power law spectral behavior in the linear and circular components (see Equation~\ref{eqn:LV_fit}). $\gamma_L$ and $\gamma_V$ correspond to the linear and circular spectral parameters, respectively. $p$ and $p_V$ refer to the polarized fraction of the linear and circular components at the bottom of the band ($\nu \approx 400.391$ MHz).} 
\label{fig:example_refine}
\end{center}
\end{figure*}

In Appendix~\ref{sec:systematics}, we provide details on how the QU-fitting routine is modified to successfully fit for additional parameters that characterize instrumental systematics. Here, we explore how the simple Faraday model of Figure~\ref{fig:sim_example1} can be extended to fit for additional features that are intrinsic to the polarized signal. Figure~\ref{fig:example_refine} shows the fitted spectrum of FRB 20191219F, where the model has been extended to fit for spectral parameters that define the linear and circular polarized signal over the CHIME band. In particular, a power-law spectrum is assumed for the two polarized components, such that, 

\begin{align}
\begin{split}
&p(\nu)=p_0(\frac{\nu}{\nu_0})^{\gamma_L}, \\
&p_V(\nu)=p_{V,0}(\frac{\nu}{\nu_0})^{\gamma_V}. 
\end{split}
\label{eqn:LV_fit}
\end{align}
Here, $p_0$ and $p_{V,0}$ are the linear and circular polarized fractions at the bottom of the burst sub-band. 

Unlike the model currently implemented in the pipeline, this model allows for a non-zero circularly polarized component that is intrinsic to the source and allows both circular and linear polarized fractions to vary across the burst sub-band. Comparing Figure~\ref{fig:example2} with Figure~\ref{fig:example_refine}, we see that the refined model results in a substantially improved fit, particularly at frequencies above 600 MHz, where the default model does a poor job of simultaneously fitting Stokes $U$ and $V$. 

The 2D posterior distributions for the fit parameter of the refined model show substantial degeneracy between ($\mathrm{p_0}$, $\mathrm{p_{V,0}}$) and their respective indices, ($\mathrm{\gamma_L}$, $\mathrm{\gamma_V}$). As is the case for other model parameters, uniform priors are assumed here. 
Inspection of the corner plot reveals that QU-fitting of this refined model leads to $p_0$ converging on an unrealistic value, $p_0>1$. This is likely an artefact of coupling between the $X$, $Y$ polarizations in individual feeds leading to mixing between linear and circular polarized signal. This effect is likely only noticeable for extremely bright events such as the one analyzed here.

\begin{figure}
	\centering
\begin{center}
    \includegraphics[width=0.45\textwidth]{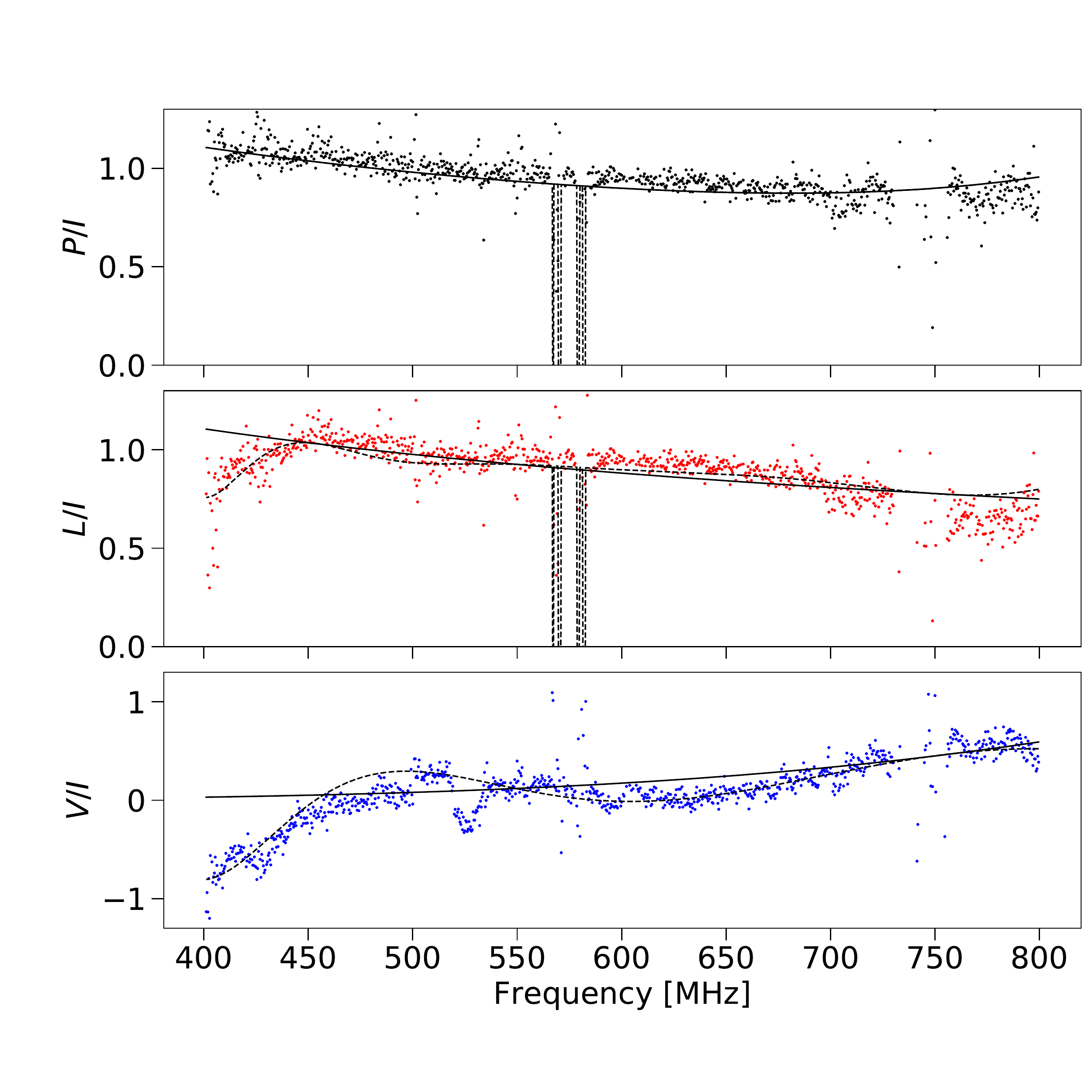}
    \caption{The total (black), linear (red) and circular (blue) polarized fractions across the CHIME band for FRB 20191219F. Solid lines represent the best-fit model of the intrinsic spectrum. Dashed lines correspond to model fits convolved with the systematic of a non-zero cable delay.} 
\label{fig:example_refine2}
\end{center}
\end{figure}

Figure~\ref{fig:example_refine2} shows the total (black), linear (red) and circular (blue) polarized fractions across the burst sub-band. Solid and dashed lines represent the intrinsic model fits before and after convolution with cable delay. Simple power-law models for the linear and circular components do a remarkable job fitting the observed spectrum. Looking at the model fits for the intrinsic spectrum, the burst appears to be $100\%$ linearly polarized near 400 MHz. The steady decrease in the linearly polarized fraction towards higher frequencies seems to indicate that this is intrinsic to the source itself and not a result of differential Faraday rotation through a scattering foreground medium. Interestingly, this loss of linear polarized component at higher frequencies is partially offset by an increase in the circular component, and possibly suggests some relation either through Faraday conversion \citep[e.g.,][]{Gruzinov2019,Vedantham2019} or some other process. 

\end{document}